\documentclass{article}

\usepackage[utf8]{inputenc}
\usepackage[a4paper, total={6in, 8in}]{geometry}
\usepackage{verbatim}
\usepackage{amsfonts}  %
\usepackage{amssymb}
\usepackage{amsmath}
\usepackage{blkarray}
\usepackage{mathtools}  %
\usepackage{xspace}

\usepackage{listings}
\usepackage[usenames,dvipsnames,table]{xcolor}

\usepackage{graphicx}
\usepackage[labelfont=bf]{caption}
\usepackage[format=hang]{subcaption}
\usepackage{floatrow}

\usepackage{paralist}

\usepackage{booktabs}
\usepackage{multirow}
\usepackage{float}
\usepackage{colortbl}
\usepackage{makecell}

\usepackage[numbers,sort,compress]{natbib}

\usepackage[colorlinks,linktoc=all]{hyperref}
\usepackage[all]{hypcap}
\hypersetup{colorlinks}
\hypersetup{linktoc=all}
\hypersetup{citecolor=MidnightBlue}
\hypersetup{linkcolor=black}
\hypersetup{urlcolor=MidnightBlue}

\usepackage[nameinlink]{cleveref}
\creflabelformat{equation}{#1#2#3}
\Crefname{equation}{Eq.}{Eqs.}
\Crefname{assumption}{Assumption}{Assumptions}
\Crefname{prop}{Proposition}{Propositions}
\Crefname{lem}{Lemma}{Lemmas}

\usepackage{algorithm}
\usepackage[noend]{algpseudocode}

\usepackage{amsthm}

\theoremstyle{plain}

\usepackage{newtxtext,newtxmath,microtype}

\usepackage{setspace}
\usepackage{xr}
\makeatletter
\newcommand*{\addFileDependency}[1]{
  \typeout{(#1)}
  \@addtofilelist{#1}
  \IfFileExists{#1}{}{\typeout{No file #1.}}
}
\makeatother

\usepackage[section]{placeins}

\usepackage[colorinlistoftodos,
            prependcaption,
            textsize=scriptsize,
            backgroundcolor=yellow,
            linecolor=lightgray,
            bordercolor=lightgray]{todonotes}

\usepackage{tikzit}
\usepackage{amsmath,amsfonts}
\usepackage{keycommand}
\usepackage{hyperref}

\usepackage{color}

\definecolor{zxredfg}{RGB}{100,0,0}
\definecolor{zxgreenfg}{RGB}{0,50,0}

\definecolor{zxred}{RGB}{232, 165, 165}
\definecolor{zxgreen}{RGB}{216, 248, 216}
\definecolor{zxhad}{RGB}{255, 255, 130}
\definecolor{zxpaulibox}{RGB}{221, 221, 255}

\newcommand{\ket}[1]{\ensuremath{|#1\rangle\xspace}}

\def\<{\langle}
\def\>{\rangle}
\mathchardef\mhyphen="2D

\newcommand{\CNOT}{\text{CNOT}\xspace}

\newcommand{\CZ}{\text{CZ}\xspace}

\newcommand{\R}{\mathbb{R}}

\usetikzlibrary{decorations.pathmorphing}
\usetikzlibrary{fadings}
\usetikzlibrary{matrix}
\usetikzlibrary{decorations.pathreplacing}
\usetikzlibrary{decorations.markings}

\newcommand{\zxgflabel}{\textbf{\textit{gf}}}

\tikzset{meter/.append style={draw, inner sep=5, rectangle, font=\vphantom{A}, minimum width=20, line width=.4,
 path picture={\draw[black] ([shift={(.1,.2)}]path picture bounding box.south west) to[bend left=70] ([shift={(-.1,.2)}]path picture bounding box.south east);\draw[black,-latex] ([shift={(0,.1)}]path picture bounding box.south) -- ([shift={(.25,-.05)}]path picture bounding box.north);}}}

\tikzstyle{gate}=[shape=rectangle, text height=1.5ex, text depth=0.25ex, yshift=0.5mm, fill=white, draw=black, minimum height=3mm, yshift=-0.5mm, minimum width=3mm, font={\small}, tikzit category=circuit, inner sep=2pt]
\tikzstyle{big gate}=[shape=rectangle, text height=1.5ex, text depth=0.25ex, yshift=0.5mm, fill=white, draw=black, minimum height=10mm, yshift=-0.5mm, minimum width=5mm, font={\small}, tikzit category=circuit]
\tikzstyle{Z dot}=[inner sep=0mm, minimum size=2mm, shape=circle, draw=black, fill={rgb,255: red,221; green,255; blue,221}, tikzit category=zx]
\tikzstyle{Z phase dot}=[minimum size=5mm, font={\footnotesize\boldmath}, shape=rectangle, rounded corners=2mm, inner sep=0.2mm, outer sep=-2mm, scale=0.8, tikzit shape=circle, draw=black, fill={rgb,255: red,221; green,255; blue,221}, tikzit draw=blue, tikzit category=zx]
\tikzstyle{X dot}=[Z dot, shape=circle, draw=black, fill={rgb,255: red,255; green,136; blue,136}, tikzit category=zx]
\tikzstyle{X phase dot}=[Z phase dot, tikzit shape=circle, tikzit draw=blue, fill={rgb,255: red,255; green,136; blue,136}, font={\footnotesize\boldmath}, tikzit category=zx]
\tikzstyle{hadamard}=[fill=yellow, draw=black, shape=rectangle, inner sep=0.6mm, minimum height=1.5mm, minimum width=1.5mm, tikzit category=zx]
\tikzstyle{paulibox}=[fill={rgb,255: red,221; green,221; blue,255}, draw=black, shape=rectangle, inner sep=0.6mm, minimum height=5mm, minimum width=5mm, font={\footnotesize}, text height=1.5ex, text depth=0.25ex, tikzit category=zx]
\tikzstyle{vertex}=[inner sep=0mm, minimum size=1mm, shape=circle, draw=black, fill=black, tikzit category=misc]
\tikzstyle{vertex set}=[inner sep=0mm, minimum size=1mm, shape=circle, draw=black, fill=white, font={\footnotesize\boldmath}, tikzit category=misc]
\tikzstyle{small black dot}=[fill=black, draw=black, shape=circle, inner sep=0pt, minimum width=1.2mm, tikzit category=circuit]
\tikzstyle{cnot ctrl}=[fill=black, draw=black, shape=circle, inner sep=0pt, minimum width=1.2mm, tikzit category=circuit]
\tikzstyle{cnot targ}=[fill=white, draw=white, shape=circle, tikzit category=circuit, label={center:$\oplus$}, inner sep=0pt, minimum width=2.1mm, tikzit fill={rgb,255: red,102; green,204; blue,255}, tikzit draw=black]
\tikzstyle{ket}=[fill=white, draw=black, shape=regular polygon, regular polygon sides=3, regular polygon rotate=-30, scale=0.7, inner sep=1pt, tikzit category=circuit, tikzit shape=rectangle, tikzit fill=green]
\tikzstyle{bra}=[fill=white, draw=black, shape=regular polygon, regular polygon sides=3, regular polygon rotate=30, scale=0.7, inner sep=1pt, tikzit category=circuit, tikzit shape=rectangle, tikzit fill=red]
\tikzstyle{scalar}=[shape=rectangle, text height=1.5ex, text depth=0.25ex, yshift=0.5mm, fill=white, draw=black, minimum height=5mm, yshift=-0.5mm, minimum width=5mm, font={\small}]
\tikzstyle{clabel}=[fill=white, draw=none, shape=rectangle, tikzit fill={rgb,255: red,56; green,255; blue,242}, font={\footnotesize}, inner sep=1pt, tikzit category=labels]
\tikzstyle{empty diagram}=[draw={gray!40!white}, dashed, shape=rectangle, minimum width=1cm, minimum height=1cm, tikzit category=misc]
\tikzstyle{amap}=[fill=white, draw=black, shape=NEbox, tikzit category=asymmetric, tikzit fill=yellow, tikzit shape=rectangle]
\tikzstyle{amap conj}=[fill=white, draw=black, shape=NWbox, tikzit category=asymmetric, tikzit fill=green, tikzit shape=rectangle]
\tikzstyle{amap adj}=[fill=white, draw=black, shape=SEbox, tikzit category=asymmetric, tikzit fill=red, tikzit shape=rectangle]
\tikzstyle{amap trans}=[fill=white, draw=black, shape=SWbox, tikzit category=asymmetric, tikzit fill=orange, tikzit shape=rectangle]
\tikzstyle{astate}=[fill=white, draw=black, shape=NEtriangle, tikzit category=asymmetric, tikzit shape=circle, tikzit fill=yellow]
\tikzstyle{astate conj}=[fill=white, draw=black, shape=NWtriangle, tikzit category=asymmetric, tikzit shape=circle, tikzit fill=green]
\tikzstyle{astate adj}=[fill=white, draw=black, shape=SEtriangle, tikzit category=asymmetric, tikzit shape=circle, tikzit fill=red]
\tikzstyle{astate trans}=[fill=white, draw=black, shape=SWtriangle, tikzit category=asymmetric, tikzit shape=circle, tikzit fill=orange]
\tikzstyle{white dot}=[inner sep=0mm, minimum size=2mm, shape=circle, draw=black, fill={rgb,255: red,250; green,250; blue,250}]
\tikzstyle{white phase dot}=[minimum size=5mm, font={\footnotesize\boldmath}, shape=rectangle, rounded corners=2mm, inner sep=0.2mm, outer sep=-2mm, scale=0.8, tikzit shape=circle, draw=black, fill={rgb,255: red,250; green,250; blue,250}, tikzit draw=blue]
\tikzstyle{hbox}=[shape=rectangle, text height=2mm, fill={rgb,255: red,255; green,235; blue,61}, draw=black, minimum height=2mm, minimum width=2mm, font={\small}, tikzit category=zh, inner sep=0pt, rounded corners=0.5mm]
\tikzstyle{Z dot (zh)}=[inner sep=0mm, minimum size=2mm, shape=circle, draw=black, fill={rgb,255: red,250; green,250; blue,250}, tikzit category=zh]
\tikzstyle{X dot (zh)}=[Z dot, shape=circle, draw=black, fill={rgb,255: red,193; green,193; blue,193}, tikzit category=zh]
\tikzstyle{triangle}=[fill={rgb,255: red,255; green,136; blue,136}, draw=black, shape=isosceles triangle, isosceles triangle apex angle=60, minimum size=2.5mm, inner sep=0mm]
\tikzstyle{labelled hbox}=[shape=rectangle, text height=1.75ex, text depth=0.5ex, fill={rgb,255: red,255; green,235; blue,61}, draw=black, minimum height=3mm, minimum width=4mm, font={\small}, tikzit category=zh, inner sep=1.3pt, rounded corners=0.5mm]
\tikzstyle{Z phase dot (zh)}=[Z phase dot, tikzit shape=circle, tikzit draw=blue, fill={rgb,255: red,250; green,250; blue,250}, font={\footnotesize\boldmath}, tikzit category=zh]
\tikzstyle{X phase dot (zh)}=[Z phase dot, tikzit shape=circle, tikzit draw=blue, fill={rgb,255: red,193; green,193; blue,193}, font={\footnotesize\boldmath}, tikzit category=zh]
\tikzstyle{W node}=[fill=black, draw=black, shape=regular polygon, regular polygon sides=3, minimum size=2mm]
\tikzstyle{Z dot (zw)}=[fill=white, draw=black, shape=circle, minimum width=1.2mm, inner sep=0pt]
\tikzstyle{Z phase dot XL}=[Z phase dot, fill={rgb,255: red,250; green,250; blue,250}, draw=black, shape=circle, tikzit draw={rgb,255: red,191; green,0; blue,64}, tikzit shape=circle, font={\large\boldmath}, inner sep=0.0mm]

\tikzstyle{hadamard edge}=[-, dashed, dash pattern=on 2pt off 0.5pt, thick, draw={rgb,255: red,68; green,136; blue,255}]
\tikzstyle{box edge}=[-, dashed, dash pattern=on 2pt off 0.5pt, thick, draw={rgb,255: red,203; green,192; blue,225}]
\tikzstyle{brace edge}=[-, tikzit draw=blue, decorate, decoration={brace,amplitude=1mm,raise=-1mm}]
\tikzstyle{diredge}=[->, thick]
\tikzstyle{double edge}=[-, double, shorten <=-1mm, shorten >=-1mm, double distance=2pt]
\tikzstyle{gray edge}=[-, {gray!60!white}]
\tikzstyle{pointer edge}=[->, very thick, gray]
\tikzstyle{boldedge}=[-, line width=1.0pt, shorten <=-0.17mm, shorten >=-0.17mm]
\tikzstyle{bidir edge}=[<->, very thick, draw={rgb,255: red,191; green,191; blue,191}]
\tikzstyle{purple edge}=[->, thick, draw={rgb,255: red,225; green,117; blue,216}]
\tikzstyle{green edge}=[->, thick, draw={rgb,255: red,167; green,231; blue,137}]
\tikzstyle{orange edge}=[->, thick, draw={rgb,255: red,245; green,170; blue,63}]
\tikzstyle{blue edge}=[->, thick, draw={rgb,255: red,68; green,136; blue,255}]
\tikzstyle{any edge}=[->, thick, draw=cyan]
\tikzstyle{red edge}=[->, thick, draw={rgb,255: red,255; green,136; blue,136}]
\tikzstyle{bidiredge}=[<->, thick]
\tikzstyle{dashed diredge}=[->, dashed, dash pattern=on 1pt off 0.5pt]
\tikzstyle{bidashed diredge}=[<->, dashed, dash pattern=on 1pt off 0.5pt]

\newcommand{\alphatensor}{AlphaTensor\xspace}
\newcommand{\game}{TensorGame\xspace}
\newcommand{\method}{AlphaTensor-Quantum\xspace}
\newcommand{\tcount}{T-count\xspace}
\newcommand{\tgate}[1]{T gate{#1}\xspace}
\newcommand{\C}{\mathbb{C}}
\newcommand{\CCZmat}{\textrm{\emph{CCZ}}}
\newcommand{\CNOTmat}{\textrm{\emph{CNOT}}}
\newcommand{\CZmat}{\textrm{\emph{CZ}}}

\newcommand{\Tofmat}{\textrm{\emph{Tof}}}
\newcommand{\CCZ}{\textrm{CCZ}\xspace}
\newcommand{\CS}{\textrm{CS}\xspace}
\newcommand\ee{\mathrm{e}}

\usepackage{tikz}
\usetikzlibrary{quantikz}

\newcommand\blfootnote[1]{%
  \begingroup
  \renewcommand\thefootnote{}\footnote{#1}%
  \addtocounter{footnote}{-1}%
  \endgroup
}

\newcommand{\prepare}{\mbox{P{\scriptsize REPARE}}}
\newcommand{\prep}{\mbox{P{\scriptsize REP}}}
\newcommand{\select}{\mbox{S\scriptsize ELECT}}
\newcommand{\sel}{\mbox{S\scriptsize EL}}

\author{%
Francisco J.\ R.\ Ruiz$^{*,1}$\and
Tuomas Laakkonen$^{*,2}$ \and
Johannes Bausch$^1$ \and
Matej Balog$^1$ \and
Mohammadamin Barekatain$^1$ \and
Francisco J.\ H.\ Heras$^1$ \and
Alexander Novikov$^1$ \and
Nathan Fitzpatrick$^3$ \and
Bernardino Romera-Paredes$^1$ \and
John van de Wetering$^4$ \and
Alhussein Fawzi$^1$ \and
Konstantinos Meichanetzidis$^2$ \and
Pushmeet Kohli$^1$ \\[0.5em] \begin{tabular}{l} 
    {\small $^1$ Google DeepMind, 6-8 Handyside Street, London N1C 4UZ, UK} \\[-0.2em] 
    {\small $^2$ Quantinuum, 17 Beaumont Street, Oxford OX1 2NA, UK} \\[-0.2em] 
    {\small $^3$ Quantinuum, Terrington House, 13–15 Hills Road, Cambridge CB2 1NL, UK} \\[-0.2em] 
    {\small $^4$ Informatics Institute, University of Amsterdam, 1098 XH Amsterdam, NL}
\end{tabular}}

\date{}

\title{Quantum Circuit Optimization with AlphaTensor}

\begin{document}

\maketitle

\begin{abstract}
    A key challenge in realizing fault-tolerant quantum computers is circuit optimization. Focusing on the most expensive gates in fault-tolerant quantum computation (namely, the \tgate{s}), we address the problem of \tcount optimization, i.e., minimizing the number of \tgate{s} that are needed to implement a given circuit. To achieve this, we develop \method, a method based on deep reinforcement learning that exploits the relationship between optimizing \tcount and tensor decomposition. Unlike existing methods for \tcount optimization, \method can incorporate domain-specific knowledge about quantum computation and leverage \emph{gadgets}, which significantly reduces the \tcount of the optimized circuits. \method outperforms the existing methods for \tcount optimization on a set of arithmetic benchmarks (even when compared without making use of gadgets). Remarkably, it discovers an efficient algorithm akin to Karatsuba's method for multiplication in finite fields. \method also finds the best human-designed solutions for relevant arithmetic computations used in Shor's algorithm and for quantum chemistry simulation, thus demonstrating it can save hundreds of hours of research by optimizing relevant quantum circuits in a fully automated way.
\end{abstract}

\blfootnote{$^*$Equal contributors.}

\section{Introduction}
\label{sec:introduction}

Quantum computation presents a fundamentally new approach to solving computational problems. Since its inception \citep{feynman1982simulating,YuriManinQuantumComputation}, many potential applications in various fields have been proposed, including cryptography \citep{Shor_1997}, drug discovery \citep{Blunt2022}, and materials science and high energy physics \citep{dalzell2023quantum}. Yet, fault-tolerant quantum computation introduce some expensive components that have a significant impact on the overall runtime and resource cost \citep{bravyi2005universal,Campbell_2017}; thus it is important to minimize the use of these components in order to enable the execution of large computations that address these real-world problems.

To be able to implement any quantum algorithm, universal quantum computers require a combination of two types of quantum gates: Clifford (e.g., Hadamard or CNOT) and non-Clifford gates (e.g., the \tgate). Non-Clifford gates are required to achieve universality \citep{gidney2019efficient}; indeed, a quantum algorithm implemented using Clifford gates alone can be simulated on a classical computer efficiently, i.e., in polynomial time (Gottesman–Knill theorem \citep{gottesman1998heisenger}).
Unfortunately, non-Clifford gates are the expensive components, as they are notoriously harder to implement than Clifford gates. This is because fault-tolerant quantum computation requires error correction schemes \citep{Calderbank_1996,aharonov1996fault}, whose implementation requires distilling magic states \citep{bravyi2005universal}, a process with high spacetime cost (qubit-seconds) \citep{Eastin_2009}. For example, the spacetime cost of the \tgate{} is about two orders of magnitude larger than the cost of a CNOT operation between two adjacent qubits \citep{gidney2019efficient}. Consequently, the cost of a quantum algorithm in the fault-tolerant era is arguably dominated by the cost of implementing the non-Clifford gates.

In this paper, we focus on the problem of \tcount optimization, i.e., minimizing the number of \tgate{s} of quantum algorithms. \tcount optimization is an NP-hard problem \citep{vandewetering2023optimising} that has been addressed in the literature using different approaches \citep{amy2013polynomial,gosset2013algorithm,heyfron2018efficient,nam2018automated,amy2019tcount,kissinger2019reducing,gheorghiu2022tcount,kissinger2022simulating,debeaudrap2019techniques,debeaudrap2020fast}. Like most of these works, and for the reasons laid out before, we consider the \tcount as the \emph{sole} metric of complexity of a quantum circuit. 

\begin{figure}[t]
    \centering
    \includegraphics[width=\textwidth]{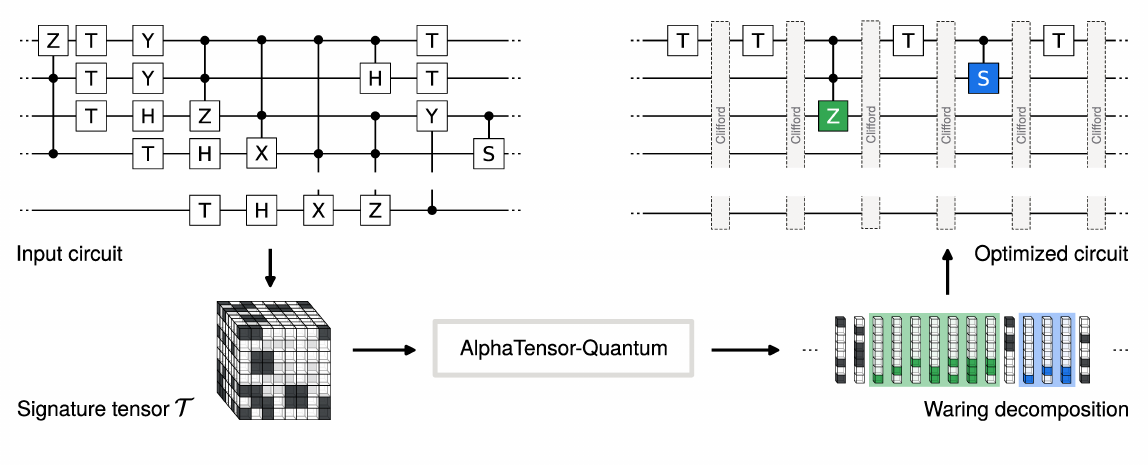}
    \caption{Pipeline of \method. We first extract the non-Clifford components of an input circuit and represent them as a symmetric \emph{signature tensor} $\mathcal{T}$, a binary tensor depicted as a cube of solid and transparent blocks. We then use \method to find a low-rank Waring decomposition of that tensor, that is, a set of factors (displayed as columns of blocks) that can be mapped back into an optimized quantum circuit with reduced \tcount. In general, there is a one-to-one correspondence between factors and \tgate{s}, but \method can also group factors into \emph{gadgets} (highlighted in green and blue)---constructions that can be equivalently implemented with fewer \tgate{s} than the sum of the factors.}
    \label{fig:alphatensor_pipeline}
\end{figure}

With that aim, we first transform the problem into an instance of tensor decomposition, exploiting the relationship between \tcount optimization and finding the symmetric tensor rank \citep{amy2019tcount,heyfron2018efficient}.
We develop \emph{\method}, an extension of \alphatensor \citep{fawzi2022discovering}, a method that finds low-rank tensor decompositions using deep reinforcement learning (RL). \method tackles the problem from the tensor decomposition point of view and is able to find efficient algorithm implementations with low \tcount. The approach is illustrated in \Cref{fig:alphatensor_pipeline}.

\method addresses three main challenges that go beyond the capabilities of AlphaTensor \citep{fawzi2022discovering} when applied on this problem. First, we need to optimize the \emph{symmetric} tensor rank, as opposed to the standard notion of tensor rank. Second, the approach must extend to \emph{large tensor sizes}, since the size of the tensor corresponds directly to the number of qubits in the circuit to be optimized. Third, AlphaTensor cannot leverage domain knowledge that falls outside of the tensor decomposition framework. \method addresses these three challenges. First, it modifies the RL environment and actions to provide symmetric decompositions (called \emph{Waring decompositions}) of the tensor; this has the beneficial side effect of reducing the action search space. Second, it scales up to larger tensors due to its innovative neural network architecture featuring \emph{symmetrization layers}. Third, it incorporates domain knowledge by leveraging so-called \emph{gadgets} (constructions that can save \tgate{s} by using auxiliary ancilla qubits); this is achieved through an efficient procedure embedded in the RL environment that exploits the Toffoli gadget \citep{jones2013lowoverhead} and the controlled S (\CS) gadget \citep{beverland2020lower}.

We show that \method is a powerful method for finding efficient quantum circuits. On a benchmark of arithmetic primitives, \method outperforms all existing methods for \tcount optimization, especially when allowed to leverage domain knowledge. For the relevant operation of multiplication in finite fields, which has applications in cryptography \citep{cheung2008design}, \method finds an efficient quantum algorithm with the same complexity as the \emph{classical} Karatsuba's method \citep{karatsuba1962multiplication}. As quantum computations are reversible by nature, naive translations of classical algorithms commonly introduce overhead \cite{Bennett1989-ct, gidney2019asymptotically}. %
\method thus finds the most efficient quantum algorithm for multiplication on finite fields reported to date.
We also optimize quantum primitives for other relevant problems, ranging from arithmetic computations used, e.g., in Shor's algorithm \citep{gidney2018halving},
to Hamiltonian simulation in quantum chemistry, e.g., FeMoco simulation \citep{reiher2017elucidating,babbush2018encoding}.
Here, \method recovers the best known hand-designed solutions, demonstrating it can effectively optimize circuits of interest in a fully automated way. We envision this approach can significantly accelerate discoveries in quantum computation as it saves the numerous hours of research invested in the design of optimized circuits.

These results show that \method can effectively exploit the domain knowledge that is provided in the form of gadgets and state-of-the-art magic state factories \citep{gidney2019efficient}, finding constructions that are optimal for the chosen metrics of interest. Because of its flexibility, \method can be readily extended in multiple ways, e.g., by considering different complexity metrics (other than the \tcount), or by incorporating new domain knowledge. Thus, we expect that \method will become instrumental for automatic circuit optimization with new advancements in quantum computing.

\section{\method for \tcount Optimization}
\label{sec:method}

We describe \method, an RL agent to optimize the \tcount of quantum circuits. It relies on the relation between \tcount optimization and tensor decomposition (see \Cref{fig:alphatensor_pipeline}), which we briefly review next.

\paragraph{Quantum gates.}
Quantum computers work with \emph{qubits}. The state of a qubit is representable as a L2-normalized vector in $\C^2$, typically denoted with the \emph{bra-ket} notation as $\ket{\psi}$, with the basis vectors being $\ket{0}$ and $\ket{1}$. An $N$-qubit state is represented by a normalized vector in $\C^{2^N}$. Quantum circuits perform linear operations over a quantum state that preserve the normalization, thus they can be represented by \emph{unitary matrices}.
Quantum circuits are implemented by \emph{quantum gates}.

\begin{table}[t]
\centering
\footnotesize
\begin{tabular}{cccc} \toprule
Quantum operator & \#Qubits & Gate symbol & Unitary matrix \\ \midrule
Controlled NOT (\CNOT) & 2 & \raisebox{-.3\height}{\includegraphics[width=15pt]{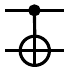}} & $\begin{pmatrix} 1 & 0 & 0 & 0 \\ 0 & 1 & 0 & 0 \\ 0 & 0 & 0 & 1 \\ 0 & 0 & 1 & 0 \end{pmatrix}$ \\ \midrule
Controlled S (\CS) & 2 & \raisebox{-.3\height}{\includegraphics[width=15pt]{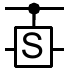}} & $\begin{pmatrix} 1 & 0 & 0 & 0 \\ 0 & 1 & 0 & 0 \\ 0 & 0 & 1 & 0 \\ 0 & 0 & 0 & i \end{pmatrix}$ \\ \midrule
Controlled Z (\CZ) & 2 & \raisebox{-.3\height}{\includegraphics[width=15pt]{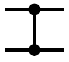}} & $\begin{pmatrix} 1 & 0 & 0 & 0 \\ 0 & 1 & 0 & 0 \\ 0 & 0 & 1 & 0 \\ 0 & 0 & 0 & -1 \end{pmatrix}$ \\ \midrule
Controlled controlled Z (\CCZ) & 3 & \raisebox{-.4\height}{\includegraphics[width=15pt]{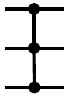}} & {\tiny $\begin{pmatrix} 1 & 0 & 0 & 0 & 0 & 0 & 0 & 0 \\ 0 & 1 & 0 & 0 & 0 & 0 & 0 & 0 \\ 0 & 0 & 1 & 0 & 0 & 0 & 0 & 0 \\ 0 & 0 & 0 & 1 & 0 & 0 & 0 & 0 \\ 0 & 0 & 0 & 0 & 1 & 0 & 0 & 0 \\ 0 & 0 & 0 & 0 & 0 & 1 & 0 & 0 \\ 0 & 0 & 0 & 0 & 0 & 0 & 1 & 0 \\ 0 & 0 & 0 & 0 & 0 & 0 & 0 & -1 \end{pmatrix}$} \\ \midrule
Toffoli (Controlled \CNOT) & 3 & \raisebox{-.4\height}{\includegraphics[width=15pt]{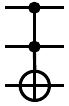}} & {\tiny $\begin{pmatrix} 1 & 0 & 0 & 0 & 0 & 0 & 0 & 0 \\ 0 & 1 & 0 & 0 & 0 & 0 & 0 & 0 \\ 0 & 0 & 1 & 0 & 0 & 0 & 0 & 0 \\ 0 & 0 & 0 & 1 & 0 & 0 & 0 & 0 \\ 0 & 0 & 0 & 0 & 1 & 0 & 0 & 0 \\ 0 & 0 & 0 & 0 & 0 & 1 & 0 & 0 \\ 0 & 0 & 0 & 0 & 0 & 0 & 0 & 1 \\ 0 & 0 & 0 & 0 & 0 & 0 & 1 & 0 \end{pmatrix}$} \\ \bottomrule
\end{tabular}
\caption{Common controlled gates used in quantum computing, along with their symbols and unitary matrices.%
\label{tab:quantum_circuits}}
\end{table}
 
Following convention, we focus on the Clifford+T gate set because it is the leading candidate for fault-tolerant implementation and it is approximately universal, meaning that any unitary matrix can be approximated arbitrarily well by a circuit consisting of these gates. In particular, we consider the following gate set: Hadamard (H) and controlled NOT (\CNOT), which are Clifford gates, and \tgate{s}, given by the unitary matrices
\begin{equation*}
    T = \begin{pmatrix} 1 & 0 \\ 0 & \ee^{i\pi/4}
    \end{pmatrix}, \qquad 
    H = \frac{1}{\sqrt{2}}\begin{pmatrix} 1 & 1 \\ 1& -1\end{pmatrix},\qquad
    \textrm{and}\qquad
    \CNOTmat = \begin{pmatrix} 1 & 0 & 0 & 0 \\ 0 & 1 & 0 & 0 \\ 0 & 0 & 0 & 1 \\ 0 & 0 & 1 & 0 \end{pmatrix}.
\end{equation*}
For convenience, we also define the following Clifford gates:
\begin{equation*}
S=T^2, \qquad Z=S^2, \qquad X=\begin{pmatrix} 0 & 1 \\ 1 & 0\end{pmatrix}, \qquad\textrm{and}\qquad Y=\begin{pmatrix} 0 & -i \\ i & 0 \end{pmatrix}.
\end{equation*}

Each of these gates, except the \CNOT, acts on a single qubit. The T, S, and Z gates perform phase operations, e.g., for a quantum state $\ket{x}$ with $x\in\{0, 1\}$, we have $T\ket{x} = \ee^{i\frac{\pi}{4}x} \ket{x}$. The \CNOT acts on two qubits: it flips the second qubit if the first qubit is in the $\ket{1}$ state, i.e., $\CNOTmat\ket{x, y} = \ket{x, x\oplus y}$, where `$\oplus$' denotes the XOR operation. Controlled gates like the \CNOT enable quantum entanglement; these gates are summarized in \Cref{tab:quantum_circuits}. For instance, the controlled \CNOT, or Toffoli gate, acts on three qubits and applies the operation $\Tofmat\ket{x,y,z} = \ket{x,y,z \oplus (xy)}$.

Quantum gates can be composed to implement arbitrary unitary operations: serial composition corresponds to matrix multiplication, and parallel composition corresponds to Kronecker product.

\paragraph{Signature tensor.}
Given a circuit composed of \CNOT and \tgate{s} alone, we can encode the information about the non-Clifford components into a symmetric \emph{signature tensor} $\mathcal{T} \in \{0, 1\}^{N\times N\times N}$, with $N$ being the number of qubits. Indeed, two
\CNOT + T circuits implement the same unitary matrix, up to Clifford gates, if and only if they have the same signature tensor \citep{heyfron2018efficient,amy2019tcount} (see \Cref{subsec:signature_tensor}).

To optimize the \tcount of a circuit, we first find its representation in terms of the signature tensor $\mathcal{T}$.
When the circuit has gates other than \CNOT and \tgate{s} alone, to obtain the signature tensor we
first split the circuit into two parts---one that consists only of Clifford gates, and a diagonal part that contains only \CNOT and \tgate{s}. We do this using the techniques of \citet{heyfron2018efficient} in combination with the algorithm by \citet{vandaele2023optimal} (see \Cref{fig:mod54} for an example and \Cref{app:subsec:compilation} for more details).

After that, we obtain $\mathcal{T}$ by first mapping each \tgate{} to a vector $u^{(r)}\in \{0, 1\}^N$, which is a straightforward step \citep{amy2013meet}, and then constructing the tensor from the \emph{Waring decomposition} (a symmetric form of the tensor rank decomposition) given by these vectors, i.e.,
\begin{equation*}
    \mathcal{T} = \sum_{r=1}^{R} u^{(r)} \otimes u^{(r)} \otimes u^{(r)},
\end{equation*}
where $R$ is the number of \tgate{s}, `$\otimes$' denotes the outer product, and the additions are under modulo $2$. %

Crucially, any \emph{alternative} Waring decomposition of $\mathcal{T}$ can \emph{also} be mapped to a quantum circuit, as each vector $u^{(r)}$ corresponds one-to-one to a \tgate. The resulting circuit is equivalent to the original one up to Clifford operations \citep{amy2019tcount}, albeit with a different number of \tgate{s}---according to the number of factors $R$ in the decomposition. Hence, we recast the problem of \tcount optimization as finding low-rank decompositions of $\mathcal{T}$.

\paragraph{\game.}
\method is an RL-based tensor factorization approach that finds low-rank Waring decompositions of tensors.
It builds upon AlphaTensor \citep{fawzi2022discovering}, which uses deep RL to find low-rank decompositions of 3-dimensional tensors. \method recasts the problem of tensor decomposition into a single-player game, called \emph{\game}, in which at each step $s$ the player observes the \emph{game state}, consisting of a tensor $\mathcal{T}^{(s)}$ and all the past played actions (initially, the tensor $\mathcal{T}^{(1)}$ is set to the signature tensor of interest $\mathcal{T}$, and there are no past actions). At step $s$, the player selects an \emph{action} consisting of a vector (or factor) $u^{(s)}\in \{0, 1\}^N$. Choosing an action affects the game state: the tensor $\mathcal{T}^{(s+1)} = \mathcal{T}^{(s)} - u^{(s)} \otimes u^{(s)} \otimes u^{(s)}$, and $u^{(s)}$ is included as the last played action. The goal of \game is to reach the all-zero tensor in as few moves as possible. The game ends after $S$ moves if that is the case and, by construction, the played actions form a rank-$S$ Waring decomposition of the signature tensor, since $\mathcal{T} = \sum_{s=1}^{S} u^{(s)} \otimes u^{(s)} \otimes u^{(s)}$. (The game also ends after a user-specified maximum number of moves to avoid infinitely long games, in which case the game is declared ``unsuccessful'' as it did not find a decomposition of the tensor.)
To encourage the RL agent to find low-rank decompositions or, equivalently, circuits with lower \tcount, games are penalized based on the number of moves taken to reach the all-zero tensor (with an additional penalty for unsuccessful games): every played move incurs a reward of $-1$.

\begin{figure}[hp!]
    \centering
    \includegraphics[width=\linewidth]{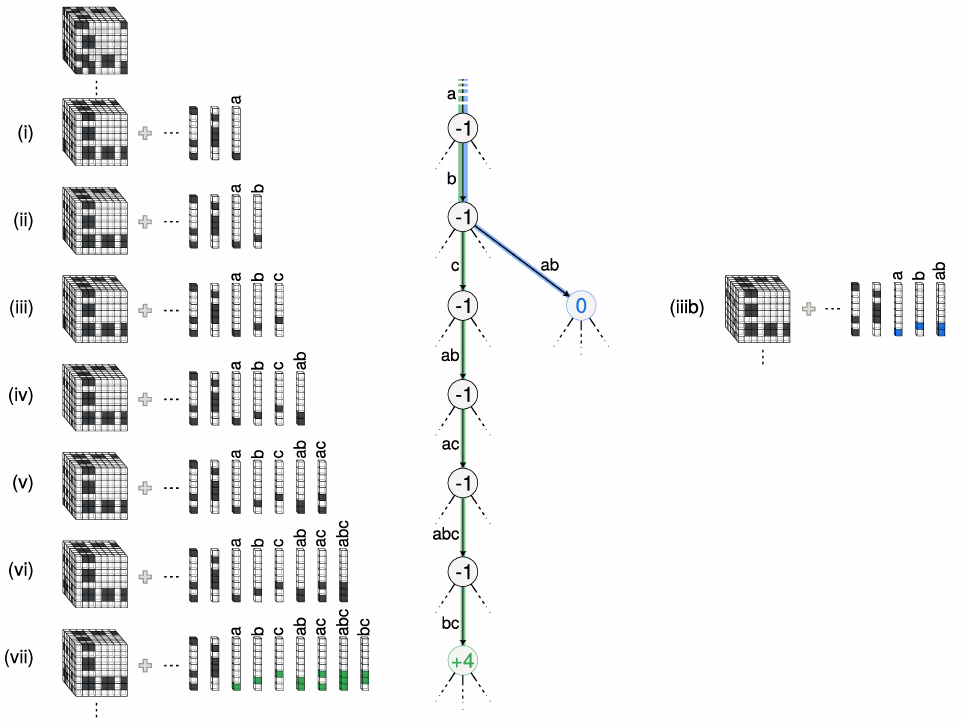}
    \caption{Illustration of gadgetization in \game. Nodes in the tree correspond to states, and the number in each node is the immediate reward associated to the action leading to that node.
    In state (i), the last played action is labeled ``a'', and state (ii) is reached after playing another action (labeled ``b''). From state (ii), playing action ``c'' leads to state (iii), incurring an additional $-1$ reward. If action ``ab'' is played instead from state (ii), the reward leading to state (iiib) is $0$ because the move completes a \CS gadget (blue path). Similarly, the sequence of moves in the green path completes a Toffoli gadget, and thus the immediate reward in state (vii) is $+4$ so that the last seven actions jointly receive a reward of $-2$ (see \Cref{subsec:system_description} for details on the gadgetization patterns). With its ability to plan, from state (i) the agent may decide to play actions in the green or blue paths and benefit from the adjusted rewards, or to play other actions that are not part of a gadget (black dashed paths). In this way, \method can automatically find trade-offs between playing actions that are part of a gadget and actions that are not.}
    \label{fig:gadgets}
\end{figure}

In \method, each action consists of a single factor $u^{(s)}$. This represents a significant advantage over standard AlphaTensor (where actions consist of  factor triplets), because for a fixed tensor size $N$, the action space reduces from $2^{3N}$ to $2^{N}$. This is one of the key properties that allows \method to scale up to tensors larger than the ones considered by \citet{fawzi2022discovering}; we consider tensors of size up to $N=72$ in \Cref{sec:results}.

\paragraph{Gadgets.}
In contrast to other state-of-the-art \tcount optimization approaches, \method can leverage \emph{gadgetization} techniques to find more efficient constructions. 
In this context, a gadget is an alternative implementation of a quantum gate that reduces the number of \tgate{s} at the cost of introducing additional (ancilla) qubits and operations such as Clifford gates and measurement-based corrections. (These extra qubits do not represent a significant limitation in practice, as they can be reused throughout the circuit.)

We consider two such gadgets.
First, the Toffoli gate, a quantum gate that needs at least seven \tgate{s} \citep{gosset2013algorithm} to implement without ancillae. The Toffoli gate admits a gadget to implement it using four \tgate{s} \citep{jones2013lowoverhead}, or when considering a state-of-the-art magic state factory \citep{gidney2019efficient}, a gadget to implement it at a cost equivalent to that of two \tgate{s}. %
Second, the \CS gate, which can be built with a cost equivalent to that of two \tgate{s} due to the magic state factory, instead of the three \tgate{s} of its direct implementation \citep[Appendix A.3]{beverland2020lower}.

Although the gadgetization techniques fall outside of the tensor decomposition framework, \method incorporates them into \game as follows. Every time that an action is played, we check whether it completes a Toffoli when combined with the six previous actions, a \CS gate when combined with the two previous actions, or none of them. (These checks are up to Clifford operations and they simply involve finding linear dependencies among the factors; see \Cref{subsec:system_description,app:subsec:gadgetization} for details.) If it completes a Toffoli, we assign a positive reward, so that the seven actions jointly incur a reward of $-2$ (corresponding to two \tgate{s}) as opposed to the reward of $-7$ corresponding to the un-gadgetized implementation. If it completes a \CS, we adjust the reward so that the three actions jointly incur a reward of $-2$. See \Cref{fig:gadgets} for an illustration of a \game with gadgetization.

Other than the reward signal, \method has no further information about the gadgets. It learns to recognize and exploit those patterns of actions from the data, and it finds good trade-offs between playing actions that are part of a gadget and actions that are not.

\section{Results}
\label{sec:results}

\Cref{subsec:benchmark} demonstrates the effectiveness of \method on a benchmark of circuits widely used in the literature of \tcount optimization, and \Cref{subsec:applications} shows applications of \method in different domains.

\subsection{Benchmark Circuits}
\label{subsec:benchmark}

We consider a benchmark set of quantum circuits originally proposed by \citet{amy2013polynomial} and widely used in the literature of \tcount optimization \citep{heyfron2018efficient,nam2018automated,amy2019tcount,debeaudrap2020fast}. We take the circuits from \citet{amy2016feynman} (we also include GF($2^2$)-mult and GF($2^3$)-mult for completeness; see \Cref{app:subsec:gfmult_circuits}) %
and apply two versions of \method: the full approach that includes gadgetization, and an alternative agent that does not consider any gadgets. We report the resulting \tcount for each approach in \Cref{tab:results_benchmarks_no_split}. 

\begin{figure}
    \centering
    \includegraphics[width=\textwidth]{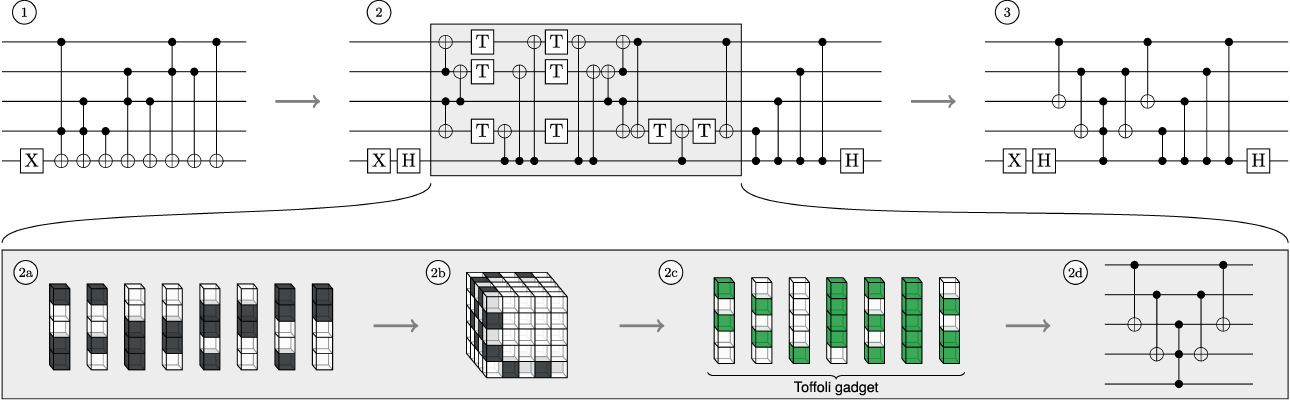}
    \caption{Results of \method on the Mod $5_4$ circuit. \textbf{(1)} The original quantum circuit, which can be implemented with $4$ Toffoli gates (with an equivalent cost of $8$ \tgate{s}). \textbf{(2)} The compiled circuit, which has been split into Clifford gates and a subcircuit of just CNOT and \tgate{s} (in the highlighted box). \textbf{(2a)} The Waring decomposition corresponding to the highlighted CNOT+T circuit: each \tgate{} corresponds precisely to one of the factors displayed as columns of blocks. \textbf{(2b)} The signature tensor formed from this decomposition. \textbf{(2c)} An alternative Waring decomposition of the tensor found by the RL agent. Again, each factor corresponds one-to-one to a \tgate{}; however in this case the seven factors are grouped into a single Toffoli gadget. \textbf{(2d)} The quantum circuit obtained from this new decomposition. \textbf{(3)} The optimized circuit after having applied \method, which can be implemented using Clifford gates and a single Toffoli gate (i.e., its equivalent \tcount is $2$).}
    \label{fig:mod54}
\end{figure}

\paragraph{Circuit compilation.}
We apply a standard compilation technique
\citep{heyfron2018efficient,debeaudrap2020fast} with an improvement \citep{vandaele2023optimal} (see \Cref{app:subsec:compilation} for details and \Cref{fig:mod54} 
for an example). 
To make the circuits amenable to the tensor decomposition approach, we replace the Hadamard gates by extra ancilla qubits and Clifford gates (we refer to this as \emph{Hadamard gadgetization} \cite{heyfron2018efficient} to distinguish it from the gadgetization described in \Cref{sec:method}). For circuits exceeding $60$ qubits after this process, we split them into smaller sub-circuits of fewer than $60$ qubits each, and apply \method on each sub-circuit independently. We split the circuits at the boundaries corresponding to Hadamard gates via a random-greedy algorithm designed to minimize the number of splits (see \Cref{app:subsec:splitting}).

\paragraph{Results.}
In \Cref{tab:results_benchmarks_no_split}, we compare the results of the non-gadgetized version of \method against the best \tcount reported in the literature \citep{amy2013polynomial,heyfron2018efficient,nam2018automated,amy2019tcount,debeaudrap2020fast,kissinger2019reducing,debeaudrap2019techniques,zhang2019optimizing,munson2020andgates}, as the baselines do not consider any gadgets either.
For the circuits that did not require splitting, \method matches or outperforms all the existing \tcount optimization methods. For all small circuits (at most $20$ \tgate{s}), we confirmed with the Z3 theorem prover \cite{demoura2008z3} that \method obtains the best possible \tcount achievable with a tensor decomposition approach.
For the circuits that exceed $60$ qubits, \method outperforms the baselines for Hamming$_{15}$ (high) and Mod-Adder$_{1024}$, but not for the others, likely due to the lack of optimality of the splitting technique.

When considering gadgets, \method leverages the Toffoli and \CS gates to \emph{further drastically reduce} the \tcount. When compared against the original construction of the circuits, consisting mostly of Toffoli gates, \method reduces the \tcount in all except two cases (likely due to the splitting). When compared against the baselines, \method reduces the \tcount for all circuits except Grover$_5$. Remarkably, for the GF($2^m$)-mult circuits, \method finds constructions with significantly smaller \tcount than the baselines or the original constructions; we discuss this further in \Cref{subsec:applications}.

\begin{table}[H]
\centering
\footnotesize
\rowcolors{2}{gray!25}{white}
\hspace*{-5mm}
\begin{tabular}{c|cc|cc|cc} \toprule
& \multicolumn{2}{c|}{\#Qubits} & \multicolumn{2}{c|}{\tcount without gadgets} & \multicolumn{2}{c}{\tcount with gadgets} \\
\multirow{-2}{*}{Circuit} & Original & Compiled & Baselines & \method & Original & \method \\ \midrule  %
$8$-bit adder & $24$ & $\star$ & $\mathbf{129}$ {\scriptsize\citep{heyfron2018efficient}} & $139$ & $114$ \scriptsize{($57$Tof)} & $\mathbf{94}$ \scriptsize{($33$Tof + $28$T)} \\
Barenco Toff$_{3}$ & $5$ & $8$ & $13$ {\scriptsize\citep{debeaudrap2020fast,debeaudrap2019techniques}} & $13^*$ & $8$ \scriptsize{($4$Tof)} & $\mathbf{4}$ \scriptsize{($2$Tof)} \\
Barenco Toff$_{4}$ & $7$ & $14$ & $24$ {\scriptsize\citep{heyfron2018efficient,debeaudrap2019techniques}} & $\mathbf{23}$ & $16$ \scriptsize{($8$Tof)} & $\mathbf{8}$ \scriptsize{($4$Tof)} \\
Barenco Toff$_{5}$ & $9$ & $20$ & $34$ {\scriptsize\citep{heyfron2018efficient,debeaudrap2019techniques}} & $\mathbf{33}$ & $24$ \scriptsize{($12$Tof)} & $\mathbf{12}$ \scriptsize{($6$Tof)} \\
Barenco Toff$_{10}$ & $19$ & $50$ & $84$ {\scriptsize\citep{heyfron2018efficient}} & $\mathbf{83}$ & $64$ \scriptsize{($32$Tof)} & $\mathbf{32}$ \scriptsize{($16$Tof)} \\
CSLA-MUX$_{3}$ & $15$ & $21$ & $40$ {\scriptsize\citep{debeaudrap2019techniques}} & $\mathbf{39}$ & $20$ \scriptsize{($10$Tof)} & $\mathbf{16}$ \scriptsize{($8$Tof)} \\
CSUM-MUX$_{9}$ & $30$ & $42$ & $72$ {\scriptsize\citep{heyfron2018efficient,kissinger2019reducing}} & $\mathbf{71}$ & $56$ \scriptsize{($28$Tof)} & $\mathbf{28}$ \scriptsize{($14$Tof)} \\
GF($2^{2}$)-mult & $6$ & $6$ & $17^\dagger$ & $17^*$ & $8$ \scriptsize{($4$Tof)} & $\mathbf{6}$ \scriptsize{($3$Tof)} \\
GF($2^{3}$)-mult & $9$ & $9$ & $31^\dagger$ & $\mathbf{29}$ & $18$ \scriptsize{($9$Tof)} & $\mathbf{12}$ \scriptsize{($6$Tof)} \\
GF($2^{4}$)-mult & $12$ & $12$ & $47$ {\scriptsize\citep{debeaudrap2019techniques}} & $\mathbf{39}$ & $32$ \scriptsize{($16$Tof)} & $\mathbf{18}$ \scriptsize{($9$Tof)} \\
GF($2^{5}$)-mult & $15$ & $15$ & $84$ {\scriptsize\citep{debeaudrap2019techniques}} & $\mathbf{59}$ & $50$ \scriptsize{($25$Tof)} & $\mathbf{26}$ \scriptsize{($13$Tof)} \\
GF($2^{6}$)-mult & $18$ & $18$ & $118$ {\scriptsize\citep{debeaudrap2019techniques}} & $\mathbf{77}$ & $72$ \scriptsize{($36$Tof)} & $\mathbf{36}$ \scriptsize{($18$Tof)} \\
GF($2^{7}$)-mult & $21$ & $21$ & $167$ {\scriptsize\citep{debeaudrap2020fast}} & $\mathbf{104}$ & $98$ \scriptsize{($49$Tof)} & $\mathbf{44}$ \scriptsize{($22$Tof)} \\
GF($2^{8}$)-mult & $24$ & $24$ & $214$ {\scriptsize\citep{kissinger2019reducing}} & $\mathbf{123}$ & $128$ \scriptsize{($64$Tof)} & $\mathbf{58}$ \scriptsize{($29$Tof)} \\
GF($2^{9}$)-mult & $27$ & $27$ & $295$ {\scriptsize\citep{heyfron2018efficient}} & $\mathbf{161}$ & $162$ \scriptsize{($81$Tof)} & $\mathbf{70}$ \scriptsize{($35$Tof)} \\
GF($2^{10}$)-mult & $30$ & $30$ & $350$ {\scriptsize\citep{heyfron2018efficient}} & $\mathbf{196}$ & $200$ \scriptsize{($100$Tof)} & $\mathbf{92}$ \scriptsize{($46$Tof)} \\
Grover$_{5}$ & $9$ & $\star$ & $\mathbf{44}$ {\scriptsize\citep{heyfron2018efficient}} & $152$ & $96$ \scriptsize{($48$Tof)} & $\mathbf{66}$ \scriptsize{($27$Tof + $12$T)} \\
Hamming$_{15}$ (high) & $20$ & $\star$ & $787^\dagger$ \scriptsize{($1010$ \cite{heyfron2018efficient})} & $\mathbf{773}$ & $702$ \scriptsize{($351$Tof)} & $\mathbf{440}$ \scriptsize{($173$Tof + $2$\CS + $90$T)} \\
Hamming$_{15}$ (low) & $17$ & $42$ & $75$ {\scriptsize\citep{heyfron2018efficient}} & $\mathbf{73}$ & $46$ \scriptsize{($23$Tof)} & $\mathbf{34}$ \scriptsize{($17$Tof)} \\
Hamming$_{15}$ (med) & $17$ & $\star$ & $156^\dagger$ \scriptsize{($162$ \cite{heyfron2018efficient})} & $156$ & $164$ \scriptsize{($82$Tof)} & $\mathbf{78}$ \scriptsize{($35$Tof + $8$T)} \\
HWB$_{6}$ & $7$ & $27$ & $51$ {\scriptsize\citep{heyfron2018efficient}} & $51$ & $30$ \scriptsize{($15$Tof)} & $\mathbf{20}$ \scriptsize{($10$Tof)} \\
Mod-Adder$_{1024}$ & $28$ & $\star$ & $798^\dagger$ \scriptsize{($978$ \cite{heyfron2018efficient})} & $\mathbf{762}$ & $570$ \scriptsize{($285$Tof)} & $\mathbf{500}$ \scriptsize{($141$Tof + $15$\CS + $188$T)} \\
Mod-Mult$_{55}$ & $9$ & $11$ & $17$ {\scriptsize\citep{heyfron2018efficient}} & $17^*$ & $14$ \scriptsize{($7$Tof)} & $\mathbf{6}$ \scriptsize{($3$Tof)} \\
Mod-Red$_{21}$ & $11$ & $28$ & $55$ {\scriptsize\citep{heyfron2018efficient,debeaudrap2019techniques}} & $\mathbf{51}$ & $34$ \scriptsize{($17$Tof)} & $\mathbf{22}$ \scriptsize{($11$Tof)} \\
Mod $5_4$ & $5$ & $5$ & $7$ {\scriptsize\citep{debeaudrap2020fast,kissinger2019reducing,debeaudrap2019techniques}} & $7^*$ & $8$ \scriptsize{($4$Tof)} & $\mathbf{2}$ \scriptsize{($1$Tof)} \\
QCLA-Adder$_{10}$ & $36$ & $\star$ & $\mathbf{116}$ {\scriptsize\citep{heyfron2018efficient}} & $135$ & $\mathbf{68}$ \scriptsize{($34$Tof)} & $94$ \scriptsize{($28$Tof + $5$\CS + $28$T)} \\
QCLA-Com$_{7}$ & $24$ & $42$ & $59$ {\scriptsize\citep{heyfron2018efficient}} & $59$ & $58$ \scriptsize{($29$Tof)} & $\mathbf{24}$ \scriptsize{($12$Tof)} \\
QCLA-Mod$_{7}$ & $26$ & $\star$ & $\mathbf{165}$ {\scriptsize\citep{heyfron2018efficient}} & $199$ & $\mathbf{118}$ \scriptsize{($59$Tof)} & $122$ \scriptsize{($43$Tof + $36$T)} \\
QFT$_{4}$ & $5$ & $43$ & $55$ {\scriptsize\citep{heyfron2018efficient}} & $\mathbf{53}$ & $59$ \scriptsize{($2$Tof + $55$T)} & $\mathbf{44}$ \scriptsize{($4$Tof + $3$\CS + $30$T)} \\
RC-Adder$_{6}$ & $14$ & $24$ & $37$ {\scriptsize\citep{heyfron2018efficient,debeaudrap2019techniques}} & $37$ & $22$ \scriptsize{($11$Tof)} & $\mathbf{12}$ \scriptsize{($6$Tof)} \\
NC Toff$_{3}$ & $5$ & $7$ & $13$ {\scriptsize\citep{heyfron2018efficient,debeaudrap2020fast,debeaudrap2019techniques}} & $13^*$ & $6$ \scriptsize{($3$Tof)} & $\mathbf{4}$ \scriptsize{($2$Tof)} \\
NC Toff$_{4}$ & $7$ & $11$ & $19$ {\scriptsize\citep{heyfron2018efficient,debeaudrap2020fast,debeaudrap2019techniques}} & $19^*$ & $10$ \scriptsize{($5$Tof)} & $\mathbf{6}$ \scriptsize{($3$Tof)} \\
NC Toff$_{5}$ & $9$ & $15$ & $25$ {\scriptsize\citep{heyfron2018efficient}} & $25$ & $14$ \scriptsize{($7$Tof)} & $\mathbf{8}$ \scriptsize{($4$Tof)} \\
NC Toff$_{10}$ & $19$ & $35$ & $55$ {\scriptsize\citep{heyfron2018efficient}} & $55$ & $34$ \scriptsize{($17$Tof)} & $\mathbf{18}$ \scriptsize{($9$Tof)} \\
VBE-Adder$_{3}$ & $10$ & $14$ & $20$ {\scriptsize\citep{heyfron2018efficient,debeaudrap2020fast,debeaudrap2019techniques}} & $\mathbf{19}^*$ & $20$ \scriptsize{($10$Tof)} & $\mathbf{6}$ \scriptsize{($3$Tof)} \\
\bottomrule
\end{tabular}
\caption{\tcount achieved by different methods on a set of benchmark circuits. Even without gadgetization, \method matches or outperforms the considered baselines for all circuits that have not been split into sub-circuits (split circuits are marked with $\star$). When considering gadgets, \method further reduces the \tcount drastically, and generally outperforms the original constructions with gadgets. The notation ``$a$Tof+$b$\CS\!+$c$T'' indicates a circuit with $a$ Toffoli gates, $b$ \CS gates, and $c$ \tgate{s} (when multiple solutions from \method achieve the same \tcount but differ in the trade-off between gadgets and \tgate{s}, we report the result with the highest number of Toffoli gates). Baseline results marked with $\dagger$ were obtained with the methods in Appendix \ref{app:subsec:baselines} and are better than the best published \tcount (provided in parentheses where applicable). Results marked with * were proven to be optimal decompositions of the given tensor using the Z3 theorem prover.
\label{tab:results_benchmarks_no_split}}
\end{table}
 
\subsection{Applications}
\label{subsec:applications}

We applied \method to circuits derived from several important use-cases for quantum computing, which fit into two broad categories: quantum implementations of classical reversible logic for arithmetic operations, and primitives important to quantum chemistry. We study the former because many quantum algorithms (for instance, Shor's algorithm and many variations of Grover's algorithm) require a quantum implementation of a classical oracle that performs some arithmetical or logical operation. We study the latter because quantum chemistry is a major application of quantum computing, and many of the state-of-the-art methods share similar circuit constructions.

\paragraph{Multiplication in finite fields.}
Multiplication in finite fields is a crucial operation for cryptography, since many elliptic-curve cryptography (ECC) implementations are built over such finite fields, and in fact these circuits are derived from a quantum algorithm that cracks ECC in polynomial time \citep{cheung2008design}.

\begin{figure}
    \centering
    \begin{subfigure}{0.49\textwidth}
        \includegraphics[trim={4mm 0 0 0},clip,height=5.2cm]{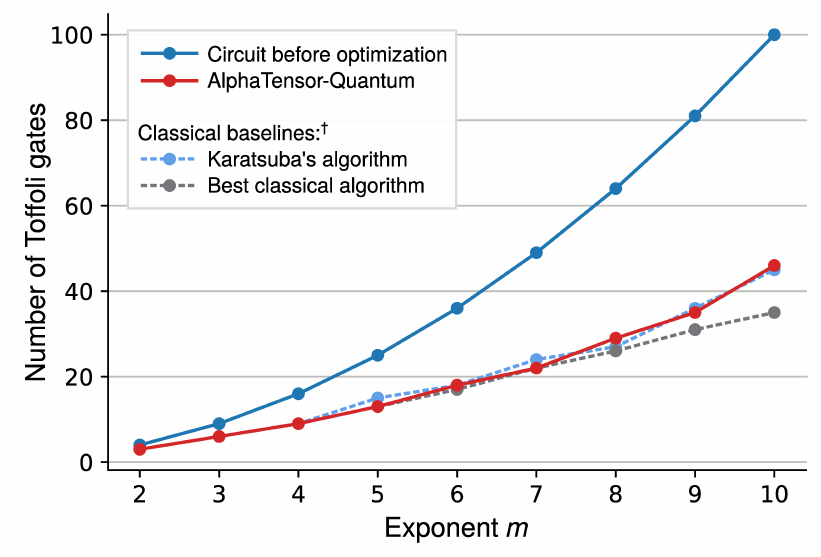}
        \caption{}
        \label{fig:gf_mult}
    \end{subfigure}
    \hfill
    \begin{subfigure}{0.49\textwidth}
        \includegraphics[height=5.2cm]{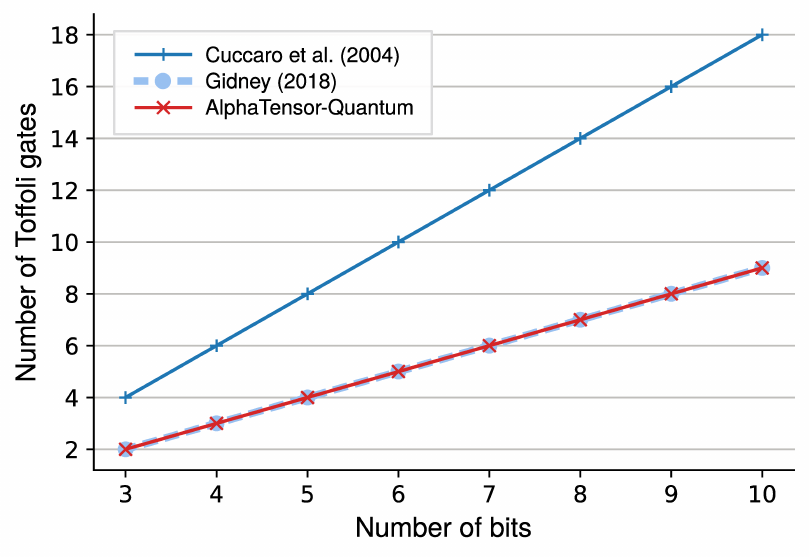}
        \caption{}
        \label{fig:cuccaro}
    \end{subfigure}
    \caption{Number of Toffoli gates of the optimized circuits found by \method. \textbf{(a)} For multiplication in finite fields of order $2^m$, \method finds efficient circuits that significantly outperform the original construction, which scales like $\mathcal{O}(m^2)$. The number of Toffoli gates matches the best known lower bound of the \emph{classical} circuits \citep{barbulescu2012finding} for some values of $m$, and scales as $\sim m^{\log_2(3)}$, showing that \method found an algorithm with the same complexity as Karatsuba's method \citep{karatsuba1962multiplication}, a classical algorithm for multiplication on finite fields for which a quantum version has not been reported in the literature. (The baselines marked with $\dagger$ are classical circuits, and hence not directly comparable, as naive translations of classical to quantum circuits commonly introduce overheads. To compare, we assume the number of effective Toffoli gates is the number of 1-bit AND gates in the classical circuit.)
    \textbf{(b)} For binary addition, \method halves the cost of the circuits from \citet{cuccaro2004new} matching the state-of-the-art circuits from \citet{gidney2018halving}. Remarkably, it does so automatically without any prior knowledge of the  measurement-based uncomputation technique, which was crucial to their results.}
\end{figure}

The results from \Cref{tab:results_benchmarks_no_split} show that \method excels in optimizing the primitives for multiplication in finite fields of order $2^m$, i.e., the circuits GF($2^m$)-mult. In \Cref{fig:gf_mult}, we show that for $m=\{2,3,4,5,7\}$, \method actually achieves the same Toffoli gate count as the best known \emph{upper bound} for the equivalent classical (non-reversible) circuits \citep{montgomery2005five,fan2007comments}. For $m=\{2,3,4,5\}$ this is also the best known \emph{lower bound} \cite{barbulescu2012finding}, which suggests that the resulting quantum circuits are likely optimal. While the circuits for these targets were constructed using the naive $\mathcal{O}(m^2)$ multiplication algorithm, the number of Toffoli gates of the optimized circuits follows $\sim m^{\log_2(3)}$ for the considered values of $m$. Thus, \method found a multiplication algorithm with the same complexity as Karatsuba's method \citep{karatsuba1962multiplication}, a \emph{classical} algorithm for multiplication in finite fields.  While it is conceivable to construct the resulting circuits by hand in principle, to the best of our knowledge no sub-quadratic constructions of quantum circuits for multiplication in finite fields have been discussed in the literature. \method finds them automatically, similar to Gidney's translation of Karatsuba's method for multiplication of integers \cite{gidney2019asymptotically}, but in contrast to the latter without requiring additionally ancilla qubits.

\paragraph{Binary addition.} We apply \method to two different quantum implementations of the binary addition operation, which is a relevant primitive in oracle construction because many other common operations are built from it (such as multiplication, comparators, and modular arithmetic). For example, controlled addition circuits are the dominant cost of Shor's algorithm. There are many types of quantum addition circuits (ripple-carry, carry-save, or block-carry-lookahead); we focus on ripple-carry addition circuits because they have fewer ancilla qubits generated by Hadamard gadgetization.

In particular, we optimize the circuits of \citet{cuccaro2004new}, which have a Toffoli count scaling as $2n + \mathcal{O}(1)$ where $n$ is the number of input bits, as well as the circuits of \citet{gidney2018halving}, with a Toffoli count of $n + \mathcal{O}(1)$. As shown in \Cref{fig:cuccaro}, \method reduces the Toffoli count of Cuccaro et al.'s circuits, matching the Toffoli count of Gidney's construction of $n + \mathcal{O}(1)$.

Prior to the publication of Gidney's results, Cuccaro et al.'s circuits attained the lowest Toffoli count of any addition circuit. To halve their cost, Gidney introduced a new approach called \emph{measurement-based uncomputation}, which allows Toffoli gates that were performed on zero-initialized ancilla qubits to be uncomputed for free under certain conditions, by introducing measurements and classically-controlled Clifford gates. This non-trivial technique was invented more than a decade after the addition circuits of \citet{cuccaro2004new}, but is now widely used for optimizing fault-tolerant quantum circuits.

\begin{figure}[t]
    \centering
    \includegraphics[width=\linewidth]{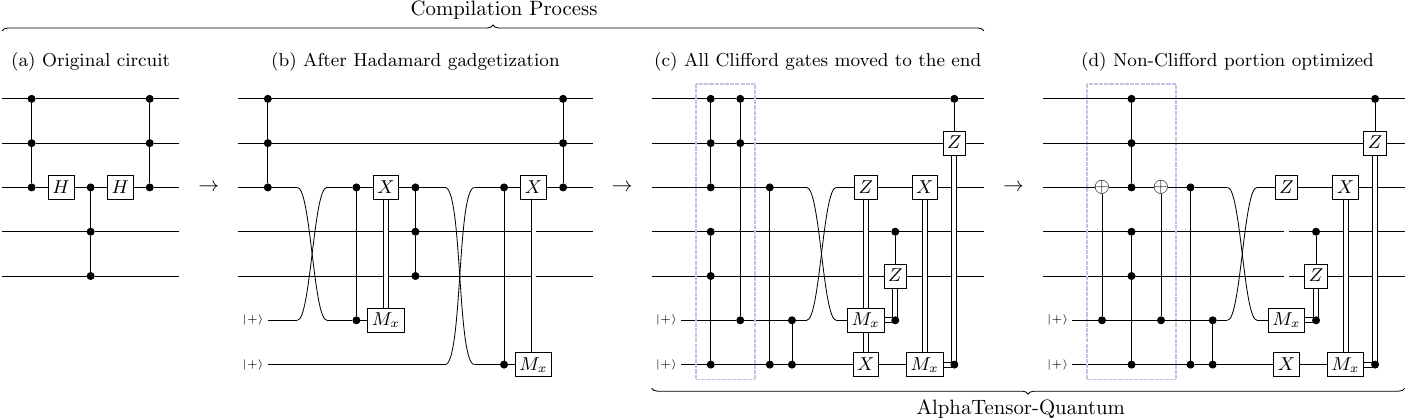}
    \caption{An example in which measurement-based uncomputation tricks can be recovered by \method. Here, a portion of the NC Toff$_3$ circuit is mapped to three \CCZ gates, a pair of which share two inputs (in the highlighted box), by the Hadamard gadgetization process (detailed in \Cref{app:subsec:compilation}). \method optimizes this pair by expressing it as a single gadget, saving one Toffoli gate with respect to the original construction. This kind of optimization happens often because \CCZ gates can usually be moved freely along the circuit, as the corresponding unitary matrix will be diagonal after compilation. This is reminiscent of the temporary logical-AND construction from \citet{gidney2018halving}, although some patterns with more than two \CCZ gates can also be combined in this way.}
    \label{fig:uncomputation-derivation}
\end{figure}

Importantly, for both circuit constructions, \method had no knowledge of measurement-based uncomputation, and all Toffoli gates were compiled naively using the standard unitary construction with seven \tgate{s}. 
Yet, \method automatically finds optimized circuits only accessible via measurement-based uncomputation. It does so with its ability to find Toffoli gadgets, by exploiting the fact that ancilla qubits from the Hadamard gadgetization allow the movement of Toffoli gates; see \Cref{fig:uncomputation-derivation} for an example.

\paragraph{Hamming weight and phase gradients.} Applying small angle rotations to qubits is a common operation in quantum algorithms, e.g., in the quantum Fourier transform, data loading, and quantum chemistry. In fault-tolerant quantum circuits, any rotation can be implemented using the Solovay-Kitaev algorithm with a sequence of Hadamard and \tgate{s}. However, if the same small angle is to be applied repeatedly (such as in quantum chemistry), it is often more efficient to use a technique called Hamming weight phasing.

We use \method to optimize the Hamming weight phasing circuits from \citet{gidney2018halving}. While the original circuits use measurement-based uncomputation, we instead compile a version where every Toffoli gate is constructed explicitly, as before. Again, \method halves the cost with respect to the naive input circuits, thus matching the complexity of the circuits with measurement-based uncomputation tricks.

\begin{figure}[t]
\centering
\includegraphics[width=0.4\textwidth]{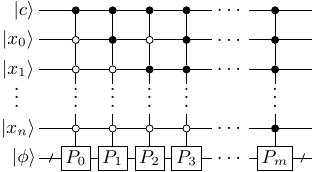}
\caption{The action implemented by the unary iteration family of circuits. If the control qubit $c$ is in the state $\ket{1}$, the circuit applies one of the Pauli operators from the set $\{P_1, \dots, P_m\}$ to the multiqubit state $\ket{\phi}$, depending on the value of the qubits $x_1$ to $x_n$. If the control qubit $c$ is in the state $\ket{0}$, no operation is applied to $\ket{\phi}$.}
\label{fgr:c-select}
\end{figure}

\paragraph{Unary iteration.} We also apply \method to a family of circuits called \emph{unary iteration} \citep{babbush2018encoding}. These circuits allow selecting which Pauli operator 
(from a set of operators) to apply on the data qubits, based on the value of some control qubits. We apply this construction to the Pauli operators $P_1 = XI\cdots I$, $P_2 = IX\cdots I$, $\dots$, $P_m = II\cdots X$ on $m = 8$, $16$, and $32$ qubits, therefore implementing the quantum equivalent of a classical $n$-to-$2^n$ demultiplexer or decoder for $n = 3, 4, 5$ (see \Cref{fgr:c-select}). 
These circuits could be used to implement the same construction for any set of $8$, $16$, or $32$ Pauli operators, without extra \tcount, and because of their generality, they appear both in oracle construction (demultiplexers are commonly used in classical logic) and in quantum chemistry (as in the linear combination of unitaries method discussed further below).

The unary iteration circuits were implemented following \citet{babbush2018encoding}, except we compiled the Toffoli gates unitarily with seven \tgate{s} instead of using measurement-based uncomputation. The largest circuit ($n = 5$) has $72$ qubits after compilation. %
As before, \method finds its own version of measurement-based uncomputation to match the best known \tcount of these circuits.

We also apply a version of \method without gadgets. Remarkably, it finds optimized circuits that, for each value of $n$, have only three \tgate{s} more than the circuits from \citet{babbush2018encoding} (where each Toffoli gate is implemented using four \tgate{s}), suggesting that the unary iteration circuits have an internal structure particularly amenable to this kind of optimization.

\paragraph{Quantum chemistry.}
Molecular simulation involves solving the Schr\"{o}dinger equation, which requires a computational cost scaling exponentially with the number of particles in the system. This makes classical methods struggle to simulate strongly correlated molecules \cite{chem-tensornet}.
Quantum computers can theoretically handle these calculations more efficiently, with one compelling use case being the simulation of the FeMoco molecule, the active site in the nitrogenase enzyme responsible for nitrogen fixation.  This use case has high potential industrial impact, but is challenging due to the highly complex electronic structure of FeMoco \cite{msft-femoco}.

FeMoco simulation motivated Lee et al.'s \citep{PRXQuantum.2.030305} fault-tolerant implementation of a quantum walk-based algorithm. In this method, the Hamiltonian associated with the molecule is encoded into the quantum circuit with a technique known as \emph{linear combination of unitaries} (LCU). This framework requires the implementation of two oracles called \textsc{Prepare} and \textsc{Select} (each being an instance of the unary iteration operation discussed earlier). These are repeated many times, making up the dominant cost of the whole algorithm, and therefore optimizing these oracles can significantly impact the total resource requirements (see \Cref{app:sec:preliminaries} for more details on the algorithm).

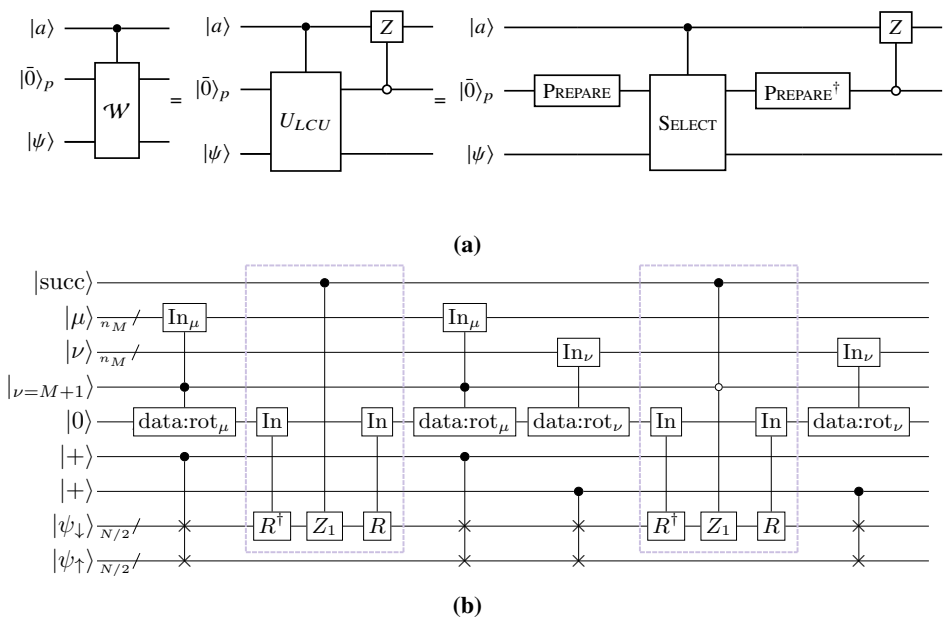
\begin{figure}
\centering
\begin{subfigure}{0.8\textwidth}
\begin{tikzpicture}
\node[scale=0.8] {
\begin{quantikz}
\lstick{$|a\rangle$}& \ctrl{1} &  \qw \\
\lstick{$|\bar{0}\rangle_p$} &\gate[2]{\mathcal{W}} & \qw\\
\lstick{$|\psi\rangle$} & \qw  & \qw  \\
\end{quantikz}
=
\begin{quantikz}
\lstick{$|a\rangle$}& \ctrl{1}& \gate{Z}  & \qw \\
\lstick{$|\bar{0}\rangle_p$} &\gate[2]{U_{LCU}}  &\octrl{-1}  &\qw\\
\lstick{$|\psi\rangle$}  & \qw & \qw & \qw\\
\end{quantikz}
=
\begin{quantikz}
\lstick{$|a\rangle$}& \qw & \ctrl{1} & \qw & \gate{Z}  & \qw  \\
\lstick{$|\bar{0}\rangle_p$} & \gate{\prepare}  &\gate[2]{\select} & \gate{\prepare^\dagger} &\octrl{-1}  &\qw\\
\lstick{$|\psi\rangle$} & \qw & & \qw & \qw& \qw \\
\end{quantikz}
\label{lcu_figure}
};
\end{tikzpicture}
\caption{}
\end{subfigure}

\begin{subfigure}{0.8\textwidth}
\includegraphics[width=\textwidth]{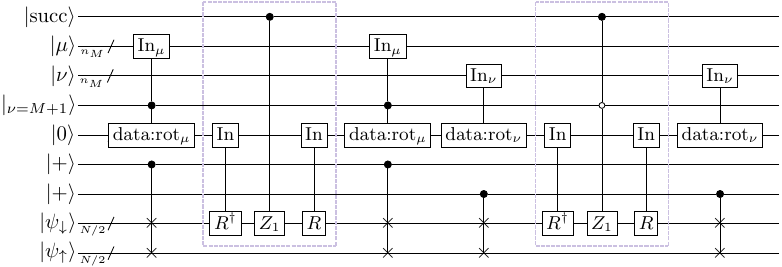}
\caption{}
\end{subfigure}

\caption{\textbf{(a)}
The quantum walk operator $\cal{W}$ at the heart of state-of-the-art quantum chemistry algorithms \cite{PRXQuantum.2.030305,Berry2019qubitizationof,babbush2018encoding} (see \Cref{fgr:qpe}) can be decomposed using the LCU technique into two oracles, \textsc{Prepare} and \textsc{Select}, which encode the details of the molecule. This operator is repeated many times, so its cost dominates the overall algorithm. \textbf{(b)} Construction of the \textsc{Select} oracle. This figure is adapted from \citet[Figure 5, pp.\ 18]{PRXQuantum.2.030305}. We compile and optimize the part in the dashed box, along with its constituent components, for a range of parameters (up to $10$ bits of precision per rotation, and for systems of up to $10$ spin-orbitals).}
\label{fgr:qubitz}
\end{figure}

We run \method on portions of Lee et al.'s \textsc{Select} oracle construction (which is itself based on \citet{vonburg2021quantum}). Specifically, we set a range of parameters to keep the total qubit count below $60$, and compile angle rotations of the state into the diagonal basis using the phase-gradient technique from \citet{gidney2018halving}; see \Cref{fgr:qubitz} for details.
As before, we remove all measurement-based uncomputation tricks and compile the Toffoli gates unitarily with seven \tgate{s}. We find that, for all parameter values, \method is able to match the \tcount of \citet{PRXQuantum.2.030305}, which relies heavily on measurement-based uncomputation.

Thus, \method automatically finds state-of-the-art constructions for multiple use cases. We expect that, as new improvements on quantum chemistry are made at a structural level, \method will be able to automatically find the best constructions (in terms of \tcount), without additional human effort.

\section{Discussion}
\label{sec:discussion}

We have demonstrated the effectiveness of \method at optimizing the \tcount, a key metric in quantum computations, showing that it significantly outperforms existing approaches, being able to automatically discover constructions as efficient as the best human-designed ones across various applications.
For those applications, the fact that very different techniques achieve the same \tcount might hint to the optimality of the discovered constructions. In fact, although \method does not come with optimality guarantees, we proved that some decompositions in \Cref{subsec:benchmark} are optimal in terms of the number of factors of the decomposition. %

Compared to previous works that use RL to optimize quantum circuits \citep{moro2021quantum,fosel2021quantum,quetschlich2023compiler}
or to improve variational quantum algorithms \citep{ostaszewski2021reinforcement,kuo2021quantum,zhu2023quantum,yao2022montecarlo,rosenhahn2023montecarlo}, 
Compared to previous works that use RL to optimize quantum circuits \citep{moro2021quantum,fosel2021quantum,quetschlich2023compiler},  \method is the the first one that scales to a large enough number of qubits to cover a wide range of applications (see \Cref{subsec:applications}), and the first one to consider fault-tolerant compilation. Compared to existing \tcount optimization approaches, which are not RL-based, \method significantly outperforms them; however this comes at the cost of increased computational complexity (e.g., it took around $8$ hours to optimize the addition circuits of \citet{cuccaro2004new}) and loss of optimality due to extra compilation steps (e.g., if the circuits need to be split into parts). A possible future research avenue to accelerate the algorithm is to improve the neural network architecture; another avenue to scale up further is to handle Hadamard gadgetization differently, so that the number of qubits does not increase as much during the compilation process, or to apply RL to alternative circuit representations that can optimize Hadamard gates natively. %
However, we argue that its comparably increased computational complexity is outweighed by its results, especially considering the cost of manually designing optimized constructions. Moreover, it is a one-off cost: once \method has been applied, the optimized circuits can be reused.

Unlike existing approaches for \tcount optimization, \method can leverage and exploit gadgets. These constructions are an active area of research \citep{jones2013lowoverhead,beverland2020lower,gidney2021cccz,amy2021phase}, and as fundamentally new gadgets are discovered, they can be readily incorporated into the RL environment, possibly allowing \method to discover even more efficient constructions for established quantum circuits. Similarly, due to its flexibility, \method can be extended to optimize metrics other than the \tcount, such as T-depth
\citep{amy2013meet} or weighted combinations of different types of gates \citep{nam2018automated}, by changing the reward signal accordingly.

We envision \method will play a pivotal role as quantum chemistry and related fields advance, and new improvements at a structural level emerge. 
Similarly to how \citet{gidney2018halving} optimized the addition circuits of \citet{cuccaro2004new} over a decade later, \method finds and exploits non-trivial patterns, thus eliminating the need for manual construction design and saving research costs.

\clearpage

\section{Methods}
\label{sec:methods_after_paper}

\subsection{The Signature Tensor}
\label{subsec:signature_tensor}

The tensor representation of a circuit relies on the concept of \emph{phase polynomials}. Consider a circuit with \CNOT and \tgate{s} alone. Its corresponding unitary matrix $U$ performs the operation $U\ket{x_{1:N}} = e^{i\phi(x_{1:N})} \ket{Ax_{1:N}}$, where $\phi(x_{1:N})$ is the phase polynomial and $A$ is an invertible matrix that can be implemented with Clifford gates only. For example, applying a \tgate{} to the second qubit after a \CNOT gives $(I \otimes T) \CNOTmat \ket{x,y} = e^{i\frac{\pi}{4}(x\oplus y)} \ket{x,x\oplus y}$, i.e., $\phi(x,y)=\frac{\pi}{4}(x\oplus y) = \frac{\pi}{4}(x+y-2xy)$. After cancelling out the terms leading to multiples of $2\pi$, the phase polynomial takes a multilinear form, i.e.,  $\phi(x_{1:N}) = \frac{\pi}{4}\left(\sum_i a_i x_i + 2\sum_{i<j} b_{ij}x_ix_j + 4\sum_{i<j<k} c_{ijk}x_ix_jx_k \right)$, where $a_i \in \{0,\ldots,7\}$, $b_{ij}\in\{0,\ldots, 3\}$, and $c_{ijk} \in \{0,1\}$. This maps directly to a circuit of T, \CS and \CCZ gates: each $a_i x_i$ term is implemented by $a_i$ \tgate{s} acting on the $i$-th qubit, each $b_{ij}x_ix_j$ term corresponds to $b_{ij}$ $\CS$ gates on qubits $i$ and $j$, and each non-zero $c_{ijk}x_ix_jx_k$ term is a \CCZ gate on qubits $i$, $j$, and $k$. Each of these terms corresponds to a non-Clifford operation only if the corresponding constant ($a_i$, $b_{ij}$, or $c_{ijk}$) is odd. Thus, two \CNOT + T circuits implement the same unitary up to Clifford gates if and only if they have the same phase polynomial after taking the modulo $2$ of their coefficients.
This information about the polynomial can then be captured in a symmetric 3-dimensional binary \emph{signature tensor} $\mathcal{T}$ of size $N\times N \times N$ \citep{heyfron2018efficient,amy2019tcount}. See \Cref{app:sec:preliminaries} for more details.

\subsection{System Description}
\label{subsec:system_description}

\method builds on AlphaTensor, which is in turn based on AlphaZero \citep{silver2018general}. It is implemented over a distributed computing architecture, composed of one \emph{learner} and multiple \emph{actors} that communicate asynchronously. The learner is responsible for hosting a neural network and updating its parameters, while the main role of the actors is to play games in order to generate data for the learner.  When the learner updates the network parameters by gradient descent steps, it sends them to the actors. The actors employ Monte Carlo tree search (MCTS), using queries to the network to guide their policy, and play games whose actions tend to improve over the policy dictated by the neural network. Played games are sent to the learner and stored in a buffer, and they will serve as future data points to train the network parameters.

\begin{figure}[t]
\begin{subfigure}{.5\textwidth}
  \centering
  \includegraphics[width=\linewidth]{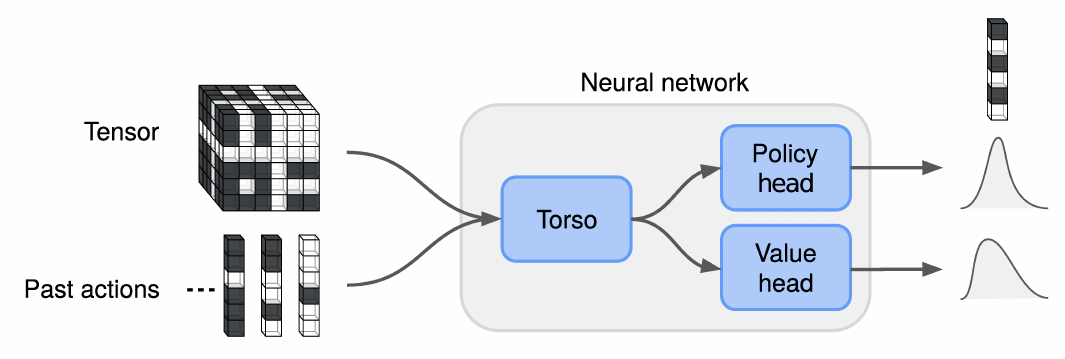}
  \caption{}%
  \label{fig:nn_overview}
\end{subfigure}%
\hspace*{5pt}
\begin{subfigure}{.49\textwidth}
  \centering
  \includegraphics[width=\linewidth]{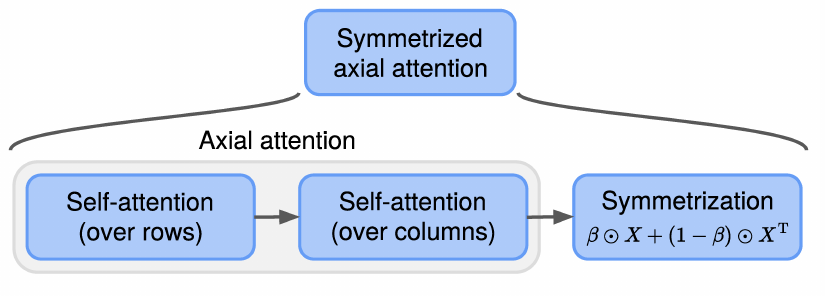}
  \caption{}%
  \label{fig:symmetrized_axial_attn}
\end{subfigure}
\caption{\textbf{(a)} An overview of the neural network, where the input is the current state (tensor and past actions) and the outputs are distributions over the next action to play (given by the policy head) and over the estimated return (given by the value head). \textbf{(b)} A symmetrized axial attention layer, the building block of the neural network torso. Symmetrized axial attention incorporates a symmetrization operation after each axial attention layer to favour exchange of information between rows and columns.}
\label{fig:nn_main_paper}
\end{figure}

\paragraph{Neural network.}
The neural network at the core of \method takes as input the game state and outputs: (i) a policy, i.e., a probability distribution over the next action to play, and (ii) a probability distribution over the estimated return (i.e., the sum of rewards from the current state until the end of the game); see \Cref{fig:nn_overview}. The neural network has a key component, the \emph{torso}, where a set of attention operations \citep{vaswani2017attention} are applied. In \method, the torso interleaves two types of layers: axial attention \citep{ho2019axial} and \emph{symmetrization layers} (\Cref{fig:symmetrized_axial_attn}). Symmetrization layers are inspired by the empirical observation that, for a symmetric input $X \in \mathbb{R}^{N\times N}$ (with $X = X^{\top}$), preserving the symmetry of the output in-between axial attention layers significantly improves performance. Therefore, after each axial attention layer, we add a symmetrization operation, defined as $X\leftarrow \beta \odot X + (1-\beta)  \odot X^{\top}$, where `$\odot$' denotes element-wise product. For $\beta=\frac{1}{2}$, this makes the output $X$ symmetric. %
In practice, we found that letting $\beta$ be a learnable $N\times N$ matrix improves performance even further, even though $X$ is no longer symmetric, due to the ability of the network to exchange information between rows and columns after each axial attention layer. (When the input is a tensor, $X\in \mathbb{R}^{N\times N\times C}$, we apply the symmetrization operations independently across the dimensions $C$, using the same matrix $\beta$.) We refer to this architecture as \emph{symmetrized axial attention}. See \Cref{app:subsec:neural_network} for more details of the full neural network architecture.

Symmetrized axial attention performs better than the neural network of standard AlphaTensor (which requires attention operations across every pair of modes of the tensor) \citep{fawzi2022discovering}. It is also one of the key ingredients that allow \method to scale up to larger tensors: for $N=30$ and the same number of layers, symmetrized axial attention runs about $4$x faster and reduces the memory consumption by a factor of $3$x. 
(See also \Cref{app:subsec:nn_ablations} for further ablations.) We believe symmetrized axial attention may be useful for other applications beyond \method.

\paragraph{Gadgetization patterns.}
In \game, every time an action is played, we check whether it completes a gadget to adjust the reward signal accordingly. If the action $u^{(s)}$ at step $s$ does not complete a gadget, the immediate reward is simply $-1$. Otherwise, the reward is $+4$ if it completes a Toffoli, or $0$ if it completes a \CS gadget.
These checks are done up to Clifford operations, and they involve finding certain linear dependencies between $u^{(s)}$ and the previous actions:
\begin{compactitem}
    \item If $s\geq 7$, we check whether the action $u^{(s)}$ completes a \emph{Toffoli gadget}. Let $a\triangleq u^{(s-6)}$, $b\triangleq u^{(s-5)}$, and $c\triangleq u^{(s-4)}$. If the vectors $a$, $b$, and $c$ are linearly independent and all the following relationships hold, then we declare that $u^{(s)}$ completes a Toffoli gadget:
    \begin{equation*}
        u^{(s-3)}=a+b, \qquad u^{(s-2)}=a+c, \qquad u^{(s-1)}=a+b+c, \qquad\textrm{and}\qquad u^{(s)}=b+c.
    \end{equation*}
    \item If $s\geq 3$, we check whether the action $u^{(s)}$ completes a \emph{\CS gadget}. Let $a\triangleq u^{(s-2)}$ and $b\triangleq u^{(s-1)}$. If the vectors $a$ and $b$ are linearly independent and $u^{(s)}=a+b$, then $u^{(s)}$ completes a \CS gadget.
\end{compactitem}

We refer to the relationships above as \emph{gadgetization patterns} (see \Cref{app:subsec:gadgetization} for a justification of these steps). Each factor (action) may belong to at most one gadget; thus, if the action at step $s$ completes a gadget of any type, we only check for a new \CS gadget from step $s+3$ onward, and for a new Toffoli gadget from step $s+7$ onward.

\paragraph{Synthetic demonstrations.}
Like its predecessor, \method uses a data set of randomly generated tensor/factorization pairs to train the neural network. Specifically, the network is trained on a mixture of a supervised loss (which encourages it to imitate the synthetic demonstrations) and a standard RL loss (which encourages it to learn to decompose the target quantum signature tensor).
For each experiment, we generate $5$ million synthetic demonstrations.
To generate each demonstration, we first randomly sample the number of factors $R$ from a uniform distribution in $[1, 125]$ and then sample $R$ binary factors $u^{(r)}$, such that each element has a probability of $0.75$ of being set to $0$.

After having generated the factors, we randomly overwrite some of them to incorporate the patterns of the Toffoli and \CS gadgets, in order to help the network learn and identify those patterns. For each demonstration, we include at least one gadget with probability $0.9$. The number of gadgets is uniformly sampled in $[1, 15]$ (the value is also truncated when $R$ is not large enough), and for each gadget we sample the starting index $r$ and the type of gadget (Toffoli with probability $0.6$ and \CS with probability $0.4$), overwriting the necessary factors according to the gadget's pattern. We ensure gadgets do not overlap throughout the demonstration.

\paragraph{Change of basis.}
To increase the diversity of the played games, the actors apply a random change of basis to the target tensor $\mathcal{T}$ at the beginning of each game. From the quantum computation point of view, a change of basis can be implemented by a Clifford-only circuit, and therefore this procedure does not affect the resulting number of \tgate{s}.

A change of basis is represented by a matrix $M \in \{0, 1\}^{N\times N}$ such that $\det(M)=1$ under modulo $2$. We randomly generate $50\,000$ such basis changes and use them throughout the experiment, i.e., the actor first randomly picks a matrix $M$ and then applies the following transformation to the target tensor $\mathcal{T}$:
\begin{equation*}
    \widetilde{\mathcal{T}}_{ijk} = \sum_{a=1}^N \sum_{b=1}^N \sum_{c=1}^N M_{ia}M_{jb}M_{kc} \mathcal{T}_{abc}.
\end{equation*}
The actor then plays the game in that basis (i.e., it aims at finding a decomposition of $\widetilde{\mathcal{T}}$). After the game has finished, we can map the played actions back to the standard basis by applying the inverse change of basis.

The diversity injected by the change of basis facilitates the agent's task, because it suffices to find a low-rank decomposition in any basis. Note that gadgets are not affected by the change of basis, since their patterns correspond to linear relationships that are preserved after linear transformations.

\paragraph{Data augmentation.}
For each game played by the actors, we obtain a new game by swapping two of the actions in it. Specifically, we swap the last action that is not part of a gadget with a randomly chosen action that is also not part of a gadget.
Mathematically, swapping actions is a valid operation due to commutativity. From the quantum computation point of view, it is a valid operation because the considered unitary matrices are diagonal.

Besides the action permutation, we also permute the inputs at acting time every time we query the neural network. In particular, we apply a random permutation to the input tensor and past actions that represent the current state, and then invert this permutation at the output of the network's policy head. Similarly, at training time, we apply a random permutation to both the input and the policy targets, and train the neural network on this transformed version of the data. In practice, we sample $100$ permutations at the beginning of an experiment, and use them thereafter.

\paragraph{Sample-based MCTS.}
In \method, the action space is huge, and therefore it is not possible to enumerate all the possible actions from a given state. Instead, \method relies on sample-based MCTS, similarly to sampled AlphaZero \citep{hubert2021learning}.

Before committing to an action, the agent uses a \emph{search tree} to explore multiple actions to play. Each node in the tree represents a state, and each edge represents an action. For each state-action pair $(s, a)$, we store a set of statistics: the visit count $N(s,a)$, the action value $Q(s,a)$, and the empirical policy probability $\widehat{\pi}(s,a)$. The search consists of multiple simulations; each simulation traverses the tree from the root state $s_0$ until a leaf state $s_L$ is reached, by recursively selecting in each state $s$ an action $a$ that has high empirical policy probability and high value but has not been frequently explored, i.e., we choose $a$ according to
\begin{equation*}
    \arg\max_{a} Q(s,a) + c(s) \widehat{\pi}(s,a)\frac{\sqrt{\sum_b N(s,b)}}{1+N(s,a)},
\end{equation*}
where $c(s)$ is an exploration factor. The MCTS simulation keeps choosing actions until a leaf state $s_L$ is reached; when that happens, the network is queried on $s_L$ --- obtaining the policy probability $\pi(s_L, \cdot)$ and the return probability $z(s_L)$. We sample $K$ actions $a_k$ from the policy to initialize $\widehat{\pi}(s_L,a) = \frac{1}{K}\sum_k \delta_{a, a_k}$, and initialize $Q(s_L,a)$ from $z(s_L)$ using a risk-seeking value \citep{fawzi2022discovering}. After the node $s_L$ has been expanded, we update the visit counts and values on the simulated trajectory in a backward pass \citep{hubert2021learning}.
We simulate $800$ trajectories using MCTS; after that, we use the visit counts of the actions at the root of the search tree to form a sample-based improved policy, following \citet{fawzi2022discovering}, and commit to an action sampled from that improved policy.

\paragraph{Training regime.}
We train \method  on a TPU with 256 cores, with a total batch size of $2\,048$, training for $1$ million iterations on average (the number of training iterations varies depending on the application, ranging between $80\,000$ and $3$ million iterations). We use $3\,600$ actors on average per experiment, each running on a standalone TPU.
Each experiment is ran on a set of target tensors, which we group by application (i.e., one experiment for addition, one for unary iteration, etc.); this allows the RL agent to recognize and exploit the decomposition patterns that might be present across multiple tensors within the same experiment. We find that the experiment variance is very low: a single experiment is enough for \method to find the best constructions for each application.

\paragraph{Other versions of \method.}
We found empirically that the GF($2^m$)-mult circuits were particularly challenging to optimize, and they required additional training iterations. To reduce the size of the search space and accelerate the training procedure, we developed a variant of \method that significantly favours Toffoli gates. Specifically, every time that three consecutive and linearly independent actions (factors) are played, the environment automatically applies the remaining four actions that would complete the Toffoli gadget. We applied this version of \method to optimize these targets only.

\subsection{Verification}
\label{subsec:verification}

In order to verify that the results of \method are correct, we used the \texttt{feynver} tool developed by Amy \cite{amy2016feynman} after each step of the compilation process discussed in \Cref{app:subsec:compilation}. It uses the sum-over-paths formalism \cite{dawson2005quantum} to produce a proof of equality and can be applied to any circuit. However, it is a sound but not complete method, so it may fail both to prove and to disprove equality for some pairs of circuits. Since the tool runs slowly for larger circuits, we ran it wherever practical (limiting each run to several hours). Each circuit was verified assuming that any intermediate measurements introduced by Hadamard gadetization always return zero, because our software pipeline does not generate the classically-controlled correction circuits (these can be easily constructed but are irrelevant to \method). 

This allowed us to verify inputs up to the point where a tensor is obtained. In order to verify that the decompositions constructed by \method were correct, we computed the corresponding tensor and checked that it was equal to the original input tensor. While we did not use the optimized decompositions to produce circuits, they could be verified in the same way as discussed above.

\subsection*{Acknowledgements}

The authors thank Steve Clark, Alexander L.\ Gaunt, Alexandre Krajenbrink, Luca Mondada, and Richie Yeung
for their feedback on the paper; and Ryan Babbush, Sergio Boixo, Craig Gidney, Matt Harrigan, Tanuj Khattar, and Nicholas Rubin for insightful discussions and ideas.

\bibliography{main_arxiv_stripped}

\begin{thebibliography}{75}
\providecommand{\natexlab}[1]{#1}
\providecommand{\url}[1]{\texttt{#1}}
\expandafter\ifx\csname urlstyle\endcsname\relax
  \providecommand{\doi}[1]{doi: #1}\else
  \providecommand{\doi}{doi: \begingroup \urlstyle{rm}\Url}\fi

\bibitem[Feynman(1982)]{feynman1982simulating}
Richard~P Feynman.
\newblock Simulating physics with computers.
\newblock \emph{International journal of theoretical physics}, 21\penalty0
  (6/7):\penalty0 467--488, 1982.

\bibitem[Manin(1980)]{YuriManinQuantumComputation}
Yuri Manin.
\newblock \emph{Computable and noncomputable}.
\newblock Sovetskoe Radio, 1980.

\bibitem[Shor(1997)]{Shor_1997}
Peter~W. Shor.
\newblock Polynomial-time algorithms for prime factorization and discrete
  logarithms on a quantum computer.
\newblock \emph{SIAM Journal on Computing}, 26\penalty0 (5):\penalty0
  1484–1509, October 1997.
\newblock ISSN 1095-7111.
\newblock \doi{10.1137/s0097539795293172}.

\bibitem[Blunt et~al.(2022)Blunt, Camps, Crawford, Izsák, Leontica, Mirani,
  Moylett, Scivier, S\"{u}nderhauf, Schopf, Taylor, and Holzmann]{Blunt2022}
Nick~S. Blunt, Joan Camps, Ophelia Crawford, Róbert Izsák, Sebastian
  Leontica, Arjun Mirani, Alexandra~E. Moylett, Sam~A. Scivier, Christoph
  S\"{u}nderhauf, Patrick Schopf, Jacob~M. Taylor, and Nicole Holzmann.
\newblock Perspective on the current state-of-the-art of quantum computing for
  drug discovery applications.
\newblock \emph{Journal of Chemical Theory and Computation}, 18\penalty0
  (12):\penalty0 7001–7023, November 2022.
\newblock ISSN 1549-9626.
\newblock \doi{10.1021/acs.jctc.2c00574}.

\bibitem[Dalzell et~al.(2023)Dalzell, McArdle, Berta, Bienias, Chen, Gilyén,
  Hann, Kastoryano, Khabiboulline, Kubica, Salton, Wang, and
  Brandão]{dalzell2023quantum}
Alexander~M. Dalzell, Sam McArdle, Mario Berta, Przemyslaw Bienias, Chi-Fang
  Chen, András Gilyén, Connor~T. Hann, Michael~J. Kastoryano, Emil~T.
  Khabiboulline, Aleksander Kubica, Grant Salton, Samson Wang, and Fernando G.
  S.~L. Brandão.
\newblock Quantum algorithms: A survey of applications and end-to-end
  complexities, 2023.

\bibitem[Bravyi and Kitaev(2005)]{bravyi2005universal}
Sergey Bravyi and Alexei Kitaev.
\newblock Universal quantum computation with ideal {C}lifford gates and noisy
  ancillas.
\newblock \emph{Physical Review A}, 71:\penalty0 022316, 2005.

\bibitem[Campbell et~al.(2017)Campbell, Terhal, and Vuillot]{Campbell_2017}
Earl~T. Campbell, Barbara~M. Terhal, and Christophe Vuillot.
\newblock Roads towards fault-tolerant universal quantum computation.
\newblock \emph{Nature}, 549\penalty0 (7671):\penalty0 172–179, September
  2017.
\newblock ISSN 1476-4687.
\newblock \doi{10.1038/nature23460}.

\bibitem[Gidney and Fowler(2019)]{gidney2019efficient}
Craig Gidney and Austin~G. Fowler.
\newblock Efficient magic state factories with a catalyzed {$|CCZ\rangle$} to
  {$2|T\rangle$} transformation.
\newblock \emph{Quantum}, 3:\penalty0 135, 2019.

\bibitem[Gottesman(1998)]{gottesman1998heisenger}
Daniel Gottesman.
\newblock The {H}eisenberg representation of quantum computers.
\newblock \emph{arXiv preprint arXiv:quant-ph/9807006}, 1998.

\bibitem[Calderbank and Shor(1996)]{Calderbank_1996}
A.~R. Calderbank and Peter~W. Shor.
\newblock Good quantum error-correcting codes exist.
\newblock \emph{Physical Review A}, 54\penalty0 (2):\penalty0 1098–1105,
  August 1996.
\newblock ISSN 1094-1622.
\newblock \doi{10.1103/physreva.54.1098}.

\bibitem[Aharonov and Ben-Or(1996)]{aharonov1996fault}
Dorit Aharonov and Michael Ben-Or.
\newblock Fault tolerant quantum computation with constant error, 1996.

\bibitem[Eastin and Knill(2009)]{Eastin_2009}
Bryan Eastin and Emanuel Knill.
\newblock Restrictions on transversal encoded quantum gate sets.
\newblock \emph{Physical Review Letters}, 102\penalty0 (11), March 2009.
\newblock ISSN 1079-7114.
\newblock \doi{10.1103/physrevlett.102.110502}.

\bibitem[van~de Wetering and Amy(2023)]{vandewetering2023optimising}
John van~de Wetering and Matthew Amy.
\newblock Optimising {T}-count is {NP}-hard.
\newblock \emph{arXiv preprint arXiv:2310.05958}, 2023.

\bibitem[Amy et~al.(2013{\natexlab{a}})Amy, Maslov, and
  Mosca]{amy2013polynomial}
Matthew Amy, Dmitri Maslov, and Michele Mosca.
\newblock Polynomial-time {T}-depth optimization of clifford+{T} circuits via
  matroid partitioning.
\newblock \emph{arXiv preprint arXiv:1303.2042}, 2013{\natexlab{a}}.

\bibitem[Gosset et~al.(2014)Gosset, Kliuchnikov, Mosca, and
  Russo]{gosset2013algorithm}
David Gosset, Vadym Kliuchnikov, Michele Mosca, and Vincent Russo.
\newblock An algorithm for the {T}-count.
\newblock \emph{Quantum Information \& Computation}, 14\penalty0
  (15–16):\penalty0 1261--1276, nov 2014.

\bibitem[Heyfron and Campbell(2018)]{heyfron2018efficient}
Luke~E. Heyfron and Earl~T. Campbell.
\newblock An efficient quantum compiler that reduces {T} count.
\newblock \emph{Quantum Science and Technology}, 4\penalty0 (1):\penalty0
  015004, sep 2018.

\bibitem[Nam et~al.(2018)Nam, Ross, Su, Childs, and Maslov]{nam2018automated}
Yunseong Nam, Neil~J. Ross, Yuan Su, Andrew~M. Childs, and Dmitri Maslov.
\newblock Automated optimization of large quantum circuits with continuous
  parameters.
\newblock \emph{npj Quantum Information}, 4\penalty0 (23), 2018.

\bibitem[Amy and Mosca(2019)]{amy2019tcount}
Matthew Amy and Michele Mosca.
\newblock {T}-count optimization and {R}eed-{M}uller codes.
\newblock \emph{IEEE Transactions on Information Theory}, 65\penalty0
  (8):\penalty0 4771--4784, 2019.

\bibitem[Kissinger and van~de Wetering(2019)]{kissinger2019reducing}
Aleks Kissinger and John van~de Wetering.
\newblock Reducing {T}-count with the {ZX}-calculus.
\newblock \emph{arXiv preprint arXiv:1903.10477}, 2019.

\bibitem[Gheorghiu et~al.(2022)Gheorghiu, Mosca, and
  Mukhopadhyay]{gheorghiu2022tcount}
Vlad Gheorghiu, Michele Mosca, and Priyanka Mukhopadhyay.
\newblock {T-}count and {T-}depth of any multi-qubit unitary.
\newblock \emph{npj Quantum Information}, 8\penalty0 (141), 2022.

\bibitem[Kissinger and van~de Wetering(2022)]{kissinger2022simulating}
Aleks Kissinger and John van~de Wetering.
\newblock Simulating quantum circuits with {ZX}-calculus reduced stabiliser
  decompositions.
\newblock \emph{Quantum Science and Technology}, 7:\penalty0 044001, 2022.

\bibitem[{de Beaudrap} et~al.(2019){de Beaudrap}, Bian, and
  Wang]{debeaudrap2019techniques}
Neil {de Beaudrap}, Xiaoning Bian, and Quanlong Wang.
\newblock Techniques to reduce $\pi/4$-parity-phase circuits, motivated by the
  {ZX} calculus.
\newblock \emph{arXiv preprint arXiv:1911.09039}, 2019.

\bibitem[{de Beaudrap} et~al.(2020){de Beaudrap}, Bian, and
  Wang]{debeaudrap2020fast}
Niel {de Beaudrap}, Xiaoning Bian, and Quanlong Wang.
\newblock Fast and effective techniques for {T}-count reduction via spider nest
  identities.
\newblock In \emph{Conference on the Theory of Quantum Computation,
  Communication and Cryptography}, volume 158, pages 11:1--11:23, 2020.

\bibitem[Fawzi et~al.(2022)Fawzi, Balog, Huang, Hubert, Romera-Paredes,
  Barekatain, Novikov, Ruiz, Schrittwieser, Swirszcz, Silver, Hassabis, and
  Kohli]{fawzi2022discovering}
Alhussein Fawzi, Matej Balog, Aja Huang, Thomas Hubert, Bernardino
  Romera-Paredes, Mohammadamin Barekatain, Alexander Novikov, Francisco J.~R.
  Ruiz, Julian Schrittwieser, Grzegorz Swirszcz, David Silver, Demis Hassabis,
  and Pushmeet Kohli.
\newblock Discovering faster matrix multiplication algorithms with
  reinforcement learning.
\newblock \emph{Nature}, 610:\penalty0 47--53, 2022.

\bibitem[Jones(2013)]{jones2013lowoverhead}
Cody Jones.
\newblock Low-overhead constructions for the fault-tolerant {T}offoli gate.
\newblock \emph{Physical Review A}, 87\penalty0 (2), 2013.

\bibitem[Beverland et~al.(2020)Beverland, Campbell, Howard, and
  Kliuchnikov]{beverland2020lower}
Michael Beverland, Earl Campbell, Mark Howard, and Vadym Kliuchnikov.
\newblock Lower bounds on the non-{C}lifford resources for quantum
  computations.
\newblock \emph{Quantum Science and Technology}, 5\penalty0 (3), 2020.

\bibitem[Cheung et~al.(2008)Cheung, Maslov, Mathew, and
  Pradhan]{cheung2008design}
Donny Cheung, Dmitri Maslov, Jimson Mathew, and Dhiraj~K. Pradhan.
\newblock On the design and optimization of a quantum polynomial-time attack on
  elliptic curve cryptography.
\newblock In \emph{Theory of Quantum Computation, Communication, and
  Cryptography}, volume 5106, pages 96--104, 2008.

\bibitem[Karatsuba and Ofman(1962)]{karatsuba1962multiplication}
Anatoly Karatsuba and Yuri Ofman.
\newblock Multiplication of many-digital numbers by automatic computers.
\newblock \emph{Proceedings of the USSR Academy of Sciences}, 145:\penalty0
  293--294, 1962.

\bibitem[Bennett(1989)]{Bennett1989-ct}
Charles~H Bennett.
\newblock Time/space trade-offs for reversible computation.
\newblock \emph{SIAM J. Comput.}, 18\penalty0 (4):\penalty0 766--776, August
  1989.

\bibitem[Gidney(2019)]{gidney2019asymptotically}
Craig Gidney.
\newblock Asymptotically efficient quantum {K}aratsuba multiplication, 2019.

\bibitem[Gidney(2018)]{gidney2018halving}
Craig Gidney.
\newblock Halving the cost of quantum addition.
\newblock \emph{Quantum}, 2\penalty0 (74), 2018.

\bibitem[Reiher et~al.(2017{\natexlab{a}})Reiher, Wiebe, Svore, Wecker, and
  Troyer]{reiher2017elucidating}
Markus Reiher, Nathan Wiebe, Krysta~M. Svore, Dave Wecker, and Matthias Troyer.
\newblock Elucidating reaction mechanisms on quantum computers.
\newblock \emph{Proceedings of the National Academy of Sciences}, 114\penalty0
  (29):\penalty0 7555--7560, 2017{\natexlab{a}}.

\bibitem[Babbush et~al.(2018)Babbush, Gidney, Berry, Wiebe, McClean, Paler,
  Fowler, and Neven]{babbush2018encoding}
Ryan Babbush, Craig Gidney, Dominic~W. Berry, Nathan Wiebe, Jarrod McClean,
  Alexandru Paler, Austin Fowler, and Hartmut Neven.
\newblock Encoding electronic spectra in quantum circuits with linear t
  complexity.
\newblock \emph{Physical Review X}, 8:\penalty0 041015, Oct 2018.

\bibitem[Vandaele et~al.(2023)Vandaele, Martiel, Perdrix, and
  Vuillot]{vandaele2023optimal}
Vivien Vandaele, Simon Martiel, Simon Perdrix, and Christophe Vuillot.
\newblock Optimal {H}adamard gate count for {C}lifford$+${T} synthesis of
  {P}auli rotations sequences.
\newblock \emph{arXiv preprint arXiv:2302.07040}, 2023.

\bibitem[Amy et~al.(2013{\natexlab{b}})Amy, Maslov, Mosca, and
  Roetteler]{amy2013meet}
Matthew Amy, Dmitri Maslov, Michele Mosca, and Martin Roetteler.
\newblock A meet-in-the-middle algorithm for fast synthesis of depth-optimal
  quantum circuits.
\newblock \emph{IEEE Transactions on Computer-Aided Design of Integrated
  Circuits and Systems}, 32\penalty0 (6):\penalty0 818--830,
  2013{\natexlab{b}}.

\bibitem[Amy(2016)]{amy2016feynman}
Matthew Amy.
\newblock Feynman.
\newblock \url{https://github.com/meamy/feynman}, 2016.

\bibitem[Zhang and Chen(2019)]{zhang2019optimizing}
Fang Zhang and Jianxin Chen.
\newblock Optimizing {T} gates in {C}lifford$+${T} circuit as $\pi/4$ rotations
  around {P}aulis.
\newblock \emph{arXiv preprint arXiv:quant-ph/1903.12456}, 2019.

\bibitem[Munson et~al.(2020)Munson, Coecke, and Wang]{munson2020andgates}
Anthony Munson, Bob Coecke, and Quanlong Wang.
\newblock {AND}-gates in {ZX}-calculus: Spider nest identities and
  {QBC}-completeness.
\newblock In \emph{Quantum Physics and Logic}, 2020.

\bibitem[de~Moura and Bj{\o}rner(2008)]{demoura2008z3}
Leonardo de~Moura and Nikolaj Bj{\o}rner.
\newblock Z3: An efficient {SMT} solver.
\newblock In C.~R. Ramakrishnan and Jakob Rehof, editors, \emph{Tools and
  Algorithms for the Construction and Analysis of Systems}, pages 337--340,
  Berlin, Heidelberg, 2008. Springer Berlin Heidelberg.
\newblock ISBN 978-3-540-78800-3.

\bibitem[Montgomery(2005)]{montgomery2005five}
Peter~L. Montgomery.
\newblock Five, six, and seven-term {K}aratsuba-like formulae.
\newblock \emph{IEEE Transactions on Computers}, 54\penalty0 (3):\penalty0
  362--369, 2005.

\bibitem[Fan and Hasan(2007)]{fan2007comments}
Haining Fan and M.~Anwar Hasan.
\newblock Comments on ``five, six, and seven-term {K}aratsuba-like formulae''.
\newblock \emph{IEEE Transactions on Computers}, 56\penalty0 (5):\penalty0
  716--717, 2007.
\newblock \doi{10.1109/TC.2007.1024}.

\bibitem[Barbulescu et~al.(2012)Barbulescu, Detrey, Estibals, and
  Zimmermann]{barbulescu2012finding}
Razvan Barbulescu, J{\'e}r{\'e}mie Detrey, Nicolas Estibals, and Paul
  Zimmermann.
\newblock Finding optimal formulae for bilinear maps.
\newblock In Ferruh {\"O}zbudak and Francisco Rodr{\'\i}guez-Henr{\'\i}quez,
  editors, \emph{Arithmetic of Finite Fields}, pages 168--186, Berlin,
  Heidelberg, 2012. Springer Berlin Heidelberg.
\newblock ISBN 978-3-642-31662-3.

\bibitem[Cuccaro et~al.(2004)Cuccaro, Draper, Kutin, and
  Moulton]{cuccaro2004new}
Steven~A. Cuccaro, Thomas~G. Draper, Samuel~A. Kutin, and David~Petrie Moulton.
\newblock A new quantum ripple-carry addition circuit.
\newblock \emph{arXiv preprint arXiv:quant-ph/0410184}, 2004.

\bibitem[Or{\'u}s(2019)]{chem-tensornet}
Rom{\'a}n Or{\'u}s.
\newblock Tensor networks for complex quantum systems.
\newblock \emph{Nature Reviews Physics}, 1\penalty0 (9):\penalty0 538--550,
  2019.
\newblock \doi{10.1038/s42254-019-0086-7}.

\bibitem[Reiher et~al.(2017{\natexlab{b}})Reiher, Wiebe, Svore, Wecker, and
  Troyer]{msft-femoco}
Markus Reiher, Nathan Wiebe, Krysta~M. Svore, Dave Wecker, and Matthias Troyer.
\newblock Elucidating reaction mechanisms on quantum computers.
\newblock \emph{Proceedings of the National Academy of Sciences}, 114\penalty0
  (29):\penalty0 7555--7560, 2017{\natexlab{b}}.
\newblock \doi{10.1073/pnas.1619152114}.

\bibitem[Lee et~al.(2021)Lee, Berry, Gidney, Huggins, McClean, Wiebe, and
  Babbush]{PRXQuantum.2.030305}
Joonho Lee, Dominic~W. Berry, Craig Gidney, William~J. Huggins, Jarrod~R.
  McClean, Nathan Wiebe, and Ryan Babbush.
\newblock Even more efficient quantum computations of chemistry through tensor
  hypercontraction.
\newblock \emph{PRX Quantum}, 2:\penalty0 030305, Jul 2021.
\newblock \doi{10.1103/PRXQuantum.2.030305}.

\bibitem[von Burg et~al.(2021)von Burg, Low, H\"aner, Steiger, Reiher,
  Roetteler, and Troyer]{vonburg2021quantum}
Vera von Burg, Guang~Hao Low, Thomas H\"aner, Damian~S. Steiger, Markus Reiher,
  Martin Roetteler, and Matthias Troyer.
\newblock Quantum computing enhanced computational catalysis.
\newblock \emph{Phys. Rev. Res.}, 3:\penalty0 033055, Jul 2021.
\newblock \doi{10.1103/PhysRevResearch.3.033055}.

\bibitem[Berry et~al.(2019)Berry, Gidney, Motta, McClean, and
  Babbush]{Berry2019qubitizationof}
Dominic~W. Berry, Craig Gidney, Mario Motta, Jarrod~R. McClean, and Ryan
  Babbush.
\newblock Qubitization of {A}rbitrary {B}asis {Q}uantum {C}hemistry
  {L}everaging {S}parsity and {L}ow {R}ank {F}actorization.
\newblock \emph{{Quantum}}, 3:\penalty0 208, December 2019.
\newblock ISSN 2521-327X.
\newblock \doi{10.22331/q-2019-12-02-208}.

\bibitem[Moro et~al.(2021)Moro, Paris, Restelli, and Prati]{moro2021quantum}
Lorenzo Moro, Matteo G.~A. Paris, Marcello Restelli, and Enrico Prati.
\newblock Quantum compiling by deep reinforcement learning.
\newblock \emph{Communications Physics}, 4\penalty0 (1):\penalty0 178, 2021.
\newblock \doi{10.1038/s42005-021-00684-3}.

\bibitem[F\"{o}sel et~al.(2021)F\"{o}sel, Niu, Marquardt, and
  Li]{fosel2021quantum}
Thomas F\"{o}sel, Murphy~Y. Niu, Floriand Marquardt, and Li~Li.
\newblock Quantum circuit optimization with deep reinforcement learning.
\newblock \emph{arXiv preprint arXiv:quant-ph/2103.07585}, 2021.

\bibitem[Quetschlich et~al.(2023)Quetschlich, Burgholzer, and
  Wille]{quetschlich2023compiler}
Nils Quetschlich, Lukas Burgholzer, and Robert Wille.
\newblock Compiler optimization for quantum computing using reinforcement
  learning, 2023.

\bibitem[Ostaszewski et~al.(2021)Ostaszewski, Trenkwalder, Masarczyk, Scerri,
  and Dunjko]{ostaszewski2021reinforcement}
Mateusz Ostaszewski, Lea~M. Trenkwalder, Wojciech Masarczyk, Eleanor Scerri,
  and Vedran Dunjko.
\newblock Reinforcement learning for optimization of variational quantum
  circuit architectures.
\newblock In M.~Ranzato, A.~Beygelzimer, Y.~Dauphin, P.S. Liang, and J.~Wortman
  Vaughan, editors, \emph{Advances in Neural Information Processing Systems},
  volume~34, pages 18182--18194. Curran Associates, Inc., 2021.

\bibitem[Kuo et~al.(2021)Kuo, Fang, and Chen]{kuo2021quantum}
En-Jui Kuo, Yao-Lung~L. Fang, and Samuel Y.-C. Chen.
\newblock Quantum architecture search via deep reinforcement learning.
\newblock \emph{arXiv preprint arXiv:quant-ph/2104.07715}, 2021.

\bibitem[Zhu and Hou(2023)]{zhu2023quantum}
Xianchao Zhu and Xiaokai Hou.
\newblock Quantum architecture search via truly proximal policy optimization.
\newblock \emph{Scientific Reports}, 13\penalty0 (1):\penalty0 5157, 2023.
\newblock \doi{10.1038/s41598-023-32349-2}.

\bibitem[Yao et~al.(2022)Yao, Li, Bukov, Lin, and Ying]{yao2022montecarlo}
Jiahao Yao, Haoya Li, Marin Bukov, Lin Lin, and Lexing Ying.
\newblock {M}onte {C}arlo tree search based hybrid optimization of variational
  quantum circuits.
\newblock In Bin Dong, Qianxiao Li, Lei Wang, and Zhi-Qin~John Xu, editors,
  \emph{Proceedings of Mathematical and Scientific Machine Learning}, volume
  190 of \emph{Proceedings of Machine Learning Research}, pages 49--64. PMLR,
  15--17 Aug 2022.

\bibitem[Rosenhahn and Osborne(2023)]{rosenhahn2023montecarlo}
Bodo Rosenhahn and Tobias~J. Osborne.
\newblock Monte {C}arlo graph search for quantum circuit optimization.
\newblock \emph{Phys. Rev. A}, 108:\penalty0 062615, Dec 2023.
\newblock \doi{10.1103/PhysRevA.108.062615}.

\bibitem[Gidney and Jones(2021)]{gidney2021cccz}
Craig Gidney and Cody Jones.
\newblock A {CCCZ} gate performed with 6 {T} gates.
\newblock \emph{arXiv preprint arXiv:2106.11513}, 2021.

\bibitem[Amy and Ross(2021)]{amy2021phase}
Matthew Amy and Neil~J. Ross.
\newblock Phase-state duality in reversible circuit design.
\newblock \emph{Physical Review A}, 104:\penalty0 052602, Nov 2021.

\bibitem[Silver et~al.(2018)Silver, Hubert, Schrittwieser, Antonoglou, Lai,
  Guez, Lanctot, Sifre, Kumaran, Graepel, Lillicrap, Simonyan, and
  Hassabis]{silver2018general}
David Silver, Thomas Hubert, Julian Schrittwieser, Ioannis Antonoglou, Matthew
  Lai, Arthur Guez, Marc Lanctot, Laurent Sifre, Dharshan Kumaran, Thore
  Graepel, Timothy Lillicrap, Karen Simonyan, and Demis Hassabis.
\newblock A general reinforcement learning algorithm that masters chess, shogi,
  and {G}o through self-play.
\newblock \emph{Science}, 362\penalty0 (6419):\penalty0 1140--1144, 2018.

\bibitem[Vaswani et~al.(2017)Vaswani, Shazeer, Parmar, Uszkoreit, Jones, Gomez,
  Kaiser, and Polosukhin]{vaswani2017attention}
Ashish Vaswani, Noam Shazeer, Niki Parmar, Jakob Uszkoreit, Llion Jones,
  Aidan~N Gomez, {\L}ukasz Kaiser, and Illia Polosukhin.
\newblock Attention is all you need.
\newblock In \emph{Advances in Neural Information Processing Systems}, 2017.

\bibitem[Ho et~al.(2019)Ho, Kalchbrenner, Weissenborn, and
  Salimans]{ho2019axial}
Jonathan Ho, Nal Kalchbrenner, Dirk Weissenborn, and Tim Salimans.
\newblock Axial attention in multidimensional transformers.
\newblock \emph{arXiv preprint arXiv:1912.12180}, 2019.

\bibitem[Hubert et~al.(2021)Hubert, Schrittwieser, Antonoglou, Barekatain,
  Schmitt, and Silver]{hubert2021learning}
Thomas Hubert, Julian Schrittwieser, Ioannis Antonoglou, Mohammadamin
  Barekatain, Simon Schmitt, and David Silver.
\newblock Learning and planning in complex action spaces.
\newblock In \emph{International Conference on Machine Learning}, pages
  4476--4486, 2021.

\bibitem[Dawson et~al.(2005)Dawson, Hines, Mortimer, Haselgrove, Nielsen, and
  Osborne]{dawson2005quantum}
Christopher~M. Dawson, Andrew~P. Hines, Duncan Mortimer, Henry~L. Haselgrove,
  Michael~A. Nielsen, and Tobias~J. Osborne.
\newblock Quantum computing and polynomial equations over the finite field
  $z$2.
\newblock \emph{Quantum Info. Comput.}, 5\penalty0 (2):\penalty0 102--112, mar
  2005.
\newblock ISSN 1533-7146.

\bibitem[Su et~al.(2021)Su, Berry, Wiebe, Rubin, and
  Babbush]{PRXQuantum.2.040332}
Yuan Su, Dominic~W. Berry, Nathan Wiebe, Nicholas Rubin, and Ryan Babbush.
\newblock Fault-tolerant quantum simulations of chemistry in first
  quantization.
\newblock \emph{PRX Quantum}, 2:\penalty0 040332, Nov 2021.
\newblock \doi{10.1103/PRXQuantum.2.040332}.

\bibitem[Steudtner et~al.(2023)Steudtner, Morley-Short, Pol, Sim, Cortes,
  Loipersberger, Parrish, Degroote, Moll, Santagati, and
  Streif]{Steudtner2023faulttolerant}
Mark Steudtner, Sam Morley-Short, William Pol, Sukin Sim, Cristian~L. Cortes,
  Matthias Loipersberger, Robert~M. Parrish, Matthias Degroote, Nikolaj Moll,
  Raffaele Santagati, and Michael Streif.
\newblock Fault-tolerant quantum computation of molecular observables.
\newblock \emph{{Quantum}}, 7:\penalty0 1164, November 2023.
\newblock ISSN 2521-327X.
\newblock \doi{10.22331/q-2023-11-06-1164}.

\bibitem[de~Wolf(2019)]{dewolf2019quantum}
Ronald de~Wolf.
\newblock Quantum computing: Lecture notes, 2019.

\bibitem[Bauer et~al.(2020)Bauer, Bravyi, Motta, and Chan]{Bauer_2020}
Bela Bauer, Sergey Bravyi, Mario Motta, and Garnet Kin-Lic Chan.
\newblock Quantum algorithms for quantum chemistry and quantum materials
  science.
\newblock \emph{Chemical Reviews}, 120\penalty0 (22):\penalty0 12685–12717,
  October 2020.
\newblock ISSN 1520-6890.
\newblock \doi{10.1021/acs.chemrev.9b00829}.
\newblock URL \url{http://dx.doi.org/10.1021/acs.chemrev.9b00829}.

\bibitem[Lee et~al.(2023)Lee, Lee, Zhai, Tong, Dalzell, Kumar, Helms, Gray,
  Cui, Liu, Kastoryano, Babbush, Preskill, Reichman, Campbell, Valeev, Lin, and
  Chan]{chan-nature-advantage}
Seunghoon Lee, Joonho Lee, Huanchen Zhai, Yu~Tong, Alexander~M. Dalzell,
  Ashutosh Kumar, Phillip Helms, Johnnie Gray, Zhi-Hao Cui, Wenyuan Liu,
  Michael Kastoryano, Ryan Babbush, John Preskill, David~R. Reichman, Earl~T.
  Campbell, Edward~F. Valeev, Lin Lin, and Garnet Kin-Lic Chan.
\newblock Evaluating the evidence for exponential quantum advantage in
  ground-state quantum chemistry.
\newblock \emph{Nature Communications}, 14\penalty0 (1):\penalty0 1952, 2023.
\newblock \doi{10.1038/s41467-023-37587-6}.

\bibitem[Childs and Wiebe(2012)]{lcu_wiebe}
Andrew~M. Childs and Nathan Wiebe.
\newblock Hamiltonian simulation using linear combinations of unitary
  operations.
\newblock \emph{Quantum Info. Comput.}, 12\penalty0 (11–12):\penalty0
  901–924, nov 2012.
\newblock ISSN 1533-7146.

\bibitem[Low et~al.(2018)Low, Kliuchnikov, and Schaeffer]{low2018trading}
Guang~Hao Low, Vadym Kliuchnikov, and Luke Schaeffer.
\newblock Trading {T}-gates for dirty qubits in state preparation and unitary
  synthesis.
\newblock \emph{arXiv preprint arXiv:1812.00954}, 2018.

\bibitem[Low and Chuang(2019)]{Low2019hamiltonian}
Guang~Hao Low and Isaac~L. Chuang.
\newblock Hamiltonian {S}imulation by {Q}ubitization.
\newblock \emph{{Quantum}}, 3:\penalty0 163, July 2019.
\newblock ISSN 2521-327X.
\newblock \doi{10.22331/q-2019-07-12-163}.
\newblock URL \url{https://doi.org/10.22331/q-2019-07-12-163}.

\bibitem[van~de Wetering et~al.(2024)van~de Wetering, Yeung, Laakkonen, and
  Kissinger]{vandewetering2023parameterized}
John van~de Wetering, Richie Yeung, Tuomas Laakkonen, and Aleks Kissinger.
\newblock Optimal compilation of parametrised quantum circuits.
\newblock \emph{arXiv preprint arXiv:2401.12877}, 2024.

\bibitem[Vandaele et~al.(2022)Vandaele, Martiel, and
  de~Brugi{\`{e}}re]{vandaele2022phase}
Vivien Vandaele, Simon Martiel, and Timoth{\'{e}}e~Goubault de~Brugi{\`{e}}re.
\newblock Phase polynomials synthesis algorithms for {NISQ} architectures and
  beyond.
\newblock \emph{Quantum Science and Technology}, 7\penalty0 (4):\penalty0
  045027, sep 2022.

\bibitem[Meijer-van~de Griend and Duncan(2020)]{degriend2020architectureaware}
Arianne Meijer-van~de Griend and Ross Duncan.
\newblock Architecture-aware synthesis of phase polynomials for {NISQ} devices.
\newblock \emph{arXiv preprint arXiv:quant-ph/2004.06052}, 2020.

\bibitem[Meuli et~al.(2018)Meuli, Soeken, and De~Micheli]{meuli2018sat}
Giulia Meuli, Mathias Soeken, and Giovanni De~Micheli.
\newblock Sat-based {\{}cnot, t{\}} quantum circuit synthesis.
\newblock In Jarkko Kari and Irek Ulidowski, editors, \emph{Reversible
  Computation}, pages 175--188, Cham, 2018. Springer International Publishing.

\end{thebibliography}

\clearpage

\appendix

\section{Additional Background}
\label{app:sec:preliminaries}

\subsection{Quantum Computing}

\paragraph{Qubits.} While a classical computer works on bits $0$ and $1$, a quantum computer works with qubits, which are superpositions of $0$ and $1$, representable by a normalized vector in the vector space $\C^2$. We write $\ket{0},\ket{1} \in \C^2$ for the standard basis vectors. Generally, we adopt \emph{bra-ket} notation, and write $\ket{\psi}=a \ket{0} + b\ket{1}$ (with $a,b\in \C$) for any normalized complex vector (i.e., with $|a|^2 + |b|^2 = 1$). When we have two qubits, its state space is given by tensor product of the spaces for a single qubit: $\C^2\otimes \C^2 \cong \C^{2^2}$. Generally, the $N$-qubit state space consists of the normalized vectors in $\C^{2^N}$. A basis for this space is given by the states $\ket{\vec x}$ given by bit strings $\vec x\in \{0,1\}^N$.

\paragraph{Quantum gates.}
State transformations in quantum computing are linear operations that preserve the normalization, and hence map quantum states to quantum states. These operations are fully represented by unitary matrices. Besides the common examples from the main paper, such as the Hadamard gate or the X gate, we define the Z[$\alpha$] phase gate as
\begin{equation*}
    Z[\alpha] \ = \ \begin{pmatrix} 1 & 0 \\ 0 & \ee^{i\alpha}
    \end{pmatrix}.
\end{equation*}
We use the notation Z[$\alpha$] to explicitly indicate that the gate is parameterized by $\alpha\in\mathbb{R}$, so that $Z[\alpha]\ket{\psi} = a Z[\alpha] \ket{0} + b Z[\alpha] \ket{1} = a\ket{0} + b \ee^{i\alpha} \ket{1}$.
In contrast, when we refer to the (unparameterized) Z gate, we assume the special case $\alpha=\pi$, i.e., $Z\equiv Z[\pi]$, such that $Z\ket{\psi} = a\ket{0} - b\ket{1}$. The \tgate{} and the S gate are also phase gates, with $\alpha=\pi/4$ and $\alpha=\pi/2$, respectively.

\paragraph{Controlled gates.}
As explained in the main paper, the controlled gates are two-qubit gates that only perform a non-trivial action on the second qubit if the first qubit is in the $\ket{1}$ state. For instance, the \CNOT gate performs a NOT gate on the second qubit if the first qubit is $\ket{1}$. We can write this succinctly as $\CNOTmat\ket{x,y} = \ket{x,x\oplus y}$.
Similarly, the \CZ gate implements a Z phase gate on the second qubit if the first is in the $\ket{1}$ state, i.e., $\CZmat\ket{x,y} = (-1)^{xy} \ket{x,y}$, where here we write a juxtaposition of bits $xy$ to denote their AND. More generally, the controlled $Z[\alpha]$ phase gate acts as $\CZmat[\alpha]\ket{x,y} = \ee^{i\alpha xy} \ket{x,y}$.
We also consider the controlled \CNOT gate, also called Toffoli gate, whose action is $\Tofmat\ket{x,y,z} = \ket{x,y,(xy)\oplus z}$. Similarly, the controlled \CZ gate acts as $\CCZmat\ket{x,y,z} = (-1)^{xyz} \ket{x,y,z}$.

\paragraph{Universal gate sets.} The Hadamard $H$, \CNOT, and the Z phase gates $Z[\alpha]$ form a \emph{universal} gate set, meaning that any unitary matrix, and hence any quantum computation, can be decomposed into these gates. For instance, we have $X = HZ[\pi]H$.

In fault-tolerant quantum computation we cannot use a continuous family of gates like $Z[\alpha]$, and instead work with a small discretized set of quantum gates. A particularly often used one is the \tgate{}, $T\triangleq Z[\frac\pi4]$. The gate set consisting of $H$, \CNOT and $T$ is an \emph{approximately universal} gate set, meaning that any unitary can be approximated arbitrarily well by a circuit consisting of these gates.

\paragraph{Clifford + T.}
While quantum computing is generally believed to be hard to classically simulate, there is a useful subset of quantum computations for which there is an efficient classical simulation algorithm. These are known as \emph{Clifford} computations, and they are built out of Clifford gates. The Clifford gates are H, \CNOT, and S. For this reason, an often-used approximately universal gate set is \{H, \CNOT, S,  T\}, which is called the Clifford+\tgate{} set. As the Clifford gates are efficiently simulable, the number of \tgate{s} is a measure of how hard it is to perform the computation classically.

More concretely, \tgate{s} are also the hardest gates to execute: in most fault-tolerant architectures for quantum computation, the Clifford gates are relatively easy to perform, while \tgate{s} cost orders-of-magnitude more resources. This is because Clifford gates can be executed ``natively'' in many error-correcting codes, while \tgate{s} are implemented by injecting a magic state that first has to be distilled to a desired precision.

\paragraph{\tgate{} optimization.}
For this reason it is important to reduce the number of \tgate{s} required for a computation as much as possible. There are several ways to do this.
For instance, since $S=T^2$, whenever two \tgate{s} appear in a row, they combine to form a Clifford gate.
More generally, if we have $T^k$, then this costs just one \tgate{} if $k$ is odd, and is Clifford otherwise. %
This technique for combining \tgate{s} can be generalized to settings where they do not appear directly after one another on the same qubit ---which is a restrictive consideration. The generalization can be used to combine \tgate{s} that are applied to the same \emph{parity}.

Recall the actions of the Z[$\alpha$] and \CNOT gates. Considering $x,y \in\{0,1\}$, we have $Z[\alpha] \ket{x} = \ee^{i\alpha x}\ket{x}$ and $\CNOTmat\ket{x,y} = \ket{x, x\oplus y}$.
Thus, if we apply a $Z[\alpha]$ gate to the second qubit after a \CNOT, it applies a phase to the parity $x\oplus y$, i.e., $(I\otimes Z[\alpha])\CNOTmat\ket{x,y} = (I\otimes Z[\alpha]) \ket{x,x\oplus y} = \ee^{i\alpha (x\oplus y)} \ket{x,x\oplus y}$.

More generally, when we have a circuit consisting of just \CNOT and Z[$\alpha$] phase gates, the unitary matrix $U$ that it implements acts as $U\ket{\vec x} = \ee^{i \phi(\vec x)} \ket{A \vec x}$, where $\phi: \{0,1\}^n \to \R$ is a \emph{phase polynomial}, and $A$ is an invertible matrix that can be implemented with Clifford operations. %
Using this phase polynomial representation, phase gates that are implicitly acting on the same parity get combined. When we then resynthesize a quantum circuit from this representation, we hence possibly require fewer phase gates.

\paragraph{Multilinear phase polynomials.}
When we construct a phase polynomial from a \CNOT + Z[$\alpha$] circuit in this way, we get a phase polynomial consisting of XOR terms. This representation is not unique. As a simple example, we can see that $\ee^{i \pi (x+y+x\oplus y)} =\ee^{2\pi i} = 1$ for all $x,y\in\{0,1\}$. If we were to directly synthesize the first expression, this would require three phase gates corresponding to the three terms ($x$, $y$, and $x\oplus y$), even though it implements an identity. We can rewrite the phase polynomial into a unique form by decomposing each of the XOR terms into a \emph{multilinear} form (i.e., a polynomial where the degree of each variable in each term is either $0$ or $1$), e.g.,
\begin{equation*}
    x\oplus y = x+y - 2xy.
\end{equation*}
(Note that here `$+$' denotes regular addition in the real numbers.) We can generalize this expression to decompose an expression of $N$ parities into a multilinear polynomial:
\begin{equation}
    \label{eq:xor-decompose}
    x_1\oplus\cdots\oplus x_N \ = \ - \sum_{\vec y\neq \vec 0} (-2)^{|\vec y|-1} x_1^{y_1}\cdots x_N^{y_N}.
\end{equation}
Note that the weight of a degree-$k$ term is $2^{k-1}$. This means that if we had a phase polynomial $\phi$ where the smallest constant in its XOR form is $2\pi / 2^k$, then all the terms in its multilinear form of degree greater than $k$ will have a weight that is an integer multiple of $2\pi$, and hence can be ignored. In the example above, when we write $\ee^{i\pi(x+y+x\oplus y)}$ in multilinear form we get $\ee^{i\pi (2x+2y-2xy)}$, so that it only contains expressions that are multiples of $2\pi$.

In particular, if we have a circuit of \CNOT and \tgate{s} only (i.e., $\alpha=\pi/4$), the multilinear phase polynomial involves terms of degree at most $3$, after having removed all the terms leading to multiple of $2\pi$. %
Hence, there is a one-to-one correspondence between $N$-qubit diagonal \CNOT + T circuits and multilinear polynomials of the form
\begin{equation}\label{eq:phase-polynomial-multilinear}
    \phi(x_1,\ldots, x_N) \ = \ \frac\pi4\left( \sum_i a_i x_i + 2 \sum_{i<j} b_{ij} x_ix_j + 4  \sum_{i<j<k} c_{ijk} x_ix_jx_k \right),
\end{equation}
where $a_i \in \{0,\ldots,7\}$, $b_{ij}\in\{0,\ldots, 3\}$ and $c_{ijk} \in \{0,1\}$. 

We can now check whether two diagonal \CNOT + T circuits implement the same unitary by converting them to the multilinear phase polynomial representation of \Cref{eq:phase-polynomial-multilinear}. However, since we are considering Clifford operations to be free, we only need to check whether two \CNOT + T circuits are equal up to Clifford operations.

\paragraph{Symmetric 3-tensors.} The multilinear representation corresponds directly to a circuit of T, \CS and \CCZ gates: each $a_ix_i$ term is implemented by $a_i$ \tgate{s} acting on the $i$th qubit; each $b_{ij}x_ix_j$ term corresponds to $b_{ij}$ \CS gates on qubits $i$ and $j$; and each $c_{ijk}x_ix_jx_k$ term corresponds to one \CCZ gate on qubits $i$, $j$ and $k$ if $c_{ijk}=1$. Each of these terms only corresponds to a non-Clifford operation if the constant $a_i$, $b_{ij}$ or $c_{ijk}$ is odd. Hence, to check whether two diagonal \CNOT + T circuits implement the same unitary up to Clifford operations, we can calculate their multilinear polynomials in \Cref{eq:phase-polynomial-multilinear} and take modulo $2$ of their coefficients. That is, two \CNOT + T circuits are equal up to Clifford operations if and only if they have the same polynomial (after taking modulo $2$ of their coefficients).
All this information about the multilinear polynomial can then be captured in a single symmetric 3-tensor $\mathcal{T}$ with entries in $\{0,1\}$.
Namely, we make the correspondence as follows: for $i<j<k$ an entry $\mathcal{T}_{ijk} = 1$ if we have $c_{ijk} = 1$ in the corresponding polynomial $\phi$; for $i< j$, $\mathcal{T}_{ijj} = 1$ if $b_{ij} = 1$; and for any $i$, we have $\mathcal{T}_{iii} = 1$ if $a_i = 1$.

\paragraph{Rank of a tensor.} We can write a parity expression like $x_1\oplus x_3$ as a dot product with a vector $\vec y = (1,0,1)$ via $\vec y\cdot \vec x = y_1x_1\oplus y_2x_2\oplus y_3x_3$, where we take the value of the dot product to be modulo 2 and the ones in $\vec y$ indicate which variables are part of the parity.
Now, take an XOR phase polynomial consisting of a single term $\ee^{i\frac\pi4 \vec y\cdot \vec x}$. We can also write this as $\ee^{i\frac\pi4 x_{a_1}\oplus\cdots\oplus x_{a_k}}$ where the indices $a_k$ indicate the positions wherer $y_i=1$. Using \Cref{eq:xor-decompose} to decompose this polynomial into multilinear form, we get all combinations of the terms $x_{a_i}$, $2x_{a_i}x_{a_j}$ and $4x_{a_i}x_{a_j}x_{a_k}$ ---for all the variables $x_i$ where $y_i = 1$. Hence, we set the corresponding symmetric 3-tensor $\mathcal{T}_{ijk}$ as $\mathcal{T}_{ijk} = y_iy_jy_k$. This is then a \emph{rank-one} symmetric tensor, as it is built out of a single vector $\vec y$. Let such a rank-one tensor be denoted as $\mathcal{T}_{\vec y}$. Any rank-one tensor is defined by a vector $\vec y$, and hence if a symmetric 3-tensor built from a phase polynomial has rank-one, then it corresponds to a single XOR term.

The translation from the XOR form into the 3-tensor is linear in the XOR terms. Hence, if the phase polynomial is a sum of parity terms, $\{\vec y^1\cdot \vec x$, $\vec y^2\cdot \vec x, \ldots, \vec y^k\cdot \vec x\}$, then its corresponding symmetric 3-tensor is $\mathcal{T} = \mathcal{T}_{\vec y^1} + \cdots + \mathcal{T}_{\vec y^k}$, where again the summation is under modulo $2$. 
Conversely, each symmetric 3-tensor can be \emph{decomposed} in this manner. The \emph{rank} of the tensor is the number of terms in this decomposition. Thus, if we have a symmetric 3-tensor, finding a rank-$R$ decomposition corresponds to finding a set of $R$ XOR parity terms that result in the same tensor, which hence implement the same unitary up to Clifford operations. Given that implementing the unitary matrix corresponding to a single parity phase requires one \tgate{}, the \tcount of this circuit is then exactly equal to the rank of the decomposition.

Therefore, we have reduced the problem of \tcount optimization of \CNOT + T circuits to finding Waring decompositions of symmetric $N\times N\times N$ 3-tensors with elements in $\{0,1\}$.

\subsection{Quantum Chemistry}

Due to the exponential computational cost associated with solving the Schrödinger equation for strongly correlated many-electron systems, state-of-the-art classical quantum chemistry methods struggle to accurately simulate strongly correlated molecules \cite{chem-tensornet}. Quantum computers, in theory, can handle these calculations more efficiently, due to the logarithmic scaling of qubit number with system size. The FeMoco molecule, which is the active site in the nitrogenase enzyme responsible for nitrogen fixation, presents a compelling use case for quantum computing in strongly correlated quantum chemistry, due to its highly complex electronic structure \cite{msft-femoco}. The primary function of nitrogenase is to catalyze the conversion of atmospheric nitrogen (breaking a strong triple bond) at room temperature, into ammonia. Synthetically this is done via the Haber process which is incredibly energy intensive. The FeMoco molecule contains multiple atoms, including iron and molybdenum. 

To understand complex reaction pathways such as those in FeMoCo one must first calculate the ground state energy of a molecule at a particular geometry. Of the techniques for fault-tolerant quantum computation of molecular ground states, the current most efficient method is the \emph{linear combinations of unitaries} (LCU) method combined with \emph{qubitization} for phase estimation, which leads to a linear \tgate{} complexity in the outcome precision \cite{babbush2018encoding}. This can be further improved with Coulomb operator prepossessing that exploits sparsity \cite{Berry2019qubitizationof} or the use of tensor hyper-contraction approaches \cite{PRXQuantum.2.030305}. Furthermore, beyond the Born-Oppenheimer approximation of fixed nuclei, a first quantized plane wave framework has been proposed as the optimal for molecules \cite{PRXQuantum.2.040332}. This approach has been extended to be used with pseudo potentials for simulating materials. Furthermore, these qubitized phase estimation techniques have been used to calculate molecular observables when combined with quantum signal processing \cite{Steudtner2023faulttolerant}.

The rest of this section assumes prior knowledge of the standard mathematical formulation of quantum computing and quantum algorithms. See \citet[Sections 4 \& 9]{dewolf2019quantum} for a more gentle introduction and \citet{Bauer_2020} for a more comprehensive reference on the applicability of quantum algorithms in chemistry.

\paragraph{Quantum phase estimation.}

The typical way to calculate ground state energies is to exploit the relation of a unitary matrix acting on its eigenvector $|k\rangle$, which when applied to a mixture of eigenstates $|\psi\rangle = \sum_k c_k |k\rangle$ , gives a superposition
\begin{equation*}
    \langle  \psi | U |\psi\rangle = \sum_j |c_k|^2 \ee^{i\phi_k} \langle k | k \rangle.
\end{equation*}
where $\ee^{i\phi_k}$ are the eigenvalues of $U$ (and we call $\phi_k$ the eigenphases).

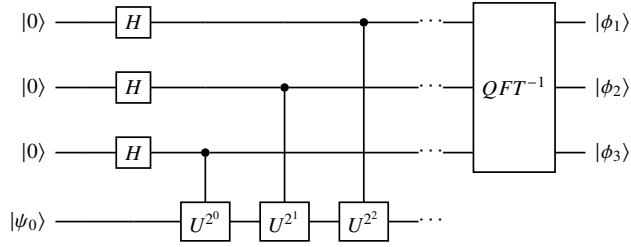
\begin{figure}[!t]
\centering
\begin{tikzpicture}
\node[scale=0.8] {
\begin{quantikz}
\lstick{$\ket{0}$} & \qw & \gate{H} & \qw & \qw & \ctrl{3} & \qw \cdots &\gate[3]{QFT^{-1}} & \qw \rstick{$\ket{\phi_1}$}\\
\lstick{$\ket{0}$} & \qw & \gate{H} & \qw & \ctrl{2} & \qw & \qw \cdots & &\qw \rstick{$\ket{\phi_2}$}\\
\lstick{$\ket{0}$} & \qw & \gate{H} & \ctrl{1} & \qw & \qw  & \qw \cdots & & \qw \rstick{$\ket{\phi_3}$} \\
\lstick{$\ket{\psi_0}$} & \qw  & \qw  & \gate{U^{2^0}}  & \gate{U^{2^1}}  & \gate{U^{2^2}} & \qw \cdots \\
\end{quantikz}
};
\end{tikzpicture}
\caption{Canonical quantum phase estimation.}
\label{fgr:qpe}
\end{figure}

The canonical phase estimation algorithm is shown in \Cref{fgr:qpe}. The ancilla register is prepared in the Hadamard basis, then phase-kickback of controlled unitaries acting on their eigenbasis $U|k\rangle = \ee^{\phi_k}|k\rangle$ is used to put the powers of eigenphases into the Fourier basis on the ancilla register. The eigenphases are then extracted into the computational basis from the Fourier basis using the inverse quantum Fourier transform. Readout of the ancilla bits gives the eigenphases in fixed-point binary representation. The eigenstates with the dominant initial overlap will have their eigenphases obtained with the largest probability. Therefore it is required that the initial state must have dominant overlap with the state of interest. Generating initial states with this property is an unresolved problem in quantum chemistry and has been discussed by \citet{chan-nature-advantage}.

In quantum chemistry applications, we encode the energies of the eigenstate $|k\rangle$ of the Hamiltonian describing the chemical system into the phase $\phi_k$ using the qubitized quantum walk unitary from \Cref{eqn:qubitz0} (see below). Qubitized quantum phase estimation \cite{babbush2018encoding} applies a controlled quantum-walk operator as opposed to a controlled time-evolution operator, where the Hamiltonian $H$ of the system is encoded in the quantum walk operator $W = \ee^{i\arccos(H) \hat{Y}}$, where $\hat{Y}$ is the Pauli $Y$ operator applied to each eigenvector independently. The eigenphases of this walk operator are proportional to the Hamiltonian eigenvalues $E_k$ and can be extracted when applied to eigenstates $|k\rangle$ of the Hamiltonian. Consider any state $|\psi\rangle = \sum_{k}c_k|k\rangle$, then:
\begin{equation*}
    \langle  \psi | W
    |\psi\rangle =
    \langle  \psi | \ee^{i\arccos(H)\hat{Y}}
    |\psi\rangle = \sum_k |c_k|^2 \ee^{\pm i\arccos(E_k)} 
   \langle k |k\rangle.
\end{equation*}

A naive implementation of the controlled quantum walk circuit $W$ is shown in \Cref{fgr:qubitz}. Its complexity can be improved by using the techniques employed by \citet{babbush2018encoding}, which exploit adjacent cancellation of reflection operators and a different initial ancilla state.

\paragraph{Linear combination of unitaries.}

The $U_{LCU}$ operator is a block encoding the Hamiltonian of interest, describing the molecule at a given geometry. It uses two registers, a state register and a prepare register. The block encoding is indexed by the unique bit strings of the $\prepare$ register, it can be thought of as there being a state block for each $\prepare$ register row and column bit strings. 

A Hamiltonian operator can be decomposed into a LCU by
\begin{equation*}
    H = \sum^{L-1}_{\ell=0} \alpha_\ell H_\ell,
\end{equation*}
where $H_\ell$ is a unitary matrix and $\alpha_l$ a linear expansion coefficient. We can assume $\alpha_l$ to be a real positive number, since any complex phases in the coefficients can be absorbed into $H_\ell$. The LCU circuit primitives were first introduced by \citet{lcu_wiebe}. The operator $H$ is encoded into a block of a larger unitary matrix $U_{LCU}$, and it is accessed by projecting into the block via post-selection -- that is, we measure the qubits defining the block structure, and the encoding succeeds when this returns zero. To ensure the overall matrix $U_{LCU}$ is unitary, the operator is re-scaled by $|\alpha|$, its $L_1$-norm , i.e.,
\begin{equation*}
    U_{LCU} = \begin{pmatrix}
    \frac{H}{|\alpha|} & * \\ 
    *  & *
    \end{pmatrix}, \quad \text{where} \quad |\alpha| = \sum_\ell |\alpha_\ell|.
\end{equation*}

The Hamiltonian is encoded into a quantum circuit via two oracles: $\select(H)$ and $\prepare(\alpha)$ which encode the unitaries $H_\ell$ and the coefficients $\alpha_\ell$ respectively. The $\prepare(\alpha)$ transforms the all zeros state of the control register to a state $|p\rangle$ (with positive real amplitudes) defined by the given coefficients in \Cref{equation:prepare},
\begin{equation}
  \prepare(\alpha): |0\rangle
  \mapsto
  \sum^{L-1}_{\ell=0} \sqrt{\frac{\alpha_\ell}{|\alpha|}} |\ell\rangle
  = \begin{bmatrix}
    \sqrt{\big(\frac{\alpha_{0}}{|\alpha|}} & \cdot & \hdots \\ 
    \sqrt{\big(\frac{\alpha_{1}}{|\alpha|}} & \cdot & \hdots  \\ 
    \vdots &  \ddots & \hdots   \\ 
    \sqrt{\big(\frac{\alpha_{k-1}}{|\alpha|}} & \cdot  &  \hdots
    \end{bmatrix}
\label{equation:prepare}
\end{equation}
A leading fault-tolerant primitive was proposed by \citet{low2018trading}. The signs and phases of the coefficients are pulled into the $\select$.

\begin{figure}
\centering
\begin{tikzpicture}
\node[scale=0.8] {
\begin{quantikz}
\lstick{$|p\rangle$} &  \ctrl{1} & \qw \\
\lstick{$|\psi\rangle$}  & \gate{\sel(H)}  &\qw \\ 
\end{quantikz}
\hspace{0.25cm}
=
\begin{quantikz}
\lstick{$|p\rangle_2$} &  \octrl{1} & \octrl{1} & \octrl{1} & \octrl{1} & \ \ldots\ \qw & \ctrl{1} & \qw\\
\lstick{$|p\rangle_1$} & \octrl{1} & \octrl{1} & \ctrl{1} & \ctrl{1} & \ \ldots\ \qw  & \ctrl{1} & \qw\\ 
\lstick{$|p\rangle_0$} & \octrl{1} & \ctrl{1} & \octrl{1} & \ctrl{1} & \ \ldots\ \qw & \ctrl{1} & \qw \\ 
\lstick{$|\psi\rangle$} & \gate{P_0}  &  \gate{P_1}  &  \gate{P_2}  &  \gate{P_3}  &  \ \ldots\ \qw & \gate{P_N}& \qw\\ 
\end{quantikz}
};
\end{tikzpicture}
\caption{$\select$ circuit implementing  $\sum_\ell |\ell\rangle\langle \ell| \otimes P_\ell $.}
\label{fgr:select}
\end{figure}
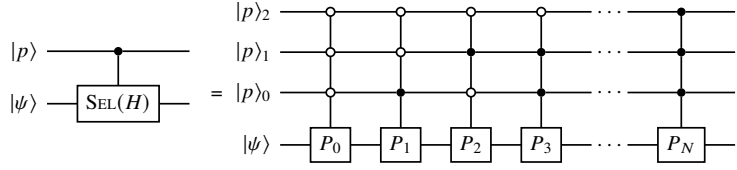

It must be noted that upon an action of the $\select$, an orthogonal complement $|p_\perp\rangle$ will also be generated, which leads to success probabilities (of the block-encoding) of less than $1$. The $\select$ oracle is typically a multi-controlled unitary matrix that applies the Hamiltonian terms to the state register via a multi-control for each term, indexed by a unique $\prepare$ register bit string. A fault tolerant approach was proposed by \citet{babbush2018encoding} built upon Gidney's work \cite{gidney2018halving}. The multi-control multiplies the weighting coefficient from the $\prepare$ to the Pauli for that bit string and applies the weighted Pauli to the state register. The $\select$ oracle is shown in \Cref{fgr:select} and given by
\begin{equation*}
    \select(H):=\sum^{L-1}_{\ell=0} |\ell\rangle \langle \ell | \otimes H_\ell
\end{equation*}

\begin{figure}[ht]
\centering
\begin{tikzpicture}
\node[scale=0.8] {
\begin{quantikz}
\lstick{$|\bar{0}\rangle_{prep}$}& \gate{\prep(\alpha)} & \ctrl{1} & \gate{\prep(\alpha)^\dagger}  &  \meterD{\bar{0}}  \\
\lstick{$|\psi\rangle$} & \qw & \gate{\sel(H)}  & \qw  & \qw \rstick{$\frac{H}{|\alpha|}|\psi\rangle$}  \\
\end{quantikz}
};
\end{tikzpicture}
\caption{LCU Circuit acting on an arbitrary quantum state $|\psi\rangle$ with a rescaled  operator $\frac{H}{|\alpha|}$, when a successful post-selection occurs.}
\label{fig:lcu_figure0}
\end{figure}
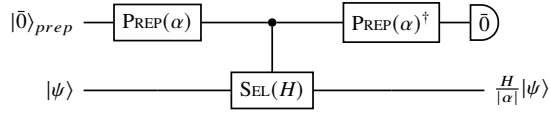

While any operator can in principle be written as a linear combination of unitary matrices, second quantized Hamiltonians in the Pauli representation are naturally expressed in this framework. The $\select(H)$ and $\prep(\alpha)$ oracles can be combined to create a probabilistic circuit for acting on an arbitrary quantum state $|\psi\rangle$ with an operator $\frac{H}{|\alpha|}$; this is shown in \Cref{fig:lcu_figure0}.

The operation $\frac{H}{\alpha}|\psi\rangle$ is successful if a measurement on the prepare register returns $|\bar{0}\rangle$. The success probability is related to the closeness of $H$ to being unitary and the initial state overlap,
\begin{equation*}
    P_{success} = \frac{1}{|\alpha|^2}\langle \psi| H^\dagger H | \psi \rangle  
\end{equation*}

The $L_1$-norm of the operator has a significant effect on the success probability and this is the dominant factor in the complexity results analysis. Therefore the majority of work on trying to improve the complexity of LCU methods typical aiming to reduce the $L_1$-norm of the operator.

\paragraph{Qubitization.}
First proposed by \citet{Low2019hamiltonian}, qubitization is a powerful method allowing one to implement polynomials of block encoded operators. In block encoding methods we are typically only interested in the $\prepare$ $|0\cdots0\rangle \langle 0\cdots0|$ block of the unitary matrix where the Hamiltonian is encoded. The rest of the rows and columns of the unitary matrix are refereed to as the orthogonal complement, $|0\cdots0_\perp\rangle$.

\begin{equation*}
    U_{LCU} = \begin{blockarray}{ccc}
       |0\cdots0\rangle|\psi\rangle & |0\cdots0_\perp\rangle|\psi\rangle &  \\
        \begin{block}{(cc)c}
        H & * & |0\cdots0\rangle|\psi\rangle  \\
        * &  * & |0\cdots0_\perp\rangle|\psi\rangle\\
        \end{block}
        \end{blockarray}
        = \bigoplus_{k}
    \begin{blockarray}{ccc}
        |0\cdots0\rangle|k\rangle & |0\cdots0_\perp\rangle |k\rangle  &  \\
         \begin{block}{(cc)c}
         \lambda_k & \sqrt{1-\lambda_k^2}  & |0\cdots0\rangle|k\rangle  \\
         \sqrt{1-\lambda_k^2} &  -\lambda_k & |0\cdots0_\perp\rangle |k\rangle  \\
         \end{block}
    \end{blockarray}.
\end{equation*}

In the eigenbasis $|\psi\rangle = \sum_k \lambda_k |k\rangle$ the $U_{LCU}$ becomes a direct sum of $2\times 2$ blocks indexed on each eigenvector. Where the orthogonal compliment $|0\cdots0_\perp\rangle$ of the prepare register is given an arbitrary vector which just has to exist to generate the qubitized 2x2 subspace via some Gram-Schmidt projection. We can then turn this into a quantum walk operator by applying a reflection, which is typically implemented as a all-$0$ \CZ gate with a minus phase,
\begin{equation*}
    W = \bigoplus_{k}
    \begin{pmatrix}
        1 & 0 \\
        0 & -1
    \end{pmatrix}
    \begin{pmatrix}
        \lambda_k & \sqrt{1-\lambda_k^2} \\
        \sqrt{1-\lambda_k^2} & -\lambda_k
    \end{pmatrix}
    = \bigoplus_{k}
    \begin{pmatrix}
        \lambda_k & \sqrt{1-\lambda_k^2} \\
        -\sqrt{1-\lambda_k^2} & \lambda_k
    \end{pmatrix}.
\end{equation*}

\begin{figure*}[!htb]
\centering
\begin{tikzpicture}
\node[scale=0.8] {
\begin{quantikz}
\lstick{$|\bar{0}\rangle_p$} &\gate[2]{\mathcal{W}} & \qw\\
\lstick{$|\psi\rangle$} & \qw  & \qw  \\
\end{quantikz}
=
\begin{quantikz}
\lstick{$|\bar{0}\rangle_p$} &\gate[2]{U_{LCU}}  &\gate{Z_{0\cdots0}} &\qw\\
\lstick{$|\psi\rangle$}  & \qw & \qw & \qw\\
\end{quantikz}
=
\begin{quantikz}
\lstick{$|\bar{0}\rangle_p$} & \gate{\prepare}  &\gate[2]{\select} & \gate{\prepare^\dagger} &\gate{Z_{0\cdots0}} &\qw\\
\lstick{$|\psi\rangle$} & \qw & & \qw & \qw& \qw \\
\end{quantikz}
\label{fgr:qubitz2}
};
\end{tikzpicture}
\end{figure*}
Therefore, powers of the quantum walk operator take the form of increments of an $R_y(\arccos(\lambda_k))$ rotation in the $2\times 2$ sub-spaces of $\{ |0\cdots0\rangle|k\rangle , |0\cdots0_\perp\rangle|k\rangle \}$. These rotation matrices can be interpreted in terms of the Chebyshev polynomials of the first $T_d(\lambda_k)$ and second $U_{d}(\lambda_k)$ kind:
\begin{equation*}
        W^d = \bigoplus_{k}
        \begin{pmatrix}
            \cos(d\theta_k) & \sin(d\theta_k) \\
            -\sin(d\theta_k) & \cos(d\theta_k) \\
        \end{pmatrix} \\
        = \bigoplus_{k}
        \begin{pmatrix}
            T_d(\lambda_k) & \sqrt{1-\lambda_k^2} U_{d-1}(\lambda_k) \\
            -\sqrt{1-\lambda_k^2} U_{d-1}(\lambda_k) & T_d(\lambda_k) \\
        \end{pmatrix} \\
        =
        \begin{pmatrix}
            T_d(H) & \cdot\\
            \cdot & \cdot \\
        \end{pmatrix}
\end{equation*}
The phase of this operator can be found in the same way as traditional phase estimation using a controlled quantum-walk:
\begin{equation}
    \langle  \psi | W |\psi \rangle = \langle \psi | \ee^{i\arccos(H) \hat{Y}}
    |\psi\rangle = \sum_k |c_k|^2 \ee^{\pm i\arccos(E_k)} 
   \langle k |k\rangle
  \label{eqn:qubitz0}
\end{equation}
This operation is the compilation target for the quantum chemistry section of this work.
\section{Baselines}
\label{app:subsec:baselines}

Methods for \tcount optimization are broadly divided into two camps: those that treat all non-Clifford phase gates equivalently, and those that utilize specific optimizations for \tgate{s}. In the algorithms in the first category \citep[e.g.,][]{kissinger2019reducing,zhang2019optimizing,nam2018automated,amy2013polynomial}, non-Clifford rotation gates are treated as distinct black boxes, without any knowledge of transformations relating them, except that they can be merged together to reduce the number of gates. While limited in this way, they also have the ability to optimize all circuits, regardless of the gates used, and in particular they can optimize circuits containing Hadamard gates. 

By contrast, algorithms in the second category \citep[e.g.,][]{amy2019tcount,heyfron2018efficient,debeaudrap2020fast} are based (either directly or indirectly) on the representation of CNOT+T circuits as Waring decompositions of the signature tensor, as they are considered in \method. These generally perform better on such circuits than the black-box methods. However, since they cannot optimize around Hadamard gates, it is beneficial to first apply a black-box optimizer that can do so \citep{vandaele2023optimal}. 

To obtain the baseline \tcount for the circuit benchmarks used in this work, we applied the phase-teleportation algorithm of \citet{kissinger2019reducing} followed by compiling the circuits to the Waring decomposition using the steps detailed in \Cref{app:subsec:compilation}. Then we apply TODD \citep{heyfron2018efficient} and STOMP \citep{debeaudrap2020fast} and take the output with better \tcount. (For STOMP, we took the best of twenty runs.) For the standard arithmetic benchmarks (i.e., the non-split circuits), we compared against the published \tcount numbers and took whichever was better.

We chose phase-teleportation because it was shown to be essentially optimal for reducing \tcount in the black-box setting \citep{vandewetering2023parameterized}, and TODD/STOMP were chosen because they consistently outperform other methods \cite{amy2019tcount} in published benchmarks. While TODD and STOMP both have methods to compile Clifford+T circuits to a Waring decomposition, we opt to use the method in \Cref{app:subsec:compilation} as it includes a more recent and better performing method from \citet{vandaele2023optimal}.

\section{Method Details}
\label{app:sec:method_details}

\subsection{Compilation of the Quantum Circuits}
\label{app:subsec:compilation}

To obtain signature tensors suitable for optimization with \method from quantum circuits, we proceed as follows:
\begin{enumerate}
    \item First, we express the input circuit over the $\{\mathrm{H}, \mathrm{Z}, \mathrm{S}, \mathrm{T}, \mathrm{CNOT}\}$ gate set.
    \item We apply the phase-teleportation optimization algorithm of \citet{kissinger2019reducing} to reduce the initial number of \tgate{s} as much as possible. As discussed in \Cref{app:subsec:baselines}, this is particularly beneficial because phase-teleportation can optimize around Hadamard gates.
    \item We rewrite the circuit to contain as few internal Hadamard gates as possible, using the algorithm of \citet{vandaele2023optimal}. A Hadamard gate is internal if there are \tgate{s} between it and both the beginning and end of the circuit that it cannot commute with (for example, in the sequence $HTHTH$ the second Hadamard is internal but the first and third ones are not).
    \item We split the circuit $U$ into alternating blocks of Clifford circuits $B_i$ and circuits $A_i$ containing only \CNOT and phase gates, such that $U = B_mA_{m - 1}B_{m -1}\cdots A_1B_1$. We accomplish this by scanning the circuit from left to right, collecting as many Clifford gates as possible into the first block $B_1$ until we encounter a \tgate{}. After this, we collect gates into the block $A_1$ until we encounter a Hadamard gate. Then we switch to collecting Clifford gates into block $B_2$, etc. We proceed in this way until the whole circuit has been consumed.
    \item We apply the Hadamard gadgetization process of \citet{heyfron2018efficient} to remove all internal Hadamard gates by replacing them with ancilla and Clifford gates. In \Cref{subsec:benchmark}, if this process generates more than 60 qubits, we take the output of the previous step and use the technique in \Cref{app:subsec:splitting} to split the circuit into smaller sections, and we repeat this compilation process on each. We discuss Hadamard gadgetization in further detail below.
    \item At this point we can partition the circuit into the form $U = U_3U_2U_1$ where $U_1$ and $U_3$ contain Clifford gates and non-internal Hadamard gates, and $U_2$ contains only \CNOT and phase gates. Let $C$ be the circuit formed by taking just the \CNOT gates in $U_2$, then consider the transformation $U = U'_3U'_2U_1 = (U_3C)(C^\dagger U_2)U_1$. Now, $U'_2$ is a diagonal operator. Let $V$ be the circuit formed by removing all \tgate{s} from $U'_2$ and $W$ be the circuit formed by removing all S and Z gates from $U'_2$. Then $V$ is Clifford, $W$ contains only \CNOT and \tgate{s}, and it can be shown that $U'_2 = VW$, so $U = (U_3CV)WU_1$.  Therefore, all the non-Clifford components of the original circuit $U$ is now contained within $W$.
    \item We map $W$ to a Waring decomposition of the signature tensor as follows. Suppose $W$ contains $R$ \tgate{s} and $N$ qubits, then let $A$ be a $N \times R$ matrix in $\mathbb{F}_2$. For the $i$th \tgate{} in $W$, set $A_{ji} = 1$ if it appears on the $j$th qubit, and $A_{ji} = 0$ otherwise. Scan over $W$ from left to right: for each \CNOT gate with control $a$ and target $b$, if there are any \tgate{s} to its right, then let $i'$ be the corresponding column of $A$ and set $A_{ai} \leftarrow A_{ai} + A_{bi}$ for all $i' \leq i \leq R$. After all \CNOT gates have been processed, $A$ will be a Waring decomposition of the signature tensor of $W$, which we can calculate as $\mathcal{T}_{ijk} = \sum_{m = 1}^R A_{im}A_{jm}A_{km}$. This is then ready to be optimized with \method. Overall, this procedure is the same as the phase polynomial method in \Cref{subsec:signature_tensor}, but phrased in terms of matrices.
\end{enumerate}

\paragraph{Hadmard gadgetization.}
To remove the internal Hadamard gates from a circuit, we can use the Hadamard gadgetization method of \citet{heyfron2018efficient}. In this method, each Hadamard gate is replaced with the following gadget circuit, consisting of an ancilla, Clifford gates, measurements, and classically-controlled gates (here, $A$ and $B$ represent the rest of the circuit):
$$\includegraphics{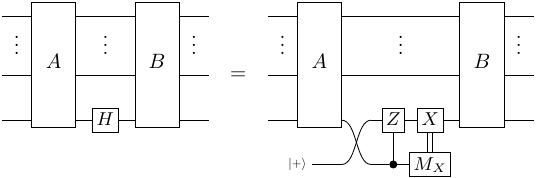}$$
However, in this implementation, we assume that the measurement always returns zero. Therefore, the corresponding gadget implementation is as follows (additionally rewritten to our target gate set):
$$\includegraphics{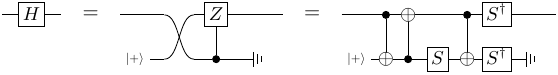}$$
This is justified in that it does not change the non-Clifford behaviour of a circuit---in fact, the classically controlled X gate correction factor can be translated into a controlled Clifford operator at the end of the circuit \cite{heyfron2018efficient, debeaudrap2020fast}, hence the \tcount is unchanged by this assumption. Because computing these correction factors is complicated, and existing quantum circuit representations have poor support for classically-controlled operations; this decision was also made for the implementations of both TODD \cite{heyfron2018efficient} and STOMP \cite{debeaudrap2020fast}. Since fault-tolerant quantum computing is still a number of years away from practical implementation, we argue that this is acceptable at this stage.

\paragraph{Reconstructing optimized circuits.}
In order to reconstruct an optimized form $W'$ of the diagonal circuit $W$ computed above, we can follow the steps in \Cref{app:subsec:gadgetization} given a new Waring decomposition of the signature tensor $\mathcal{T}$. With $W'$, we can recompose the original circuit $U = (U_3CV)WU_1$. However, note that $W'$ may differ from the original $W$ by a diagonal Clifford factor, so it must be corrected. This factor can be computed, given the original and optimized Waring decompositions along with $V$, by considering the difference between the signature tensors over $\mathbb{Z}$ rather than $\mathbb{F}_2$---see \citet{heyfron2018efficient} for more details.  As discussed above, in this process we discarded the classically-controlled Clifford gates incurred during Hadamard gadgetization. It is possible in principle (and computationally easy) to take these into account---see \citet{debeaudrap2020fast} for more details and \citet{heyfron2018efficient} for the general method. Since in this work we only determine the optimized \tcount rather than reconstructing the circuit, we do not need to implement either of these steps.

\subsection{Splitting Larger Quantum Circuits}
\label{app:subsec:splitting}

During the compilation process, each quantum circuit $U$ is split into a list of blocks alternating between circuits $A_i$ containing CNOT and phase gates, and Clifford circuits $B_i$, i.e.,
$$U = B_mA_{m - 1}B_{m - 1}\cdots A_2B_2A_1B_1.$$
For most circuits, we can then proceed with Hadamard gadgetization of the $B_i$ blocks as described above. However, for some circuits, this would result in a circuit with too many qubits. Hence, we choose some boundaries $1 = j_1 < j_2 < \cdots < j_k = m$ to break up the circuit into a set of subcircuits:
$$U_i = A_{j_{i + 1} - 1}B_{j_{i + 1} - 1}\cdots A_{j_i}B_{j_i} \text{ where } U = B_mU_k\cdots U_1.$$
We want to minimize $k$ in order to maximize the number of \tgate{s} that can be considered together in each block. To do this, we use a random-greedy algorithm: fix a threshold $t$ on the maximum number of qubits allowed in any subcircuit (e.g., $60$), and initially let $j_i = i$ so that $k = m$. For any consecutive pair $U_{i}$ and $U_{i + 1}$, we consider them to be \emph{compatible} if 
$$|\operatorname{supp} U_i \cup \operatorname{supp} U_{i + 1}| ~~+ \sum_{j_i < n < j_{i + 2}} |B_{n}|_H \leq t,$$
where $\operatorname{supp} U_i$ is the set of qubits used by gates on $U_i$, and $|C|_H$ is the number of Hadamard gates in $C$. While there is any compatible pair remaining, pick one at random and set $j_{n} \leftarrow j_{n + 1}$ for $n \geq i + 1$ and $k \leftarrow k - 1$. In this way, we merge consecutive compatible pairs until there are only a few blocks remaining, and combining any two of them would result in too many qubits being generated. We repeat this process $1\,000$ times and take the run that results in the smallest value of $k$. We can then treat each $U_i$ as its own target circuit, processing it back through the compilation pipeline, and optimizing it with \method. To composite the final circuit we combine the $U_i$ along with the Clifford circuit $B_m$ to form $U = B_mU_k\cdots U_1$.

\subsection{Gadgetization}
\label{app:subsec:gadgetization}

The conversion of phase-gadget circuits to Waring decompositions of the signature tensor as described in \Cref{app:subsec:compilation} can be inverted to obtain a phase-gadget circuit from a decomposition by placing a CNOT-fanout structure conjugating a \tgate{} for each factor, where the target corresponds to some non-zero entry in the factor, and the controls are on every other non-zero entry. For example:

\begingroup
\tikzset{every picture/.style={scale=0.6}}%
\begin{tikzpicture}
	\begin{pgfonlayer}{nodelayer}
		\node [style=none] (0) at (2.5, 0) {$\begin{pmatrix}0\\1\\1\\0\\1\end{pmatrix}$};
		\node [style=none] (1) at (5, 0) {$\begin{pmatrix}0\\1\\0\\1\\0\end{pmatrix}$};
		\node [style=none] (2) at (7.5, 0) {$\begin{pmatrix}1\\1\\0\\1\\1\end{pmatrix}$};
		\node [style=none] (3) at (9.5, 0) {$\implies$};
		\node [style=none] (56) at (11, 2) {};
		\node [style=none] (57) at (11, 1) {};
		\node [style=none] (58) at (11, 0) {};
		\node [style=none] (59) at (11, -1) {};
		\node [style=none] (60) at (11, -2) {};
		\node [style=cnot ctrl] (61) at (12, 1) {};
		\node [style=cnot ctrl] (62) at (12.75, 0) {};
		\node [style=cnot targ] (63) at (12, -2) {};
		\node [style=cnot targ] (64) at (12.75, -2) {};
		\node [style=gate] (65) at (13.75, -2) {$T$};
		\node [style=cnot ctrl] (66) at (15.5, 1) {};
		\node [style=cnot ctrl] (67) at (14.75, 0) {};
		\node [style=cnot targ] (68) at (14.75, -2) {};
		\node [style=cnot targ] (69) at (15.5, -2) {};
		\node [style=cnot ctrl] (70) at (16.5, -1) {};
		\node [style=cnot targ] (71) at (16.5, 1) {};
		\node [style=cnot targ] (72) at (18.5, 1) {};
		\node [style=cnot ctrl] (73) at (18.5, -1) {};
		\node [style=gate] (74) at (17.5, 1) {$T$};
		\node [style=cnot ctrl] (75) at (19.5, 2) {};
		\node [style=cnot ctrl] (76) at (20.25, 1) {};
		\node [style=cnot targ] (77) at (19.5, -2) {};
		\node [style=cnot targ] (78) at (20.25, -2) {};
		\node [style=gate] (79) at (22, -2) {$T$};
		\node [style=cnot ctrl] (80) at (24.5, 2) {};
		\node [style=cnot ctrl] (81) at (23.75, 1) {};
		\node [style=cnot targ] (82) at (23.75, -2) {};
		\node [style=cnot targ] (83) at (24.5, -2) {};
		\node [style=cnot ctrl] (84) at (21, -1) {};
		\node [style=cnot targ] (85) at (21, -2) {};
		\node [style=cnot ctrl] (86) at (23, -1) {};
		\node [style=cnot targ] (87) at (23, -2) {};
		\node [style=cnot ctrl] (88) at (18.5, -1) {};
		\node [style=none] (89) at (25.5, 2) {};
		\node [style=none] (90) at (25.5, 1) {};
		\node [style=none] (91) at (25.5, 0) {};
		\node [style=none] (92) at (25.5, -1) {};
		\node [style=none] (93) at (25.5, -2) {};
		\node [style=none] (94) at (19, 3.25) {};
		\node [style=none] (95) at (19, -2.5) {};
		\node [style=none] (96) at (16, -2.5) {};
		\node [style=none] (97) at (16, 3.25) {};
		\node [style=none] (98) at (2.5, 2.75) {a)};
		\node [style=none] (99) at (5, 2.75) {b)};
		\node [style=none] (100) at (7.5, 2.75) {c)};
		\node [style=none] (101) at (11.5, 3.25) {};
		\node [style=none] (102) at (11.5, -2.5) {};
		\node [style=none] (103) at (25, -2.5) {};
		\node [style=none] (104) at (25, 3.25) {};
		\node [style=none] (105) at (13.5, 2.75) {a)};
		\node [style=none] (106) at (17.5, 2.75) {b)};
		\node [style=none] (107) at (22, 2.75) {c)};
	\end{pgfonlayer}
	\begin{pgfonlayer}{edgelayer}
		\draw (60.center) to (63);
		\draw (63) to (64);
		\draw (65) to (68);
		\draw (68) to (69);
		\draw (65) to (64);
		\draw (57.center) to (61);
		\draw (61) to (66);
		\draw (58.center) to (62);
		\draw (61) to (63);
		\draw (62) to (64);
		\draw (62) to (67);
		\draw (67) to (68);
		\draw (66) to (69);
		\draw (71) to (70);
		\draw (66) to (71);
		\draw (71) to (74);
		\draw (74) to (72);
		\draw (72) to (73);
		\draw (70) to (73);
		\draw (70) to (59.center);
		\draw (77) to (78);
		\draw (82) to (83);
		\draw (75) to (80);
		\draw (75) to (77);
		\draw (76) to (78);
		\draw (76) to (81);
		\draw (81) to (82);
		\draw (80) to (83);
		\draw (84) to (85);
		\draw (84) to (86);
		\draw (86) to (87);
		\draw (78) to (85);
		\draw (85) to (79);
		\draw (79) to (87);
		\draw (87) to (82);
		\draw (76) to (72);
		\draw (84) to (88);
		\draw (77) to (69);
		\draw (75) to (56.center);
		\draw (92.center) to (86);
		\draw (93.center) to (83);
		\draw (91.center) to (67);
		\draw (90.center) to (81);
		\draw (89.center) to (80);
		\draw [style=box edge] (97.center) to (96.center);
		\draw [style=box edge] (95.center) to (94.center);
		\draw [style=box edge] (102.center) to (96.center);
		\draw [style=box edge] (101.center) to (102.center);
		\draw [style=box edge] (101.center) to (97.center);
		\draw [style=box edge] (97.center) to (94.center);
		\draw [style=box edge] (95.center) to (96.center);
		\draw [style=box edge] (104.center) to (94.center);
		\draw [style=box edge] (95.center) to (103.center);
		\draw [style=box edge] (103.center) to (104.center);
	\end{pgfonlayer}
\end{tikzpicture}
\endgroup

Note that this is not optimal with respect to the number of CNOT gates, and many strategies have been developed to improve this \citep{vandaele2022phase,degriend2020architectureaware,meuli2018sat}. In this work we ignore the cost of Clifford gates, and so do not attempt to optimize this step. If we are constrained to CNOT+T circuits, this method is optimal in terms of \tgate{s}, but some patterns of factors can be implemented with fewer non-Clifford resources if we make use of ancilla qubits, measurements, and classically-controlled Clifford gates.

\begin{figure}
    \centering
    \begingroup
    \tikzset{every picture/.style={scale=0.5}}%
    \begin{tikzpicture}
	\begin{pgfonlayer}{nodelayer}
		\node [style=none] (0) at (0, 0) {};
		\node [style=none] (1) at (0, -1.5) {};
		\node [style=none] (2) at (0, -3) {};
		\node [style=cnot targ] (3) at (1, -3) {};
		\node [style=cnot ctrl] (4) at (1, -1.5) {};
		\node [style=cnot ctrl] (5) at (1, 0) {};
		\node [style=none] (6) at (2, 0) {};
		\node [style=none] (7) at (2, -1.5) {};
		\node [style=none] (8) at (2, -3) {};
		\node [style=none] (9) at (3.5, -1.5) {$=$};
		\node [style=none] (10) at (5, 0) {};
		\node [style=none] (11) at (5, -1.5) {};
		\node [style=none] (12) at (5, -3) {};
		\node [style=none] (16) at (19.5, 0) {};
		\node [style=none] (17) at (19.5, -1.5) {};
		\node [style=none] (18) at (19.5, -3) {};
		\node [style=none] (19) at (5.5, 1.5) {\tiny $|+\rangle$};
		\node [style=none] (20) at (6, 1.5) {};
		\node [style=cnot targ] (21) at (7.75, 1.5) {};
		\node [style=cnot targ] (22) at (8.5, 1.5) {};
		\node [style=cnot ctrl] (23) at (7.75, 0) {};
		\node [style=cnot ctrl] (24) at (8.5, -1.5) {};
		\node [style=gate] (25) at (6.75, 1.5) {$T$};
		\node [style=cnot ctrl] (26) at (9.25, 1.5) {};
		\node [style=cnot ctrl] (27) at (10, 1.5) {};
		\node [style=cnot targ] (28) at (9.25, -1.5) {};
		\node [style=cnot targ] (29) at (10, 0) {};
		\node [style=gate] (30) at (11, 1.5) {$T$};
		\node [style=gate] (31) at (11, -1.5) {$T^\dagger$};
		\node [style=gate] (32) at (11, 0) {$T^\dagger$};
		\node [style=cnot ctrl] (33) at (12, 1.5) {};
		\node [style=cnot ctrl] (34) at (12.75, 1.5) {};
		\node [style=cnot targ] (35) at (12.75, -1.5) {};
		\node [style=cnot targ] (36) at (12, 0) {};
		\node [style=gate] (37) at (13.75, 1.5) {$H$};
		\node [style=gate] (38) at (14.75, 1.5) {$S$};
		\node [style=cnot ctrl] (39) at (15.75, 1.5) {};
		\node [style=cnot targ] (40) at (15.75, -3) {};
		\node [style=gate] (41) at (16.75, 1.5) {$H$};
		\node [style=cnot ctrl] (42) at (18.5, -1.5) {};
		\node [style=gate] (43) at (18.5, 0) {$Z$};
		\node [style=gate] (44) at (18.5, 1.5) {$M_Z$};
		\node [style=none] (45) at (0, -10) {};
		\node [style=none] (46) at (0, -11.5) {};
		\node [style=none] (47) at (0, -13) {};
		\node [style=cnot ctrl] (48) at (1, -10) {};
		\node [style=cnot ctrl] (49) at (1, -11.5) {};
		\node [style=cnot targ] (50) at (1, -13) {};
		\node [style=none] (51) at (2, -10) {};
		\node [style=none] (52) at (2, -11.5) {};
		\node [style=none] (53) at (2, -13) {};
		\node [style=none] (54) at (3.5, -11.5) {$=$};
		\node [style=none] (55) at (5, -10) {};
		\node [style=none] (56) at (5, -11.5) {};
		\node [style=none] (57) at (5, -13) {};
		\node [style=none] (58) at (19.5, -10) {};
		\node [style=none] (59) at (19.5, -11.5) {};
		\node [style=none] (60) at (19.5, -13) {};
		\node [style=gate] (61) at (7.75, -13) {$H$};
		\node [style=gate] (62) at (18, -13) {$H$};
		\node [style=cnot ctrl] (63) at (9.25, -13) {};
		\node [style=cnot ctrl] (64) at (8.5, -11.5) {};
		\node [style=cnot ctrl] (65) at (7.75, -10) {};
		\node [style=cnot targ] (66) at (7.75, -5.5) {};
		\node [style=cnot targ] (67) at (8.5, -7) {};
		\node [style=cnot targ] (68) at (9.25, -8.5) {};
		\node [style=gate] (69) at (15.5, -5.5) {$M_Z$};
		\node [style=gate] (70) at (13, -7) {$M_Z$};
		\node [style=gate] (71) at (10.75, -8.5) {$M_Z$};
		\node [style=none] (72) at (6.75, -8.5) {};
		\node [style=none] (73) at (6.75, -7) {};
		\node [style=none] (74) at (6.75, -5.5) {};
		\node [style=none] (75) at (6.25, -8.75) {};
		\node [style=none] (76) at (6.25, -5.25) {};
		\node [style=none] (77) at (5.5, -7) {\tiny $|$CCZ$\rangle$};
		\node [style=cnot ctrl] (78) at (15.5, -13) {};
		\node [style=gate] (79) at (15.5, -11.5) {$Z$};
		\node [style=gate] (80) at (16.75, -13) {$X$};
		\node [style=gate] (81) at (18, -11.5) {$X$};
		\node [style=cnot ctrl] (82) at (13, -13) {};
		\node [style=gate] (83) at (13, -10) {$Z$};
		\node [style=gate] (84) at (14.25, -11.5) {$X$};
		\node [style=cnot ctrl] (85) at (10.75, -11.5) {};
		\node [style=gate] (86) at (10.75, -10) {$Z$};
		\node [style=gate] (87) at (12, -13) {$X$};
		\node [style=small black dot] (88) at (12, -8.5) {};
		\node [style=small black dot] (90) at (14.25, -7) {};
		\node [style=small black dot] (91) at (18, -7) {};
		\node [style=small black dot] (92) at (16.75, -8.5) {};
		\node [style=none] (93) at (-2, -1.5) {a)};
		\node [style=none] (94) at (-2, -11.5) {b)};
	\end{pgfonlayer}
	\begin{pgfonlayer}{edgelayer}
		\draw (2.center) to (3);
		\draw (3) to (8.center);
		\draw (7.center) to (4);
		\draw (4) to (1.center);
		\draw (0.center) to (5);
		\draw (5) to (6.center);
		\draw (5) to (4);
		\draw (4) to (3);
		\draw (23) to (21);
		\draw (24) to (22);
		\draw (22) to (21);
		\draw (11.center) to (24);
		\draw (23) to (10.center);
		\draw (25) to (21);
		\draw (25) to (20.center);
		\draw (28) to (26);
		\draw (27) to (29);
		\draw (36) to (33);
		\draw (34) to (35);
		\draw (28) to (31);
		\draw (31) to (35);
		\draw (28) to (24);
		\draw (22) to (26);
		\draw (26) to (27);
		\draw (29) to (23);
		\draw (29) to (32);
		\draw (32) to (36);
		\draw (33) to (30);
		\draw (30) to (27);
		\draw (33) to (34);
		\draw (38) to (37);
		\draw (37) to (34);
		\draw (39) to (40);
		\draw (38) to (39);
		\draw (39) to (41);
		\draw (40) to (18.center);
		\draw (40) to (12.center);
		\draw (42) to (43);
		\draw (44) to (41);
		\draw [style=double edge] (43) to (44);
		\draw (47.center) to (50);
		\draw (50) to (53.center);
		\draw (52.center) to (49);
		\draw (49) to (46.center);
		\draw (45.center) to (48);
		\draw (48) to (51.center);
		\draw (48) to (49);
		\draw (49) to (50);
		\draw (60.center) to (62);
		\draw (61) to (57.center);
		\draw (66) to (65);
		\draw (67) to (64);
		\draw (68) to (63);
		\draw (72.center) to (68);
		\draw (68) to (71);
		\draw (70) to (67);
		\draw (67) to (73.center);
		\draw (66) to (74.center);
		\draw (66) to (69);
		\draw (56.center) to (64);
		\draw (61) to (63);
		\draw (65) to (55.center);
		\draw [style=brace edge] (75.center) to (76.center);
		\draw (78) to (79);
		\draw (82) to (83);
		\draw (85) to (86);
		\draw [style=double edge] (88) to (87);
		\draw [style=double edge] (90) to (84);
		\draw [style=double edge] (91) to (81);
		\draw [style=double edge] (92) to (80);
		\draw (81) to (79);
		\draw (79) to (84);
		\draw (84) to (85);
		\draw (85) to (64);
		\draw (81) to (59.center);
		\draw (62) to (80);
		\draw (80) to (78);
		\draw (78) to (82);
		\draw (82) to (87);
		\draw (87) to (63);
		\draw (65) to (86);
		\draw (86) to (83);
		\draw [style=double edge] (71) to (88);
		\draw [style=double edge] (71) to (86);
		\draw [style=double edge] (69) to (79);
		\draw [style=double edge] (90) to (91);
		\draw (83) to (58.center);
		\draw (16.center) to (43);
		\draw (43) to (36);
		\draw (35) to (42);
		\draw (42) to (17.center);
		\draw [style=double edge] (90) to (70);
		\draw [style=double edge] (83) to (70);
		\draw [style=double edge] (88) to (92);
	\end{pgfonlayer}
\end{tikzpicture}%
    \endgroup
    \caption{The Toffoli gate can be implemented efficiently using ancilla qubits, measurements,  classically-controlled Clifford gates, and either four \tgate{s} (a) as in \citet{jones2013lowoverhead}, or a \CCZ magic state (b) which are produced natively by the magic state factories of \citet{gidney2019efficient} at a cost equivalent to two \tgate{s}.}
    \label{app:fig:toffoli-gadget}
\end{figure}
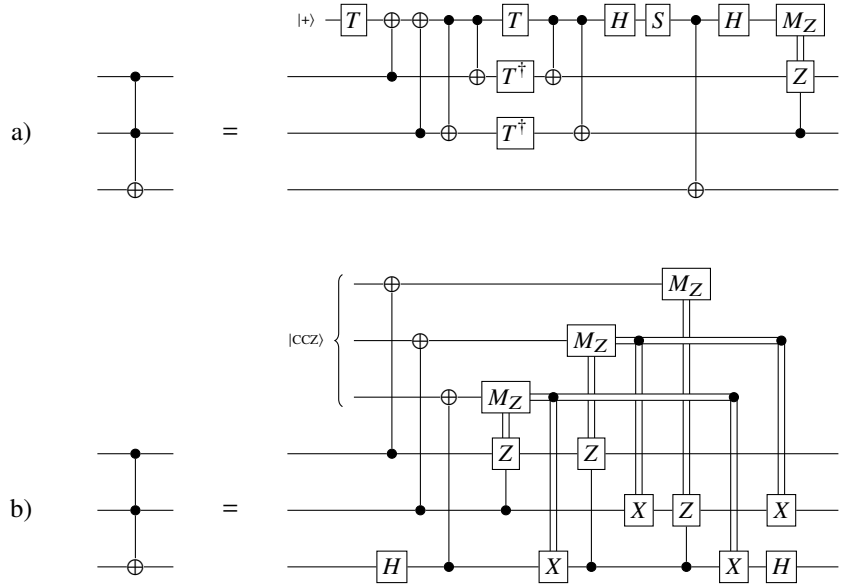

In particular, suppose we have factors $u^{(1)} \cdots u^{(7)}$ such that $u^{(1)}$, $u^{(2)}$ and $u^{(3)}$ are linearly independent, and the following equations are satisfied:
\begin{align*}
    u^{(4)} &= u^{(1)} + u^{(2)} \\
    u^{(5)} &= u^{(1)} + u^{(3)} \\
    u^{(6)} &= u^{(1)} + u^{(2)} + u^{(3)} \\
    u^{(7)} &= u^{(2)} + u^{(3)}
\end{align*}
where addition is considered elementwise over $\mathbb{F}_2$. We call such patterns \emph{Toffoli gadgets}. They can be implemented using a single Toffoli gate conjugated by \CNOT and Hadamard gates as follows,
$$\includegraphics{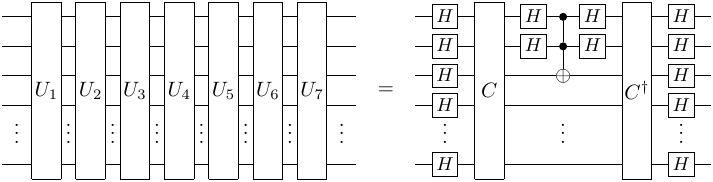}$$
where $U_i$ denotes the \CNOT-fanout circuit given above for factor $u^{(i)}$, and $C$ is a \CNOT circuit such that $C\ket{u^{(1)}} = \ket{100\cdots}$, $C\ket{u^{(2)}} = \ket{010\cdots}$, $C\ket{u^{(3)}} = \ket{001\cdots}$. This can be proven by considering this circuit in the sum-over-paths formalism \citep{dawson2005quantum}. For example:
$$\includegraphics{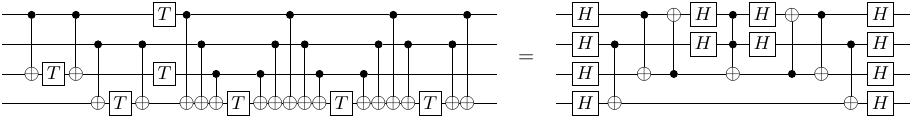}$$
Since Toffoli gates can be implemented efficiently using either four \tgate{s} or a \CCZ magic state which using state of the art factories costs the same as two \tgate{s} (see \Cref{app:fig:toffoli-gadget}), each Toffoli gadget can be implemented more efficiently than the seven \tgate{s} required to represent it as a phase-gadget circuit. 

Because $u^{(1)}, u^{(2)}, u^{(3)}$ are linearly independent, such a circuit $C$ must exist. To construct it, we first build an invertible matrix $M$ over $\mathbb{F}_2$ such that the first three columns of $M$ are $u^{(1)}, u^{(2)}, u^{(3)}$ respectively. The remaining columns of $M$ can be computed as a basis for the kernel of the first three via Gaussian elimination. Then we can synthesize a \CNOT circuit $C$ such that $C\ket{x} = \ket{M^{-1}x}$ using the Patel-Markov-Hayes algorithm. By definition, this must send $\ket{u^{(1)}}$ to $\ket{100\cdots}$, etc., as required.  
Therefore, to encourage Toffoli gadgets to be played by \method, we assign a positive reward to moves which complete a group. To identify when this happens, we set $u^{(1)}$ to be the factor at step $s-6$, $u^{(2)}$ to be the factor at step $s-5$, and so on, so that $u^{(7)}$ is the last played action (at step $s$).\footnote{Any permutation of the seven factors could be absorbed in $C$ and therefore it would still correspond to a Toffoli gadget. Additionally, the seven factors do not need to be consecutive in order to form a Toffoli gadget. However we impose a specific ordering of the (consecutive) factors $u^{(1)},\ldots,u^{(7)}$ for two reasons. First, the sequence of seven factors does not include the pattern of the \CS gadget as a subsequence. Second, this makes it easier for the RL agent to automatically detect and exploit the gadgetization patterns.} %
Finally, to synthesize a quantum circuit from a Waring decomposition of the signature tensor, we apply this construction to any Toffoli gadgets that were identified by \method, and synthesize the remaining factors with the \CNOT-fanout approach from above.

\begin{figure}[th]
    \centering
    \includegraphics{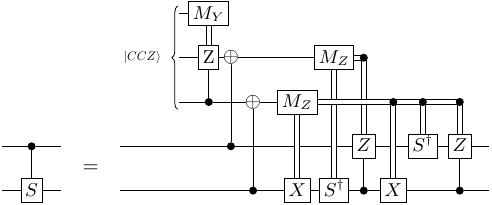}
    \caption{A \CS gate can implemented efficiently \cite[Appendix A]{beverland2020lower} using ancilla qubits, measurements, classically-controlled Clifford gates and a \CCZ magic state which is produced natively by the magic state factories of \citet{gidney2019efficient} at a cost equivalent to two \tgate{s}.}
    \label{app:fig:cs-gadget}
\end{figure}

In the same way, we call groups of three factors $u^{(1)}, u^{(2)}, u^{(3)}$ such that $u^{(1)}$ and $u^{(2)}$ are linearly independent and $u^{(3)} = u^{(1)} + u^{(2)}$ \emph{\CS gadgets}. In this case, such a group can be implemented by a single \CS gate conjugated by \CNOT{} gates:
$$\includegraphics{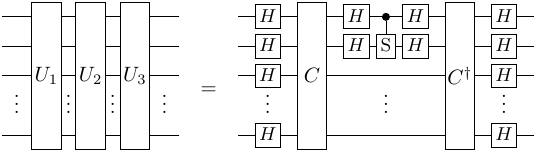}$$
The \CNOT circuit $C$ is such that $C\ket{u^{(1)}} = \ket{10\dots}$ and $C\ket{u^{(2)}} = \ket{01\cdots}$, and it can be generated in exactly the same way as for the Toffoli gadget case. For example:
$$\includegraphics{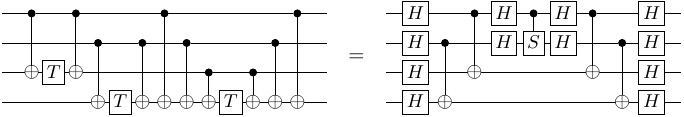}$$
The central \CS gate can be implemented using a \CCZ magic state (see \Cref{app:fig:cs-gadget}), which costs the same as two \tgate{s}, and so the whole gadget can be implemented more efficiently than representing it as a phase-gadget circuit with three \tgate{s}. Note that in both of these gadgets, most of the Hadamard gates can be eliminated by commuting them with the \CNOT{} gates in $C$. 

\subsection{Neural Network Architecture}
\label{app:subsec:neural_network}

As shown in \Cref{fig:nn_overview}, the neural network has three main components: the torso, the policy head, and the value head.

\begin{figure}[ht]
\centering
\includegraphics[width=\textwidth]{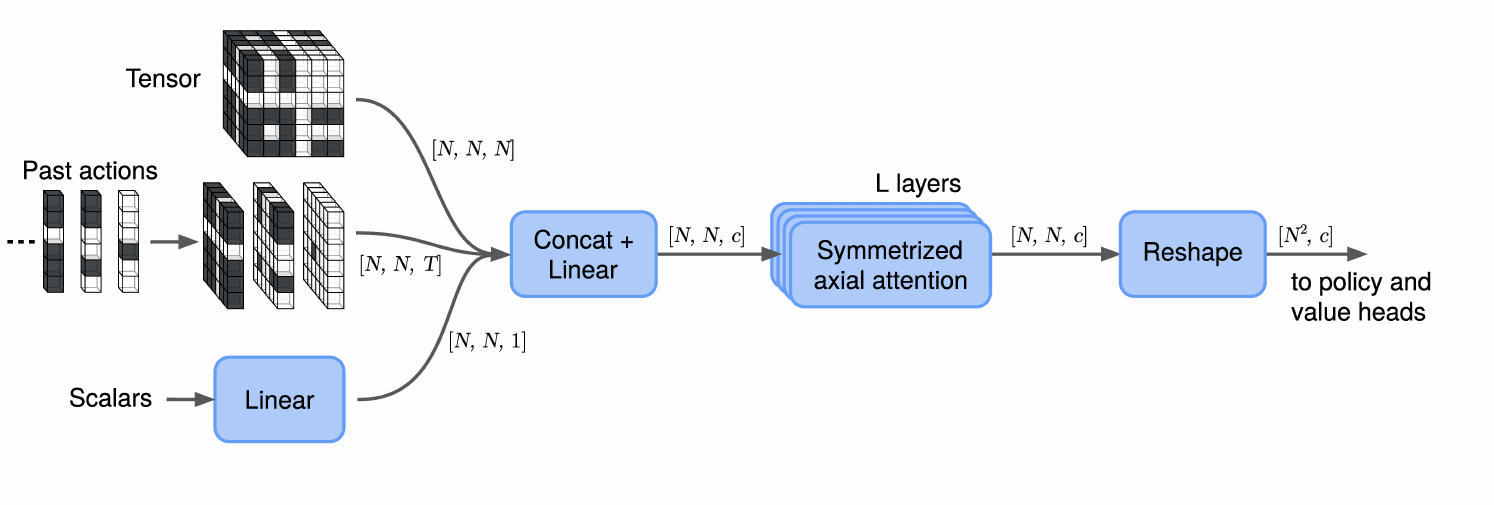}
\caption{Overview of the neural network torso. Its main component is a stack of symmetrized axial attention layers.}
\label{app:fig:torso_overview}
\end{figure}

The input to the neural network is formed from the current game state, which, at step $s$ of TensorGame, is composed of a (residual) tensor $\mathcal{T}^{(s)}$ and a list of all the previously played actions $u^{(s^\prime)}$, with $s^\prime < s$. To form the input to the network, we only consider the last $T=20$ actions; specifically we stack them into a tensor of size $N\times N \times T$, where each $N\times N$ slice is obtained by taking the outer product $u^{(s^\prime)} \otimes u^{(s^\prime)}$. This tensor is concatenated with $\mathcal{T}^{(s)}$, as well as with the output of a linear layer that takes as input a single scalar: the square root of the quotient between $s$ and the maximum number of allowed moves, which is $250$. \Cref{app:fig:torso_overview} shows an overview of the torso.

\begin{figure}[ht]
\centering
\includegraphics[width=0.6\textwidth]{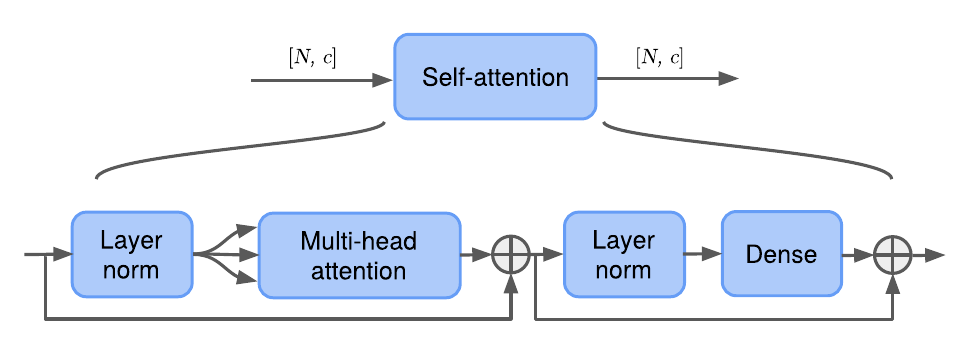}
\caption{Overview of the self-attention operation. It features the standard components of a self-attention layer \citep{vaswani2017attention}, including skip connections, layer normalization, multi-head attention, and a dense (fully connected) network with one hidden layer. When the input is a tensor of size $N\times N\times c$, the self-attention layer is applied independently over each row (or column) of the input.}
\label{app:fig:self_attention}
\end{figure}

The concatenated inputs are first passed through a linear layer that maps from $\mathbb{R}^{N+T+1}$ to $\mathbb{R}^{c}$, with $c=512$. The output of this layer is passed through $L=4$ layers of symmetrized axial attention (see \Cref{fig:symmetrized_axial_attn}), which alternates self-attention operations with symmetrization layers. Each attention operation is depicted in \Cref{app:fig:self_attention} and uses multi-head attention with $16$ heads and head depth $32$. The dense layer in  \Cref{app:fig:self_attention} consists in a feed-forward neural network with a single hidden layer that has twice as many hidden units as the input (and that applies the ReLU non-linearity).

Finally, the output of the torso is reshaped and passed to the policy and value heads (see \Cref{app:fig:torso_overview}). The architecture of both the policy and value heads is taken from \citet{fawzi2022discovering}. In particular, for the policy head, we use an autoregressive model that predicts $\widetilde{N}$ entries of $u^{(s)}$ at a time, where $\widetilde{N}$ divides $N$. For instance, we use $\widetilde{N}=10$ for the experiments from \Cref{subsec:benchmark} for which $N=60$, and $\widetilde{N}=12$ for the experiments on unary iteration for which $N=72$.

\section{Additional Results}
\label{app:sec:additional_results}

\subsection{Additional Benchmark Circuits}
\label{app:subsec:gfmult_circuits}

The benchmark circuits GF($2^2$)-mult and GF($2^3$)-mult are derived from \citet{cheung2008design} in the same way as the rest of the GF($2^m$)-mult ($m \geq 4$) series of circuits available from \citet{amy2016feynman}, using the irreducible polynomials $X^2 + X + 1$ and $X^3 + X + 1$ respectively. We include them to complete the series, since $m = 2$ is the smallest interesting instance ($m = 1$ corresponds to a single Toffoli gate). For reference, the source code is given in OpenQASM 2.0 format as follows:

\begin{minipage}[t]{0.5\textwidth}
    \begin{lstlisting}[caption=GF($2^2$)-mult]
OPENQASM 2.0;
include "qelib1.inc";
qreg a[2];
qreg b[2];
qreg c[2];
ccx a[1], b[1], c[0];
cx c[0], c[1];
ccx a[0], b[0], c[0];
ccx a[1], b[0], c[1];
ccx a[0], b[1], c[1];
    \end{lstlisting}
\end{minipage}\begin{minipage}[t]{0.45\textwidth}
    \begin{lstlisting}[caption=GF($2^3$)-mult]
OPENQASM 2.0;
include "qelib1.inc";
qreg a[3];
qreg b[3];
qreg c[3];
ccx a[1], b[2], c[0];
ccx a[2], b[1], c[0];
ccx a[2], b[2], c[1];
cx c[1], c[2];
cx c[0], c[1];
ccx a[0], b[0], c[0];
ccx a[1], b[0], c[1];
ccx a[0], b[1], c[1];
ccx a[2], b[0], c[2];
ccx a[1], b[1], c[2];
ccx a[0], b[2], c[2];
    \end{lstlisting}
\end{minipage}

\subsection{Benchmarks Circuits with a Larger Number of Qubits}
\label{app:subsec:larger_benchmark_circuits}

\Cref{tab:results_benchmarks_split} reports the \tcount obtained on the circuits from \Cref{subsec:benchmark} that were split into sub-circuits (i.e., the ones marked with $^\star$ in \Cref{tab:results_benchmarks_no_split}), together with the \tcount for each of the sub-circuits. For each sub-circuit, the baseline \tcount was obtained as described in \Cref{app:subsec:baselines}.

\begin{table}[th]
\centering
\footnotesize
\begin{tabular}{l | c | ccc} \toprule
& & & \multicolumn{2}{c}{\method} \\
\multirow{-2}{*}{Circuit} & \multirow{-2}{*}{\#Qubits} & \multirow{-2}{*}{Baselines} & Without gadgets & With gadgets \\ \midrule
\rowcolor{gray!25} $8$-bit adder &  & $129$ {\scriptsize\citep{heyfron2018efficient}} & $139$ & $\mathbf{94}$ \scriptsize{($33$Tof + $28$T)} \\
\hspace{2pt}block 1 & $31$ & $50$ & $\mathbf{48}$ & $\mathbf{36}$ \scriptsize{($11$Tof + $14$T)} \\
\hspace{2pt}block 2 & $50$ & $95$ & $\mathbf{91}$ & $\mathbf{58}$ \scriptsize{($22$Tof + $14$T)} \\
\rowcolor{gray!25} Grover$_{5}$ &  & $44$ {\scriptsize\citep{heyfron2018efficient}} & $152$ & $66$ \scriptsize{($27$Tof + $12$T)} \\
\hspace{2pt}block 1 & $58$ & $108$ & $\mathbf{107}$ & $\mathbf{44}$ \scriptsize{($19$Tof + $6$T)} \\
\hspace{2pt}block 2 & $27$ & $46$ & $\mathbf{45}$ & $\mathbf{22}$ \scriptsize{($8$Tof + $6$T)} \\
\rowcolor{gray!25} Hamming$_{15}$ (high) &  & $787^\dagger$ \scriptsize{($1010$ \cite{heyfron2018efficient})} & $\mathbf{773}$ & $\mathbf{440}$ \scriptsize{($173$Tof + $2$\CS + $90$T)} \\
\hspace{2pt}block 1 & $58$ & $92$ & $92$ & $\mathbf{52}$ \scriptsize{($21$Tof + $10$T)} \\
\hspace{2pt}block 2 & $49$ & $78$ & $78$ & $\mathbf{44}$ \scriptsize{($18$Tof + $8$T)} \\
\hspace{2pt}block 3 & $56$ & $82$ & $\mathbf{81}$ & $\mathbf{50}$ \scriptsize{($18$Tof + $14$T)} \\
\hspace{2pt}block 4 & $57$ & $95$ & $\mathbf{94}$ & $\mathbf{55}$ \scriptsize{($20$Tof + $15$T)} \\
\hspace{2pt}block 5 & $46$ & $66$ & $66$ & $\mathbf{40}$ \scriptsize{($16$Tof + $8$T)} \\
\hspace{2pt}block 6 & $50$ & $75$ & $77$ & $\mathbf{45}$ \scriptsize{($18$Tof + $9$T)} \\
\hspace{2pt}block 7 & $53$ & $88$ & $\mathbf{86}$ & $\mathbf{50}$ \scriptsize{($20$Tof + $10$T)} \\
\hspace{2pt}block 8 & $58$ & $106$ & $\mathbf{100}$ & $\mathbf{53}$ \scriptsize{($20$Tof + $2$\CS + $9$T)} \\
\hspace{2pt}block 9 & $59$ & $105$ & $\mathbf{99}$ & $\mathbf{51}$ \scriptsize{($22$Tof + $7$T)} \\
\rowcolor{gray!25} Hamming$_{15}$ (med) &  & $156^\dagger$ \scriptsize{($162$ \cite{heyfron2018efficient})} & $156$ & $\mathbf{78}$ \scriptsize{($35$Tof + $8$T)} \\
\hspace{2pt}block 1 & $25$ & $38$ & $38$ & $\mathbf{20}$ \scriptsize{($8$Tof + $4$T)} \\
\hspace{2pt}block 2 & $59$ & $118$ & $118$ & $\mathbf{58}$ \scriptsize{($27$Tof + $4$T)} \\
\rowcolor{gray!25} Mod-Adder$_{1024}$ &  & $798^\dagger$ \scriptsize{($978$ \cite{heyfron2018efficient})} & $\mathbf{762}$ & $\mathbf{500}$ \scriptsize{($141$Tof + $15$\CS + $188$T)} \\
\hspace{2pt}block 1 & $54$ & $77$ & $\mathbf{75}$ & $\mathbf{42}$ \scriptsize{($16$Tof + $10$T)} \\
\hspace{2pt}block 2 & $52$ & $71$ & $\mathbf{70}$ & $\mathbf{47}$ \scriptsize{($12$Tof + $23$T)} \\
\hspace{2pt}block 3 & $53$ & $80$ & $\mathbf{76}$ & $\mathbf{48}$ \scriptsize{($16$Tof + $16$T)} \\
\hspace{2pt}block 4 & $55$ & $86$ & $\mathbf{77}$ & $\mathbf{47}$ \scriptsize{($15$Tof + $17$T)} \\
\hspace{2pt}block 5 & $54$ & $75$ & $\mathbf{73}$ & $\mathbf{53}$ \scriptsize{($15$Tof + $5$\CS + $13$T)} \\
\hspace{2pt}block 6 & $52$ & $84$ & $\mathbf{83}$ & $\mathbf{58}$ \scriptsize{($13$Tof + $5$\CS + $22$T)} \\
\hspace{2pt}block 7 & $59$ & $89$ & $\mathbf{87}$ & $\mathbf{60}$ \scriptsize{($19$Tof + $22$T)} \\
\hspace{2pt}block 8 & $56$ & $80$ & $\mathbf{73}$ & $\mathbf{54}$ \scriptsize{($7$Tof + $5$\CS + $30$T)} \\
\hspace{2pt}block 9 & $56$ & $96$ & $\mathbf{89}$ & $\mathbf{58}$ \scriptsize{($17$Tof + $24$T)} \\
\hspace{2pt}block 10 & $53$ & $60$ & $\mathbf{59}$ & $\mathbf{33}$ \scriptsize{($11$Tof + $11$T)} \\
\rowcolor{gray!25} QCLA-Adder$_{10}$ &  & $116$ \scriptsize{\cite{heyfron2018efficient}} & $135$ & $\mathbf{94}$ \scriptsize{($28$Tof + $5$\CS + $28$T)} \\
\hspace{2pt}block 1 & $37$ & $53$ & $53$ & $\mathbf{43}$ \scriptsize{($12$Tof + $19$T)} \\
\hspace{2pt}block 2 & $48$ & $83$ & $\mathbf{82}$ & $\mathbf{51}$ \scriptsize{($16$Tof +$5$\CS + $9$T)} \\
\rowcolor{gray!25} QCLA-Mod$_{7}$ &  & $165$ \scriptsize{\cite{heyfron2018efficient}} & $199$ & $\mathbf{122}$ \scriptsize{($43$Tof + $36$T)} \\
\hspace{2pt}block 1 & $50$ & $95$ & $95$ & $\mathbf{60}$ \scriptsize{($21$Tof + $18$T)} \\
\hspace{2pt}block 2 & $59$ & $107$ & $\mathbf{104}$ & $\mathbf{62}$ \scriptsize{($22$Tof + $18$T)} \\
\bottomrule
\end{tabular}
\caption{%
    \tcount achieved by different methods on a set of benchmark circuits, after splitting them into sub-circuits (blocks). \method finds good trade-offs between gadgets and \tgate{s} and outperforms the baseline methods for all circuits except Grover$_5$. The \tcount from \method is obtained by additively combining all sub-circuits. The baseline \tcount for the sub-circuits is given by a combination of methods (see \Cref{app:subsec:baselines}). For the shaded rows, the baseline corresponds to the lowest between the best reported value in the literature and the sum of the \tcount across sub-circuits (the symbol $^\dagger$ indicates it is the latter case and also includes in parenthesis the best reported value in the literature, for completeness). The notation ``$a$Tof$+b$\CS+$c$T'' indicates $a$ Toffoli gates, $b$ \CS gates, and $c$ \tgate{s}. Bold font highlights results with strictly better \tcount than the baselines. %
    \label{tab:results_benchmarks_split}
}
\end{table}

\subsection{Neural Network Ablations}
\label{app:subsec:nn_ablations}

Here we compare different architectures for the torso of the neural network (see \Cref{subsec:system_description,app:subsec:neural_network}).

\paragraph{Comparison across architectures.}
We first compare the following: (i) \emph{attentive modes} \citep{fawzi2022discovering}, which uses attention operations across pairs of modes of the input tensor; (ii) \emph{axial attention} \citep{ho2019axial}, which we recover when we eliminate the symmetrization layers from our architecture in \Cref{fig:symmetrized_axial_attn}; and (iii) \emph{symmetrized axial attention}, which is the architecture that we introduce in \Cref{subsec:system_description}, featuring symmetrization layers.

We run each architecture on a \emph{supervised} experiment with synthetic data only (i.e., a system without actors), using tensors of size $N=30$, and compare the loss that they achieve.
For each of the methods, we set the number of layers $L=4$. For axial attention and symmetrized axial attention, this means $8$ (batched) attention operations with $N$ tokens each; for attentive modes, this means $12$ (batched) attention operations with $2N$ tokens each. Therefore, we expect the architecture of \citet{fawzi2022discovering} to run slower, and in fact we verify that its runtime is about $2$x that of the other architectures: attentive modes runs at about $6$ iterations per second, while axial attention and symmetrized axial attention complete $11.8$ and $11.4$ iterations per second, respectively.

\begin{figure}
    \centering
    \begin{subfigure}{0.45\textwidth}
        \includegraphics[width=\textwidth]{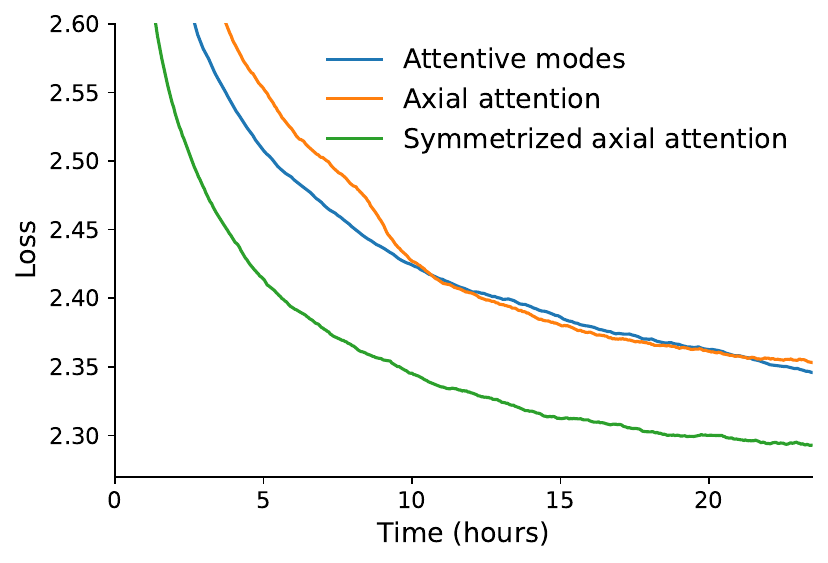}
        \caption{Policy loss (lower is better).}
        \label{app:fig:policy_loss}
    \end{subfigure}
    \hspace{0.03\textwidth}
    \begin{subfigure}{0.45\textwidth}
        \includegraphics[width=\textwidth]{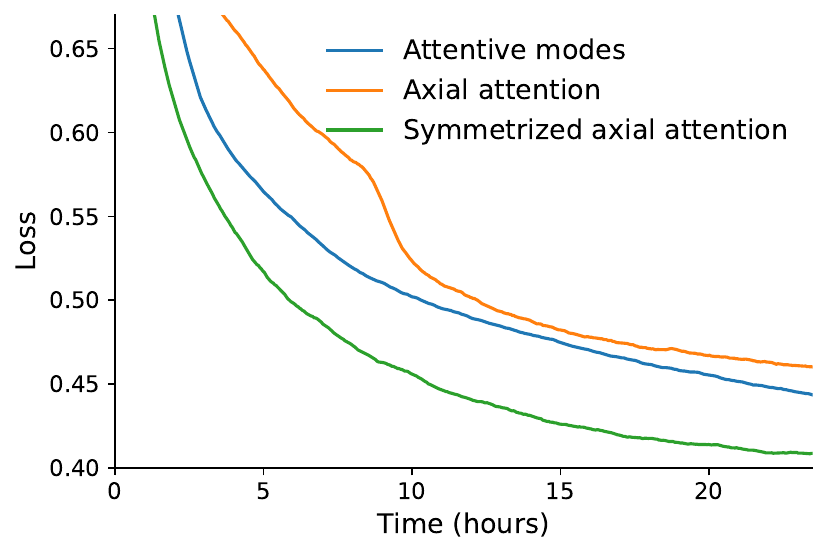}
        \caption{Value loss (lower is better).}
        \label{app:fig:value_loss}
    \end{subfigure}
    \caption{Comparison of different architectures for the torso of the neural network on a supervised experiment on $30\times 30\times 30$ tensors. Symmetrized axial attention performs best. The reported policy loss is the average cross-entropy loss for the first step of the autoregressive model. (For better visualization, all curves have been smoothed using an average moving window of size $20$.)}
    \label{app:fig:ablations}
\end{figure}

We report in \Cref{app:fig:ablations} the training loss from each head of the neural network (the loss on a test set performs similarly). While axial attention and symmetrized axial attention run at approximately the same speed, the symmetrization layers bring a significant boost in performance. In contrast, although the architecture based on attentive modes performs almost as good as symmetrized axial attention if ran for longer, its decreased speed make it an impractical choice for \method (it also presents memory issues when applied on larger tensors).

\paragraph{Comparison of symmetrization layers.} We now consider two different choices for the symmetrization layers. Recall that the symmetrization operation performs the operation $X\leftarrow \beta \odot X + (1-\beta)  \odot X^{\top}$. We refer to the choice of $\beta=\frac{1}{2}$ as \emph{fixed scalar}, in contrast to letting $\beta\in\mathbb{R}^{N\times N}$ be a \emph{learnable matrix}.

\begin{figure}
    \centering
    \begin{subfigure}{0.45\textwidth}
        \includegraphics[width=\textwidth]{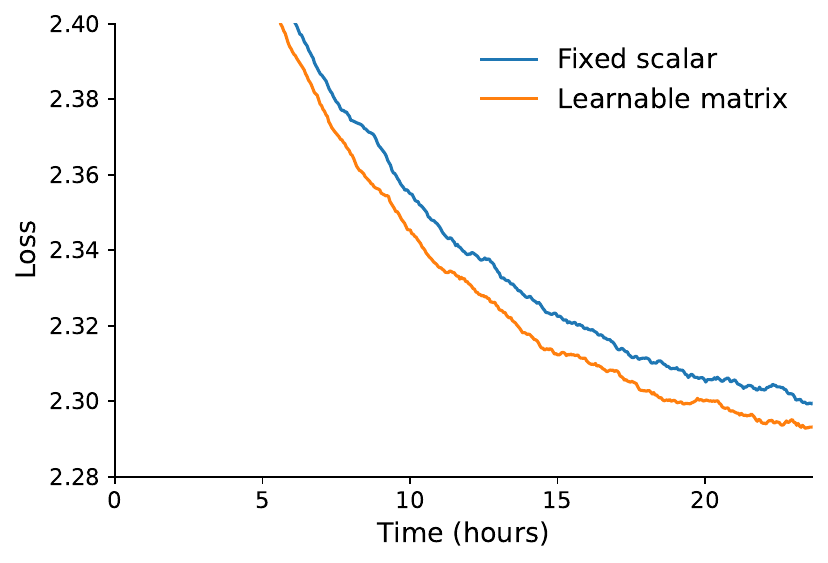}
        \caption{Policy loss (lower is better).}
        \label{app:fig:policy_loss_sym}
    \end{subfigure}
    \hspace{0.03\textwidth}
    \begin{subfigure}{0.45\textwidth}
        \includegraphics[width=\textwidth]{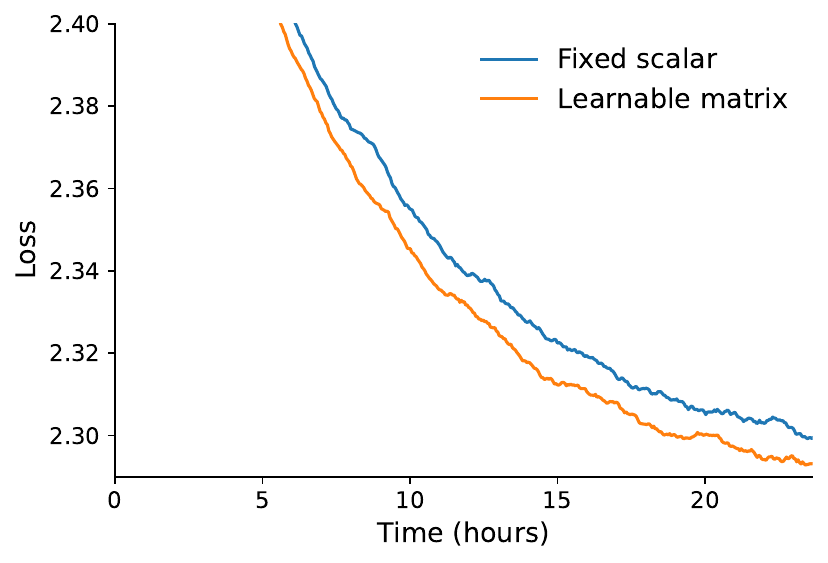}
        \caption{Value loss (lower is better).}
        \label{app:fig:value_loss_sym}
    \end{subfigure}
    \caption{Comparison of different versions of the symmetrization operation, on a supervised experiment on $30\times 30\times 30$ tensors. Symmetrization layers with learnable parameters lead to some performance improvement. The reported policy loss is the average cross-entropy loss for the first step of the autoregressive model. (For better visualization, all curves have been smoothed using an average moving window of size $20$.)}
    \label{app:fig:ablations_sym}
\end{figure}

We compare both versions of symmetrization in \Cref{app:fig:ablations_sym}, where we report the training loss. Using a learnable matrix provides a small improvement over a fixed scalar (both choices run at approximately the same speed of $11.4$ iterations per second).
Letting $\beta$ be a learnable matrix increases the number of model parameters by a negligible fraction (e.g., in this scenario, it adds $3\,600$ parameters to a model of approximately $132.7$ million parameters).

\end{document}